\documentclass[usenatbib]{mnras}

\usepackage{newtxtext,newtxmath}
\def\plus{\texttt{\textbf{+}}}
\def\cross{\textsf{\textbf{\scriptsize{x}}}}
\def\NFW{\textsc{nfw}}
\def\BMO{\textsc{bmo}}

\usepackage[T1]{fontenc}
\usepackage{ae,aecompl}

\usepackage{graphicx}	
\usepackage{amsmath}	
\usepackage[symbolgreek]{mathastext} 

\title[Halo bias in galaxy clusters]{AMICO galaxy clusters in KiDS-DR3: Measurement of the halo bias and power spectrum normalization from a stacked weak lensing analysis}

\author[Ingoglia et al.]{Lorenzo Ingoglia$^{1,2,3,}$\thanks{E-mail: \href{mailto:lorenzo.ingoglia@unina.it}{lorenzo.ingoglia@unina.it}}, 
Giovanni Covone$^{1,2,3}$, 
Mauro Sereno$^{5,6}$, 
Carlo Giocoli$^{5,4,6}$, 
\newauthor
Sandro Bardelli$^{5}$, 
Fabio Bellagamba$^{4}$, 
Gianluca Castignani$^{4,5}$, 
Samuel Farrens$^{11}$, 
\newauthor
Hendrik Hildebrandt$^{14,15}$, 
Shahab Joudaki$^{16, 17}$, 
Eric Jullo$^{10}$, 
Denise Lanzieri$^{11,1}$, 
\newauthor
Giorgio F. Lesci$^{4,5}$, 
Federico Marulli$^{4,5,6}$, 
Matteo Maturi$^{12,13}$, 
Lauro Moscardini$^{4,5,6}$, 
\newauthor
Lorenza Nanni$^{4,18}$, 
Emanuela Puddu$^{2}$, 
Mario Radovich$^{7}$, 
Mauro Roncarelli$^{4}$, 
\newauthor
Feliciana Sapio$^{8,9,1}$, 
Carlo Schimd$^{10}$
\\
$^{1}$ Department of Physics - University of Naples ``Federico II", Via Cinthia, 80126 Fuorigrotta, Napoli, Italy \\
$^{2}$ INAF - Capodimonte Astronomical Observatory, Salita Moiariello 16, 80131 Napoli, Italy \\
$^{3}$ INFN - Naples Section, Strada Comunale Cinthia, 80126 Napoli, Italy \\
$^{4}$ Department of Physics and Astronomy "A. Righi" - University of Bologna, Via Piero Gobetti 93/2, 40129 Bologna, Italy \\
$^{5}$ INAF - Astrophysics and Space Science Observatory of Bologna, Via Piero Gobetti 93/3, 40129 Bologna, Italy \\
$^{6}$ INFN - Bologna Section, Viale Berti Pichat 6/2, 40127 Bologna, Italy \\
$^{7}$ INAF - Padua Astronomical Observatory, Vicolo dell'Osservatorio, 5, 35122 Padova, Italy \\
$^{8}$ Department of Physics - University of Roma ``La Sapienza", Piazzale Aldo Moro, 2, 00185 Roma, Italy \\
$^{9}$ INAF - Istituto di Astrofisica e Planetologia Spaziali (IAPS), Via del Fosso del Cavaliere, 100, 00133 Roma, Italy \\
$^{10}$ Aix-Marseille University, CNRS, CNES, LAM, Marseille, France \\
$^{11}$AIM, CEA, CNRS, Universit\'{e} Paris-Saclay, Universit\'{e} de Paris, Sorbonne Paris Cit\'{e}, F-91191 Gif-sur-Yvette, France \\
$^{12}$ Center for Astronomy - University of Heidelberg, Albert-Ueberle-Stra{\ss}e 2, 69120 Heidelberg, Germany \\
$^{13}$ Institute of Theoretical Physics - University of Heidelberg, Albert-Ueberle-Stra{\ss}e 2, 69120 Heidelberg, Germany \\
$^{14}$ Institute of Astronomy - Ruhr University of Bochum, Universit{\"a}tsstr. 150, 44801, Bochum, Germany \\
$^{15}$ Argelander Institute of Astronomy - University of Bonn, Auf dem H{\"u}gel 71, 53121 Bonn, Germany \\
$^{16}$ Waterloo Centre for Astrophysics - University of Waterloo, 200 University Ave W, Waterloo, ON N2L 3G1, Canada \\
$^{17}$ Department of Physics and Astronomy - University of Waterloo, 200 University Ave W, Waterloo, ON N2L 3G1, Canada \\
$^{18}$ Institute of Cosmology and Gravitation - University of Portsmouth, Dennis Sciama Building, Portsmouth, PO1 3FX, UK
}

\date{Accepted XXX. Received YYY; in original form ZZZ}

\pubyear{2021}

\begin{document}
\label{firstpage}
\pagerange{\pageref{firstpage}--\pageref{lastpage}}

\maketitle

\begin{abstract}
Galaxy clusters are biased tracers of the underlying matter density field. At very large radii beyond about 10 Mpc/\textit{h}, the shear profile shows evidence of a second-halo term. This is related to the correlated matter distribution around galaxy clusters and proportional to the so-called halo bias. 
We present an observational analysis of the halo bias-mass relation based on the AMICO galaxy cluster catalog, comprising around 7000 candidates detected in the third release of the KiDS survey. We split the cluster sample into 14 redshift-richness bins and derive the halo bias and the virial mass in each bin by means of a stacked weak lensing analysis.
The observed halo bias-mass relation and the theoretical predictions based on the $\Lambda$CDM standard cosmological model show an agreement within $2\sigma$. The mean measurements of bias and mass over the full catalog give $M_{200c} = (4.9 \pm 0.3) \times 10^{13} M_{\odot}/\textit{h}$ and $b_h \sigma_8^2 = 1.2 \pm 0.1$. With the additional prior of a bias-mass relation from numerical simulations, we constrain the normalization of the power spectrum with a fixed matter density $\Omega_m = 0.3$, finding $\sigma_8 = 0.63 \pm 0.10$.
\end{abstract}




\section{Introduction}
\label{sec:Introduction} 

Clusters of galaxies occupy a special place in the hierarchy of cosmic structures as they are the most massive gravitationally bound systems in the Universe. According to the hierarchical scenario of the evolution of cosmic structure \citep{1980lssu.book.....P, 2005RvMP...77..207V}, they arise from the collapse of initial density perturbations having a typical comoving scale of about 10 Mpc/\textit{h} \citep{1993ppc..book.....P, 2008LNP...740..287B}. Above these scales, gravitational clustering is essentially in a linear regime and the dynamics are mostly driven by the Hubble flow, while the non-linear regime is prominent on smaller scales. Moreover, in the inner cluster regions, astrophysical processes such as gas cooling, star formation, feedback from supernovae and active galactic nuclei modify the evolution of the halo properties like, the density profile, the subhalo mass function, etc. \citep{rasia04,rasia06,giocoli10a,despali14,despali16,angelinelli20}. Galaxy clusters thus provide an ideal tool to study the physical mechanisms driving the formation and evolution of cosmic structures in the mildly non-linear regime \citep{tormen98a,springel01b}.
 
Massive galaxy clusters, composed of a large amount of dark matter \citep[about 85\%, see e.g.][]{1978MNRAS.183..341W}, are expected to grow at the highest peaks of the underlying matter distribution. This establishes a clear correlation between the galaxy cluster mass and the underling matter clustering amplitude. As already shown by \cite{1984ApJ...284L...9K}, the enhanced clustering of Abell galaxy clusters is explained by assuming that they form in the high-density regions. As a consequence, galaxy clusters are biased tracers of the background matter field. Several groups have further developed this idea within the framework of the \cite{PS...1974ApJ...187..425P} formalism \citep[e.g.,][]{1996MNRAS.282..347M, 1999MNRAS.308..119S, 2001MNRAS.323....1S,giocoli10b}, deriving quantitative predictions for the correlation between the halo density field and the underlying matter distribution within the hierarchical scenario for the formation of cosmic structures. 
The relation between the cluster dark matter halo density contrast, $\delta_h$, and the 
dark matter density contrast in the linear regime, $\delta_m$, is described by the so-called halo bias parameter, $b_{h}$, defined as \citep{2010ApJ...724..878T}
\begin{equation}
b_{h} = \delta_h / \delta_m \ .
\label{eq:halo bias}
\end{equation}
Measurements of the halo bias as a function of the halo mass therefore represent an important test for cosmological models. 

The total matter distribution of a galaxy cluster can be broken down in a ``one-halo" term, which determines its halo matter component on scales smaller than the halo virial radius, and a ``two-halo" term for the correlated matter of the surrounding structures, which is prominent on scales much larger than the virial radius. 
The first component is usually identified with the galaxy cluster halo and can be described by a Navarro-Frenk-White dark matter profile \citep[][]{1997ApJ...490..493N}. The second component, directly proportional to the halo bias, stems from mass elements in distinct pairs of halos. The two terms of the halo profile correlate in such a way that the bias follows an increasing function of mass \citep[][]{1984ApJ...284L...9K, 1989MNRAS.237.1127C, 1996MNRAS.282.1096M}. This relation has been shown and modeled in several studies based on \textit{N}-body numerical simulations  \citep[e.g.][]{2004MNRAS.355..129S, 2005ApJ...631...41T, 2010ApJ...724..878T}. 

Weak gravitational lensing (WL) is a suitable approach to investigate the halo model and to measure its major parameters: the mass and the bias. Gravitational lensing relates the deflection of light to the mass distribution along the line-of-sight. As gravitational lensing is based on the very well-tested theory of general relativity and does not rely on the hypothesis of dynamical equilibrium, it allows robust measurements of the mass of cosmic structures and cosmological parameters. WL by galaxy clusters is detected via statistical measurement of source galaxy shears, and provides an efficient way to derive mass density profiles without requiring any assumption about their composition or dynamical state. 
For example, WL analysis allows us to reach scales up to $\sim 30 \ Mpc/\textit{h}$ from the center and therefore to directly measure the halo bias \citep{2014ApJ...784L..25C}.

Stacking the shear measurements of cluster background galaxies is a common practice to increase the lensing signals and compensate for the typical low signal-to-noise ratio (SNR hereafter) in the shear profiles of individual galaxy clusters \citep[see for instance][]{2013MNRAS.434..878S}.
This method also makes it possible to arrange the stacked density profiles as a function of the cluster properties, such as their redshift or their richness.

Several authors have probed the dependence of the halo bias on mass \citep{2005PhRvD..71d3511S, 2007arXiv0709.1159J, 2014ApJ...784L..25C, 2015MNRAS.449.4147S, 2016A&A...586A..43V}. These studies have obtained results consistent with the theoretical predictions, but the large uncertainty in the measurements did not allow them to discriminate between different theoretical models. Moreover, recently \citet{2018NatAs...2..744S} found a peculiar galaxy cluster at $z \sim 0.62$ in the PZS2LenS sample \citep[][]{2017MNRAS.472.1946S} showing an extreme value of the halo bias, well in excess of the theoretical predictions. This result motivates further observational work in order to probe with higher accuracy the halo bias-mass relation. Large sky surveys providing deep and high-quality photometric data and reliable catalogs of galaxy clusters are essential. 

In this work we perform a novel measurement of the bias-mass relation by using the photometric data from the third data release of KiDS \citep[][]{2013ExA....35...25D, 2017A&A...604A.134D} and the galaxy cluster catalog identified using the Adaptive Matched Identifier of Clustered Objects detection algorithm \citep[AMICO,][]{2018MNRAS.473.5221B}. This catalog is optimal for a stacked WL analysis because of its large size (an effective area of 360.3 square degrees) and its dense field (an effective galaxy number density of $n_{eff} = 8.53 \ arcmin^{-2}$), which allows us to split the stacked WL signal into different bins of cluster redshift and richness while keeping a sufficiently high SNR in each of them. KiDS images are deep enough (limiting magnitudes are 24.3, 25.1, 24.9, 23.8 in \textit{ugri}, respectively) to include numerous sources (almost 15 million) and large enough to compute the profile up to the scales where the bias dominates. This study is part of a series of papers based on AMICO galaxy clusters in the third data release of KiDS. Previous and ongoing publications have presented the detection algorithm \citep{2018MNRAS.473.5221B}, the cluster catalog \citep{2019MNRAS.485..498M}, the calibration of WL masses \citep{2019MNRAS.484.1598B}, and constraints on cosmological parameters otained from cluster counts \citep{2020arXiv201212273L}, WL \citep{2021arXiv210305653G} and cluster clustering \citep{nanni}. 

Following the method explained in \cite{2019MNRAS.484.1598B}, we derive mass density profiles from almost 7000 clusters, which is among the largest cluster samples for this kind of analysis. We stack the lensing signal in richness and redshift cluster bins, calibrate the halo parameters and investigate the mass-bias relation. Throughout this paper we assume a spatially flat $\Lambda$CDM model with the following matter, dark energy and baryonic density parameters at the present time $\Omega_m=1-\Omega_{\Lambda}=0.3$, $\Omega_b=\Omega_m-\Omega_c=0.05$ and Hubble parameter $H_0 = 100\textit{h} \ km \ s^{-1} \ Mpc^{-1}$ with $\textit{h}=0.7$.

\section{Data}
\label{sec:Data}  

For an accurate lensing signal, we have to look for deep and dense source samples in such way that the statistical number of background sources increases while the contamination of foreground and cluster member galaxies is small.

Our work is based on the optical wide-field imaging Kilo-Degree Survey \citep[KiDS,][]{2013ExA....35...25D}, split into  an equatorial stripe (KiDS-N), and a second one centered around the South Galactic Pole (KiDS-S). The survey encompasses four broad-band filters (\textit{ugri}) managed by the OmegaCAM wide-field imager \citep{2011Msngr.146....8K}, presently located on the VLT Survey Telescope \citep[VST,][]{2011Msngr.146....2C}. The data set we use for this work is the Data Release 3\footnote{\url{http://kids.strw.leidenuniv.nl/DR3}} \citep[DR3,][]{2017A&A...604A.134D} and covers a total area of approximately 450 deg$^2$ in five patches following the GAMA survey convention \citep[][G9/G12/G15 within KiDS-N and G23/GS within KiDS-S]{2011MNRAS.413..971D}. This intermediate release includes one third of the final KiDS area, which will ultimately reach 1350 deg$^2$.

\subsection{Cluster catalog}
\label{sec:Cluster catalog} 

We use the galaxy cluster catalog obtained from the application of the Adaptive Matched Identifier of Clustered Objects algorithm \citep[AMICO,][]{2018MNRAS.473.5221B} on KiDS DR3 data (AK3, hereafter). AMICO was selected to form part of the \textit{Euclid} analysis pipeline \citep[][]{2019A&A...627A..23E}. The algorithm exploits the Optimal Filtering technique \citep{2005A&A...442..851M, 2011MNRAS.413.1145B} and aims at maximising the SNR for the detection of objects following a physical model for clusters. Specifically, it identifies overdensities of galaxies associated with galaxy clusters taking into account their spatial, magnitude, and photometric redshift distributions \citep{2017A&A...598A.107R}. 

The AK3 catalog is fully described in \cite{2019MNRAS.485..498M}. It contains 7988 candidate galaxy clusters covering an effective area of 377 deg$^2$. Clusters are detected above a fixed threshold of $SNR=3.5$. AK3 encompasses an intrinsic richness (defined as the sum of membership probabilities below a consistent radial and magnitude threshold across redshift) range of $2 < \lambda_{\ast} < 140$ and a redshift range $0.1 \leq z < 0.8$. 
The richness and redshift distributions are presented in Figure~\ref{fig:distribution}. From the figure we can see that the richness slightly increases with redshift.
Conversely, poor and distant clusters are not detected due to their low $SNR$.
These blank regions are usually associated to low levels of completeness (i.e. the fraction between detected and mock galaxy clusters), as shown in Figure~13 of \cite{2019MNRAS.485..498M}.

\subsection{Shear catalog}
\label{sec:Galaxy catalog} 

\begin{figure}
	\includegraphics[width=\columnwidth]{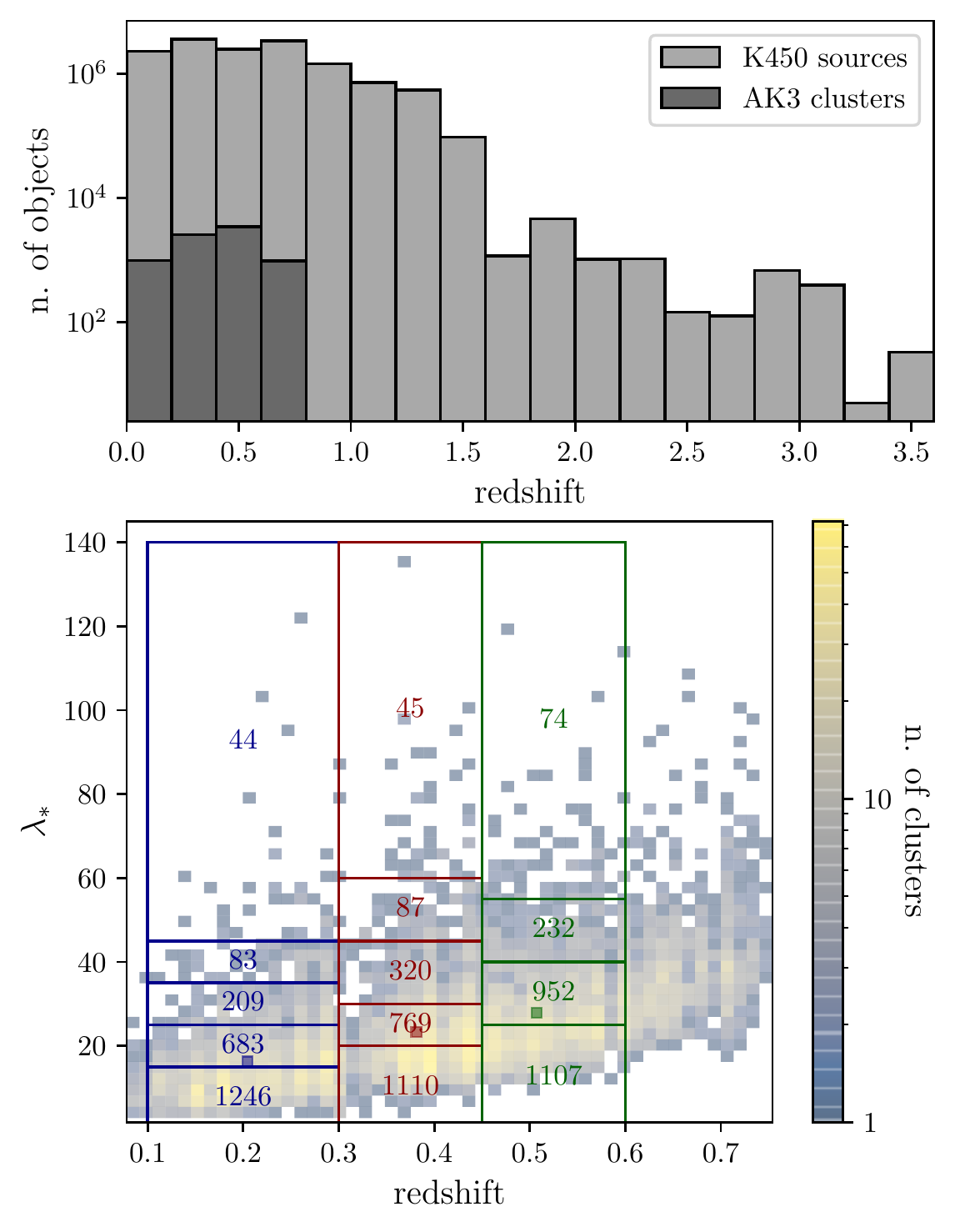}
    \caption{\textit{Top panel}: Redshift distributions of AK3 clusters (dark gray) and K450 galaxies (light gray). \textit{Bottom panel}: AK3 clusters in the redshift-richness plane with $SNR \geq 3.5$. Colored rectangles correspond to the redshift-richness bins used in the following analysis (see Section~\ref{sec:Shear data stacked in bins}); the number of clusters enclosed in each bin is displayed. Single colored squares show the mean values in each redshift bin computed as in Equation~\eqref{eq:cluster bin stacking}.}
    \label{fig:distribution}
\end{figure}

The halo lensing signal relies on the selection of background galaxies relative to galaxy clusters. 
\cite{2017MNRAS.465.1454H} presented a complete tomographic cosmic shear analysis of the KiDS-450 catalog (K450), updated from earlier works on KiDS-DR1 and -DR2 \citep{2015A&A...582A..62D, 2015MNRAS.454.3500K}. 
The shear is estimated using the \textit{lens}fit likelihood based model-fitting method \citep{2007MNRAS.382..315M, 2013MNRAS.429.2858M, 2008MNRAS.390..149K, 2017MNRAS.467.1627F} on galaxy \textit{r}-band images for which the best-seeing dark time is reserved. Photometric redshifts are derived from K450 galaxy photometry in the \textit{ugri}-bands. They are estimated with a Bayesian code \citep[BPZ,][]{2000ApJ...536..571B} following the methods used for CFHTLenS data in \cite{2012MNRAS.421.2355H}. 
The redshift distribution of the galaxies is shown on the top panel of Figure~\ref{fig:distribution} in light-gray.

The survey covers 454 tiles, which after masking overlapping tiles, provides an effective area of $360.3 \ deg^2$. It comprises 14,650,348 sources and has an effective number density \citep[as defined in][]{2012MNRAS.427..146H} of $n_{eff}=8.53 \ arcmin^{-2}$.

\section{Method}
\label{sec:Method} 

In this section we provide a short introduction to the WL formalism. We then describe the numerical method to derive the WL signal of galaxy clusters from the shapes of background sources. We discuss the selections of lens-source pairs that improve the stacked measurement and remove those for which the shear distorts the final signal. Finally, we stack the individual lens shear profiles in bins of cluster redshift and richness for an accurate measurement of the halo parameters.

\subsection{Weak-lensing formalism}
\label{sec:Weak-lensing formalism} 

In gravitational lensing, 
the matter distribution curves space-time and modifies the the path of light rays from background sources, manifesting in a distortion of their intrisic shape. Shape distortion yields isotropic or anisotropic deformation, called convergence, $\kappa$, and shear, $\gamma$, respectively. The tangential component of the shear $\gamma_{\plus}$ encodes the density of the intervening matter distributed between the source and us. Massive objects such as galaxy clusters are therefore dominant in the information that $\gamma_{\plus}$ encapsulates, as we will present later. For a review, see e.g. \cite{2001PhR...340..291B, Schneider:2005ka, 2015RPPh...78h6901K}. 

The source shape distortion can be expressed in terms of the deflection potential $\psi$. It is described by the Jacobian matrix through the second derivatives of the potential, $\psi_{ij} \equiv \partial_i \partial_j \psi$
\begin{equation}
\mathcal{A} \equiv \left( \delta_{ij} - \psi_{ij} \right) =
\begin{pmatrix}
1-\kappa-\gamma_1 & -\gamma_2 \\
-\gamma_2 & 1-\kappa+\gamma_1
\end{pmatrix}
\ ,
\label{eq:jacobian matrix}
\end{equation}
in which the convergence $\kappa$ is defined by the Poisson equation $\bigtriangledown^2 \psi \equiv 2\kappa$ and the complex shear $\gamma \equiv \gamma_1 + i \gamma_2$ is given by $\gamma_1 = \frac{1}{2} \left( \psi_{11} - \psi_{22} \right)$ and $\gamma_2 = \psi_{12}$.

Sources initially have an intrinsic unlensed ellipticity $\epsilon_s$, which is converted by cosmic shear into the observed ellipticity $\epsilon$. One describes this deformed ellipse by its minor and major axes $( a, b )$, and from the position angle $\phi$ of the source relatively to the lens, $\epsilon = \vert \epsilon \vert e^{2i\phi}$, where $\vert \epsilon \vert = (a-b) / (a+b)$. 

It is convenient to factor out the multiplicative term $(1-\kappa)$ from Equation~\eqref{eq:jacobian matrix} and thereby introduce the reduced shear observable $g \equiv \gamma / (1 - \kappa)$ and its conjugate version $g^{\ast}$. Considering $\vert g \vert \leq 1$, \cite{1997A&A...318..687S} relate shear and ellipticity by
\begin{equation}
\epsilon = \frac{\epsilon_s + g}{1 + g^{\ast} \epsilon_s} \ .
\label{eq:ellipticity-shear}
\end{equation}
In the WL limit $\gamma \ll 1$ and $\kappa \ll 1$, yielding $\epsilon \approx \epsilon_s + g$. Assuming that sources are randomly oriented, their complex intrinsic ellipticities average to zero, so $\left\langle \epsilon \right\rangle = \left\langle \gamma \right\rangle$. Therefore, the average ellipticity of background galaxies is a direct observable of the shear induced by foreground matter. 

The two components of the complex shear are defined relative to a local Cartesian space and are conveniently decomposed into a tangential and a cross component,
\begin{equation}
\begin{aligned}
& \gamma_{\plus} = - \Re \left( \gamma e^{-2i\phi} \right) = - \left( \gamma_1 \cos 2 \phi + \gamma_2 \sin 2 \phi \right) \ ,
\\ 
& \gamma_{\cross} = - \Im \left( \gamma e^{-2i\phi} \right) = - \left( \gamma_2 \cos 2 \phi - \gamma_1 \sin 2 \phi \right) \ ,
\end{aligned}
\label{eq:cross-tangential shear}
\end{equation}
respectively. Noticing the minus sign in the exponential, it is agreed that for an axially symmetric mass distribution the tangential component returns a positive value around an overdensity, while a negative value characterizes underdensities. On the other hand, the cross component of the shear does not hold any mass information, and thus averages to zero, in the absence of systematic uncertainties. It is possible to relate the shear to a physical quantity, the excess surface mass density $\Delta\Sigma$, as \citep{2004AJ....127.2544S}
\begin{equation}
\Delta\Sigma (R) \equiv \overline{\Sigma}(< R) - \Sigma (R) = \Sigma_{cr} \gamma_{\plus} (R) \ ,
\label{eq:excess surface mass density}
\end{equation}
where $\Sigma (R)$ is the surface mass density and $\overline{\Sigma}(< R)$ its mean value within the projected radius $R$, and $\Sigma_{cr}$ is the critical surface mass density, 
given by
\begin{equation}
\Sigma_{cr} \equiv \frac{c^2}{4 \pi G} \frac{D_s}{D_l D_{ls}} \ ,
\label{eq:critical density}
\end{equation}
where $c$ is the speed of light, $G$ is the gravitational constant and $D_s$, $D_l$ and $D_{ls}$ are the angular diameter distances from the observer to the source, from the observer to the lens and from the lens to the source, respectively.

The reduced shear is a more direct observable than the shear, which remains an approximation of the source ellipticities. However, the reduced shear is not directly included in the definition of the differential excess surface density, so we link these two quantities using $\kappa \equiv \Sigma / \Sigma_{cr}$ in Equation~\eqref{eq:excess surface mass density} and derive
\begin{equation}
g_{\plus} = \frac{\Delta\Sigma}{\Sigma_{cr} - \Sigma} \ .
\label{eq:g+}
\end{equation}

\subsection{Measurement of the lensing signal}
\label{sec:Measurement of the lensing signal} 

Since the ellipticity is an indirect observable of the shear, we denote the corresponding excess surface mass density for $\Sigma_{cr} \epsilon_{\plus / \cross}$ as $\widetilde{\Delta\Sigma}_{\plus / \cross}$. We compute the lensing signal at a given distance from the cluster center by stacking the radial position and the ellipticity of the $i$-th galaxy source over the $j$-th radial annulus. Thereby, we assess the two observables using their weighted mean
\begin{equation}
R_j = \left( \frac{\sum_{i \in j} w_{ls, i} R_i^{-\alpha} }{\sum_{i \in j} w_{ls, i} } \right)^{-1 / \alpha}; \ \widetilde{\Delta\Sigma}_j = \left( \frac{\sum_{i \in j} w_{ls, i} \Sigma_{cr, i} \epsilon_i}{\sum_{i \in j} w_{ls, i}} \right) \frac{1}{1 + K_j} \ , 
\label{eq:weighted radius delta sigma}
\end{equation}
where the lens-source weight of the $i$-th source  is $w_{ls, i} = w_{s, i} \Sigma_{cr, i}^{-2}$ and $w_{s, i}$ is the inverse-variance source weight as defined in \cite{2013MNRAS.429.2858M}. Here, $K_j$ is the weighted mean of the \textit{lens}fit multiplicative bias $m_i$ introduced to calibrate the shear \citep[see][]{2017MNRAS.467.1627F},
\begin{equation}
K_j= \frac{\sum_{i \in j} w_{ls, i} m_i }{\sum_{i \in j} w_{ls, i}} \ .
\label{eq:weighted multiplicative bias}
\end{equation}
The effective radius is estimated with a shear-weighted mean and computed by approximating the shear profile as a power-law, with $\alpha=1$. \cite{2017MNRAS.472.1946S}, which explored different methods to assess the mean radius, found that this configuration is less dependent on the binning scheme. We compute the average inverse surface critical density to derive the effective redshift of the background sources $z_{back}$ in each radial bin \citep{2017MNRAS.472.1946S}
\begin{equation}
\Sigma_{cr}^{-1}(z_{back}) = \frac{\sum_{i \in j} w_{s, i} \Sigma_{cr, i}^{-1}}{\sum_{i \in j} w_{s, i}} \ .
\label{eq:effective redshift}
\end{equation}
This estimate permits us to compute the modelled reduced shear in Equation~\eqref{eq:g+} as further described in Section~\ref{sec:Halo model}.

A preliminary measurement of the statistical errors of the two observables in Equation~\eqref{eq:weighted radius delta sigma} is given by the weighted standard deviation of the radial distances and by the standard error of the weighted mean, i.e.
\begin{equation}
\sigma_{R, j}^2 = \frac{\sum_{i \in j} w_{ls, i} \left( R_i - R_j \right)^2 }{\sum_{i \in j} w_{ls, i}}; \quad \sigma_{\widetilde{\Delta\Sigma}, j}^2 = \frac{1}{\sum_{i \in j} w_{ls, i}} \ ,
\label{eq:error}
\end{equation}
respectively.
A more complete way to assess the uncertainty given by the averaged signal is to compute the covariance matrix as in Appendix~\ref{sec:Covariances}. This statistical measurement of the noise includes the errors which propagate among the bins.

In the following, we provide lensing profiles sampled in 30 annuli corresponding to 31 logarithmically equi-spaced radii in the range $\left[ 0.1, 30 \right] \ Mpc/\textit{h}$. This choice is justified since our analysis both requires small and large scales to identify the two terms of the halo model. We discard the four inner annuli of the the measured shear profile to avoid contamination from cluster member galaxies and the contribution of the BCG in the resulting density profiles \citep{2019MNRAS.484.1598B}. Effects of miscentering are minimized as the lensing signal is considered only for $R \gtrsim 0.2 \ Mpc/\textit{h}$. This measurement is also repeated around random lens points to compensate for the systematic signal, as discussed in Appendix~\ref{sec:Random fields}.

\begin{figure}
	\includegraphics[width=\columnwidth]{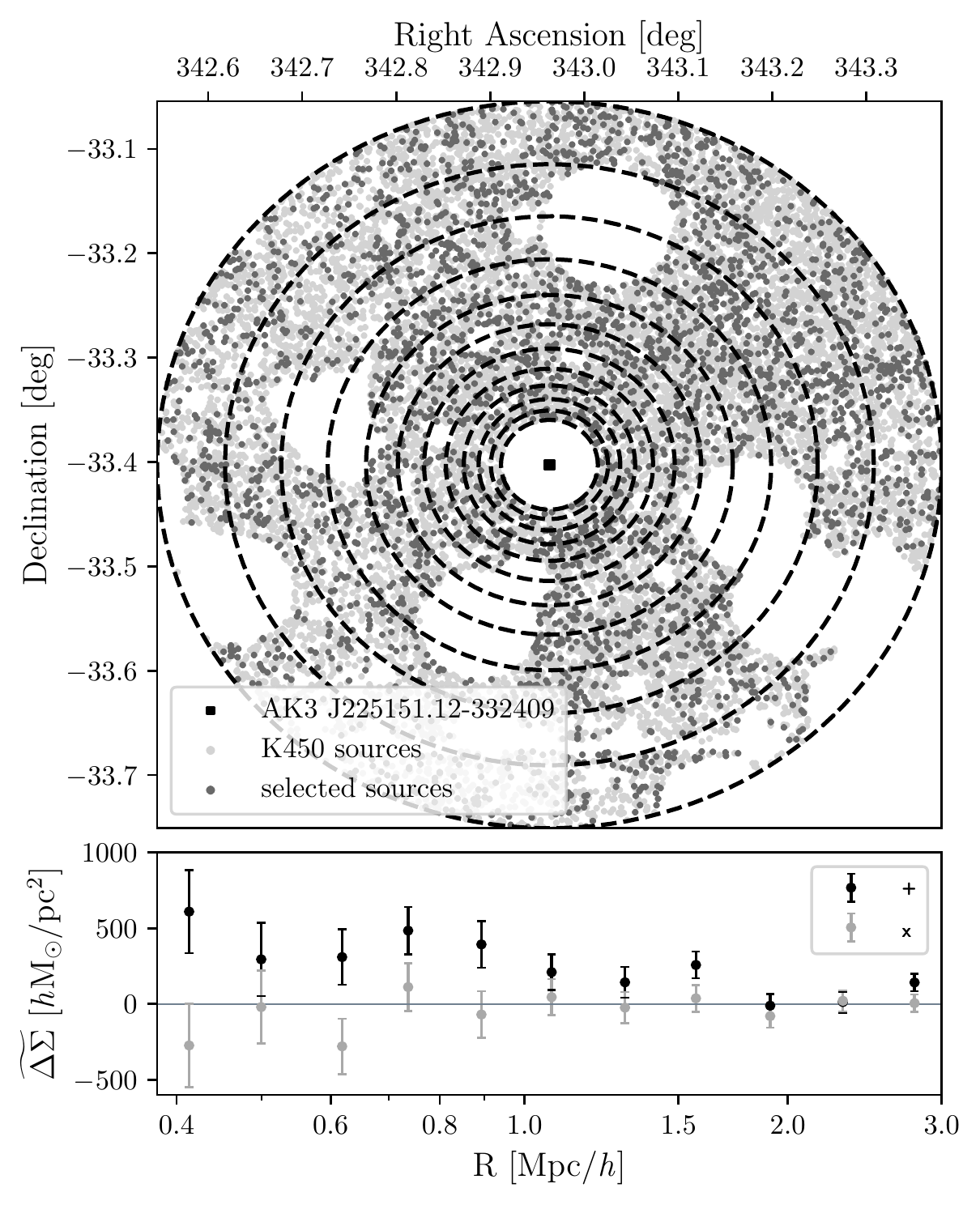}
    \caption{\textit{Top panel}: Illustration of eleven of the thirty annuli in the radial range $\left[ 0.35, 3 \right] \ Mpc/\textit{h}$,for the cluster AK3 J225151.12-332409. The sources shown are selected following the cut discussed in Section~\ref{sec:Background galaxies}. Blank regions indicate masks \citep{2017MNRAS.465.1454H}. \textit{Bottom panel}: Tangential and cross components of the excess surface mass density (Equation~\ref{eq:weighted radius delta sigma}) of J225151.12-332409. Vertical error bars are derived from Equation~\eqref{eq:error}.}
    \label{fig:shear_example}
\end{figure}

We illustrate the process of stacking the shear signal in Figure~\ref{fig:shear_example}, where a 2D distribution of selected sources around the AK3 cluster J225151.12-332409 is shown (more details in Section~\ref{sec:Selection of the sources}). For visual convenience in the illustration, we highlighted only 12 of the 31 radii in the radial range $\left[ 0.35, 3 \right] \ Mpc/\textit{h}$. The tangential and the cross components of $\widetilde{\Delta\Sigma}$ associated to the 10 annuli are additionally displayed in the bottom panel.

\subsection{Selection of lens-source pairs}
\label{sec:Selection of the sources} 

An effective discrimination between background lensed sources, and foreground and cluster member galaxies is necessary to accurately derive the halo density profile. We subsequently select background galaxies using photometric redshifts or their position in the (\textit{r}-\textit{i}) vs (\textit{g}-\textit{r})-color-color (hereafter dubbed \textit{gri}-CC) plane.

\subsubsection{Background galaxies}
\label{sec:Background galaxies}

A thorough selection of sources allows us to minimize contamination from misplaced galaxies and their incorrect shear. This step is essential as contaminated galaxies usually dilute the resulting lensing signal \citep{1538-4357-619-2-L143, 2007ApJ...663..717M}. 
We first select members in the source catalog with 
\begin{equation}
z_s > z_l + \Delta z \ ,
\label{eq:rough selection}
\end{equation}
where $z_s$ is the best-fitting BPZ photometric redshift of the source, $z_l$ is the lens redshift and $\Delta z = 0.05$ is a secure interval to balance uncertainties coming from photometric redshifts.

Then, we applied a more accurate redshift filter following the work of \cite{2019MNRAS.484.1598B} and \cite{2017MNRAS.472.1946S},
\begin{equation}
\left( 0.2 \leq z_s \leq 1 \right) \wedge \left( \texttt{ODDS} \geq 0.8 \right) \wedge \left( z_{s, min} > z_l + \Delta z \right) \ .
\label{eq:photoz selection}
\end{equation}
The \texttt{ODDS} parameter from the KiDS shear catalog accounts for the probability distribution function (PDF) of the redshift: a high value indicates a high reliability of the best photo-z estimate. The parameter $z_{s, min}$ measures the lower bound of the $2 \sigma$ confidence interval of the PDF.

A complementary approach for selecting galaxies is based on the source distribution in the \textit{gri}-CC plane. \cite{2010MNRAS.405..257M} highlight a strong correlation between the location in the (\textit{r}-\textit{i}) vs (\textit{g}-\textit{r}) diagram and the galaxy redshift. Following an original proposal by \cite{2012MNRAS.420.3213O}, \cite{2019MNRAS.484.1598B} exploit a relevant selection which filters KiDS galaxies beyond $z_s \simeq 0.7$, obtaining
\begin{equation}
\left( g-r < 0.3 \right) \vee \left( r-i > 1.3 \right) \vee \left( g-r < r-i \right) \ .
\label{eq:oguri selection}
\end{equation}
This selection was tested in \cite{2014ApJ...784L..25C}, \cite{2017MNRAS.472.1946S, 2018NatAs...2..744S} and \cite{2019MNRAS.484.1598B}, and conserves 97 percent of galaxies with CFHTLenS spectroscopic redshifts above $z_s \gtrsim 0.63$ \citep{2017MNRAS.472.1946S}. In Appendix~\ref{sec:Colour-colour selections}, we discuss the alternative color-color selection presented in \cite{2010MNRAS.405..257M} and the contamination fraction that leads the two colour-colour cuts in the COSMOS field.

Finally, we formulate the selection of the background sources by combining the following Equations as follows
\begin{equation}
\eqref{eq:rough selection} \wedge \left[ \eqref{eq:photoz selection} \vee \eqref{eq:oguri selection} \right] \ .
\label{eq:full selection}
\end{equation}
As a further restriction for the selection presented in this study, we restricted the source redshifts to the range $z_s > 0.2$. This complementary selection is assumed since a large fraction of sources are below this limit, which might increase the contamination of nearby clusters \citep{2017MNRAS.472.1946S}. 

\subsubsection{Foreground clusters}
\label{sec:Foreground clusters}

\begin{figure}
	\includegraphics[width=\columnwidth]{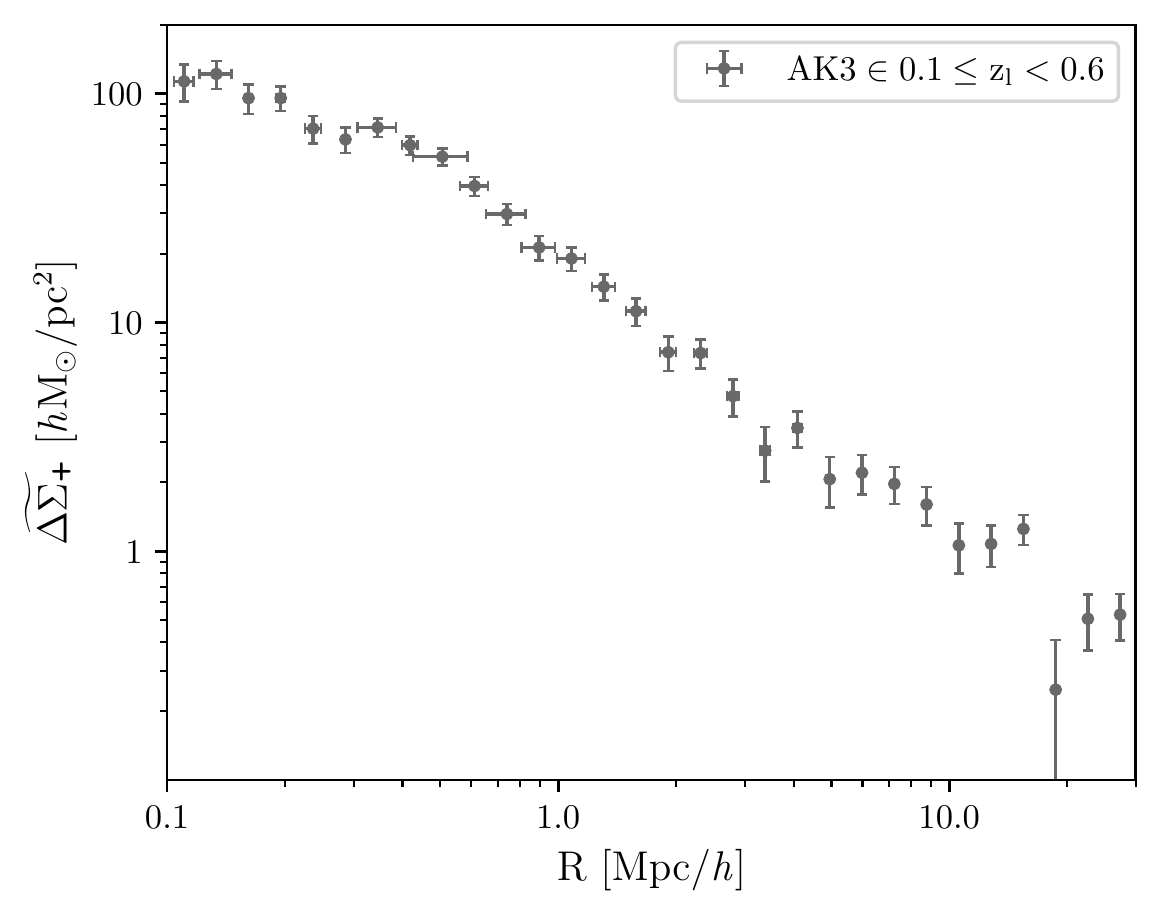}
    \caption{The stacked matter density profile of AK3 clusters with $0.1 \leq z_l < 0.6$. The signal is computed assuming the combined selections given in Equation~\eqref{eq:full selection}.
    Horizontal and vertical bars are derived from Equation~\eqref{eq:error}.}
    \label{fig:full profile}
\end{figure}

We consider galaxy clusters selected in the redshift range $z_l \in \left[ 0.1, 0.6 \right[$, as done in \cite{2019MNRAS.484.1598B}. We select clusters at $z_l < 0.6$ because the \textit{gri}-CC cut is very effective for sources at $z_s > 0.6$. Furthermore remote clusters convey a lower density of background sources. Objects at $z_l < 0.1$ are discarded because of the reduced lensing power of low mass clusters (see Figure~\ref{fig:distribution}) and the inferior photometric redshift accuracy of the sources. The final sample consists of 6961 clusters (87.1\% of the whole catalog).
In Figure~\ref{fig:full profile} we plot the mass density profile obtained for the complete cluster sample assuming the combined selection of sources given in Equation~\eqref{eq:full selection}. 

\subsection{Shear data stacked in bins}
\label{sec:Shear data stacked in bins} 

Stacking the signal permits us to constrain the two parameters of the halo model (see Section~\ref{sec:Halo model}) and derive a generic halo bias-mass relation (see Section~\ref{sec:Halo mass-bias relation}). We consider 14 cluster bins combined in redshift and richness. Table~\ref{tab:bins} shows the binning pattern, also displayed in cells in the $z_l$ vs $\lambda_\ast$ diagram in Figure~\ref{fig:distribution}.
The binning scheme mostly follows \cite{2019MNRAS.484.1598B} to provide nearly uniform WL SNR per bin. The only difference is for the last redshift bin, in which a larger number of clusters are considered for intermediate richness ranges. In this way, we compensate for the numerous galaxy clusters in the higher richness bin and homogenize the distribution of clusters in this redshift bin with the two other redshift bins.

Considering the $j$-th radial bin of the $k$-th galaxy cluster, the corresponding stacked observable in the $K$-th cluster bin is
\begin{equation}
O_{j, K} = \frac{\sum_{k \in K} W_{j, k} O_{j, k} }{\sum_{k \in K} W_{j, k}} \ ,
\label{eq:cluster radial bin stacking}
\end{equation}
with $W_{j, k} = \sum_{i \in j} w_{ls, i}$ . The shear estimate is not accurate since the correction of the multiplicative bias has already been applied via Equation~\eqref{eq:weighted radius delta sigma} to the signal of each individual galaxy cluster, while it should be corrected over the averaged measure of the bin. We compute the effective value of the cluster observable $O_k$, e.g. richness or redshift of cluster $k$, among the cluster bins $K$ through a lensing-weighted mean \citep[e.g.][]{2014ApJ...795..163U}
\begin{equation}
O_K = \frac{\sum_{k \in K} W_k O_k }{\sum_{k \in K} W_k} \ ,
\label{eq:cluster bin stacking}
\end{equation}
where $W_k = \sum_j W_{j, k}$ is the total weight of the cluster $k$ for the whole area of the cluster profile.

The analysis of covariance is performed by computing all the observable quantities using a bootstrap method with replacement and resampling the source catalog 1000 times. In addition, we combined the shear signal with a covariance matrix computed over the realizations of the bootstrap sampling. We also paid attention to the cross-covariances between the redshift-richness bins. As a final step, we subtract the signal around random points from the stacked profiles, and the corresponding error is added in quadrature. The final covariance signal can alternatively be assessed with a jackknife method, where the lensing signal is measured over regions of the sky. This way, there is no longer any need to combine cluster and random covariance matrices, since the statistical covariance is directly computed from the subtracted lensing signal \citep[][]{2017MNRAS.471.3827S}. Covariances and random signals aim to compensate for the statistical noise and the systematic effects.
We discuss these two contributions in detail in Appendices~\ref{sec:Covariances} and~\ref{sec:Random fields}.

\begin{table}
	\centering
	\caption{Redshift-richness bins for the WL analysis.}
	\begin{tabular}{cc} 
		\hline \hline
		$z_l$ & $\lambda_\ast$ \\
		\hline
		$\left[ 0.1, 0.3 \right[$ & $\left[ 0, 15 \right[ \ \left[ 15, 25 \right[ \ \left[ 25, 35 \right[ \ \left[ 35, 45 \right[ \ \left[ 45, 140 \right[$ \\
		$\left[ 0.3, 0.45 \right[$ & $\left[ 0, 20 \right[ \ \left[ 20, 30 \right[ \ \left[ 30, 45 \right[ \ \left[ 45, 60 \right[ \ \left[ 60, 140 \right[$ \\
		$\left[ 0.45, 0.6 \right[$ & $\left[ 0, 25 \right[ \ \left[ 25, 40 \right[ \ \left[ 40, 55 \right[ \ \left[ 55, 140 \right[$ \\
		\hline
	\end{tabular}
	\label{tab:bins}
\end{table}

\section{Halo model}
\label{sec:Halo model}

In this section we explore the theoretical mass density distribution of the halo, also called the halo model. A composite density profile is then fitted to the measured tangential reduced shear given in Equation~\eqref{eq:g+}. All the terms in this relation depend on the surface density $\Sigma$. It is computed by the projection over the line of sight of the excess matter density $\Delta\rho$ in a sphere centered on the halo as
\begin{equation}
\Sigma(R) = \int_{-\infty}^{\infty} \Delta\rho \left( \sqrt{R^2 + \chi^2} \right) d\chi \ .
\label{eq:Sigma}
\end{equation}
$\Delta\rho$ includes the two terms of the halo model from the halo-matter correlation function $\xi_{hm}$
\begin{equation}
\Delta\rho = \bar{\rho}_m \xi_{hm} \ ,
\label{eq:Delta rho}
\end{equation}
and the mean matter density $\bar{\rho}_m \equiv \Omega_m \rho_c$ must be computed in physical units at the redshift of the sample. The critical density $\rho_c$ is related to the first of the Friedmann equations, and is defined as
\begin{equation}
\rho_{c} = \frac{3H(z)^2}{8 \pi G} \ .
\label{eq:rho_c}
\end{equation}
In WL, we average this quantity over the disk to derive the mean surface density enclosed within the radius $R$
\begin{equation}
\overline{\Sigma}(<R) = \frac{2}{R^2} \int_0^R R^{\prime} \Sigma \left( R^{\prime} \right) dR^{\prime} \ .
\label{eq:mean Sigma}
\end{equation}

\begin{figure}
	\includegraphics[width=\columnwidth]{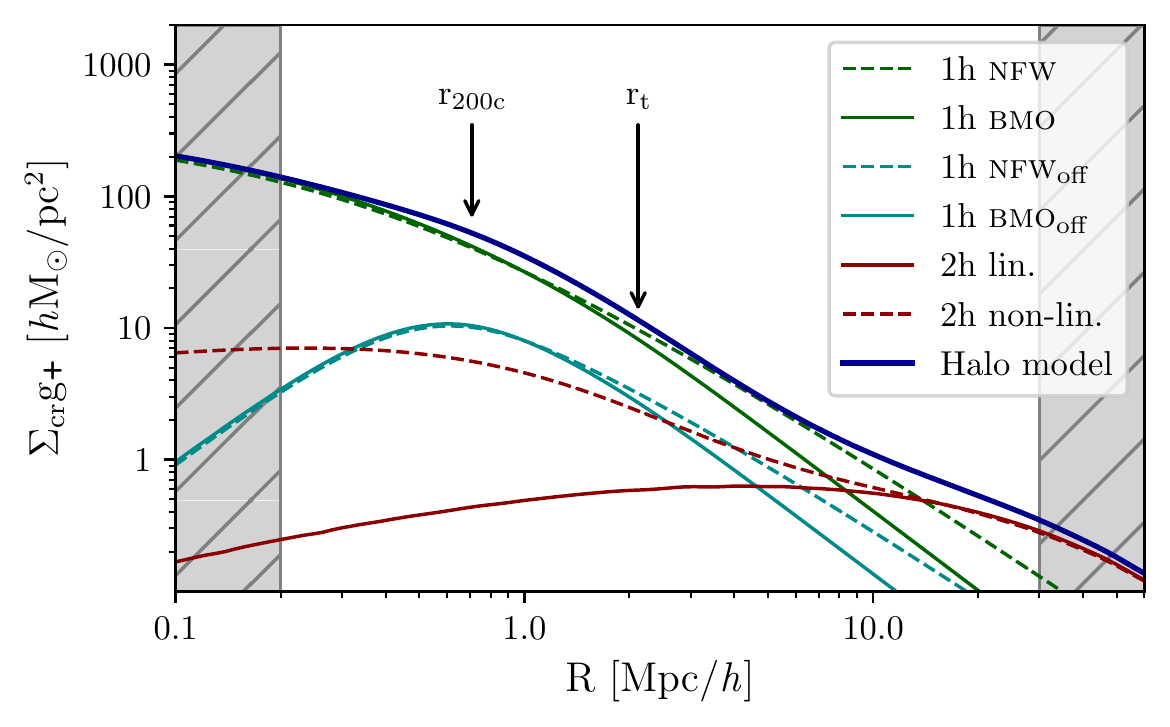}
    \caption{The halo model (blue) is composed of the BMO halo mass profile \citep[thick green,][]{2009JCAP...01..015B}, its off-centered contribution \citep[thick cyan,][]{2007ApJ...656...27J} and the second term derived from the linear matter power spectrum \citep[thick red,][]{1999ApJ...511....5E}. For comparison, we show the centered / off-centered NFW mass profile \citep[dashed green / cyan,][]{1997ApJ...490..493N} and the surrounding matter term with a non-linear power spectrum \citep[dashed red,][]{2012ApJ...761..152T}. The density profile is computed in this example for a halo at $z_l = 0.2$ with a total mass $M_{200c} = 10^{14} \ M_{\odot} / \textit{h}$, a concentration $c_{200c} = 4$ and a bias set at $b_h = 1$ (with $\sigma_8 = 0.83$). The variance and the fraction of an off-centered population contribute to the profile with $\sigma_{off} = 0.25 \ Mpc / \textit{h}$ and $f_{off} = 0.25$. Finally, the reduced shear is given for an effective source redshift $z_s = 1$, while the non-shaded area reveals the range allowed by the stacked WL analysis.}
    \label{fig:halo model}
\end{figure}

In the following and for the terms contributing to the halo model, we are interested in the main lens structure (Section~\ref{sec:Main halo component}), which comprises the total mass of the halo and its concentration. 
In addition, we include the contribution of possibly miscentered density profiles in Section~\ref{sec:Miscentering correction}. Finally, Section~\ref{sec:Correlated matter component} completes the halo model with the correlated matter component and allows the cosmological study from the analysis of the halo bias. In Figure~\ref{fig:halo model} we display, as an example, the complete model for a given mass, concentration, bias and redshift of the halo.

\subsection{Main halo component}
\label{sec:Main halo component}

The correlation between the halo and its own matter content is given by the halo matter density profile $\rho_h$
\begin{equation}
\xi_{1h} = \frac{\rho_h}{\bar{\rho}_m} - 1 \ .
\label{eq:1 halo}
\end{equation}
Analytic calculations and numerical simulations suggest that dark matter halos have a symmetric density profile in a spherical aperture \citep{1996ApJ...462..563N}. More recent studies look at the impact of the triaxiality of the halos as a new source of uncertainty in the WL signal \citep[][]{Oguri_2005, 2010A&A...514A..93M, 2011MNRAS.416.3187S}. This systematic involves a larger scatter of the mass and over-estimates the concentration when triaxial clusters are aligned with the line of sight. Several works, such as \cite{1997ApJ...490..493N, 2001MNRAS.321..559B} provided a specific analytical form for the halo distribution, also called the Navarro-Frenk-White (NFW) density profile, in which the density varies with the distance from the center $r$ as
\begin{equation}
\rho_{\NFW} = \frac{\rho_s}{(r/r_s)(1+r/r_s)^2} \ ,
\label{eq:NFW}
\end{equation}
where $\rho_s = \rho_c \delta_c$ is the scale density and $r_s$ the scale radius.
The overdensity contrast $\delta_c$ can be expressed as a function of the concentration $c$ and the overdensity factor $\Delta$ as
\begin{equation}
\delta_c = \frac{\Delta c^3}{3 m\left(c\right)} \ .
\label{eq:delta_c}
\end{equation}
The function $m(c)$ depends the choice of density profile and on the concentration parameter as in Equation\eqref{eq:m}. Therefore, we adopt the common virial value $\Delta = 200c$, relating to a spherical volume with a density 200 times higher than the critical density of the Universe. Hence, we parametrize the scale radius as $r_s = r_{200c} / c_{200c}$. We leave the concentration within that sphere free in order to study the relation between the mass and the concentration in Section~\ref{sec:Halo mass-concentration relation}. A second approach would be to consider an existing mass-concentration scaling relation, e.g. from \cite{2015ApJ...806....4M} based on X-ray selected galaxy clusters of the Cluster Lensing And Supernova Survey with Hubble \citep[CLASH,][]{2012ApJS..199...25P}, or from simulations \citep[e.g.][]{2018ApJ...859...55C}. The 3D NFW profile can be analytically converted into a 2D version and thereby extended to an excess surface mass density version following \cite{2002A&A...390..821G}. 

The NFW profile has a non-physical divergence of its total mass \citep[][]{2003MNRAS.340..580T}. The Baltz-Marshall-Oguri \citep[BMO,][]{2009JCAP...01..015B} profile is a smoothly truncated version of the NFW profile which allows to circumvent this problem with infinite mass. This profile presents the following shape
\begin{equation}
\rho_{\BMO} = \frac{\rho_s}{(r/r_s)(1+r/r_s)^2} \left( \frac{r_t^2}{r^2 + r_t^2} \right)^2 \ .
\label{eq:BMO}
\end{equation}
We set the truncation radius to $r_t = 3 r_{200c}$ in the following analysis \citep{2014ApJ...784L..25C, 2017MNRAS.472.1946S, 2019MNRAS.484.1598B}.
The BMO profile also provides less biased estimates of mass and concentration with respect to the NFW profile, and better describes the density profile at the transition scales between the one-halo and two-halo terms \citep{2011MNRAS.414.1851O}.
\cite{2009JCAP...01..015B} provide an analytical expression for the surface mass density. 
The function $m$ in Equation~\eqref{eq:delta_c} differs according to the profile as \citep{2011MNRAS.414.1851O}
\begin{equation}
\begin{aligned}
& m_{\mathrm{\NFW}} = \ln \left(1+c\right)-\frac{c}{1+c} \\
& m_{\mathrm{\BMO}} = \frac{\tau^{2}}{2(\tau^{2}+1)^{3}(1+c)(\tau^{2}+c^{2})} \\
&\quad \times\Big[c(\tau^{2}+1) \big\{c(c+1)-\tau^{2}(c-1)(2+3 x)-2 \tau^{4} \big\} \\
&\quad +\tau(c+1)(\tau^{2}+c^{2}) \big\{2(3 \tau^{2}-1) \arctan (c / \tau) \\
&\quad +\tau(\tau^{2}-3) \ln (\tau^{2}(1+c)^{2} / (\tau^{2}+c^{2})) \big\}\Big] \ ,
\end{aligned}
\label{eq:m}
\end{equation}
where $\tau \equiv r_t/r_s$. We display the NFW and BMO surface mass density profiles in Figure~\ref{fig:halo model}. We indicate $r_{200c}$ and $r_t$ locations with vertical arrows.

\subsection{Miscentering correction}
\label{sec:Miscentering correction}

The detection of clusters is based on the identification of galaxy overdensities, hence the adopted cluster center corresponds to the peak in the projected space of the galaxy distribution. This peak may not coincide with the barycenter of the DM distribution. 
In reality, we expect the detected pixel position of the cluster center to possibly be shifted with respect to the center of the halo. \cite{2011MNRAS.416.2388S} and \cite{2012ApJ...757....2G} discussed the importance of locating the centers of dark matter halos in order to properly estimate their mass profiles. Miscentering is expected to be a small with respect to the cluster radius, under the assumption that light traces dark matter
\citep{zitrin11a,zitrin11b,coe12,merten15,donahue16}. However, radial miscentering is larger for optical clusters selected in a survey with a complex mask footprint.

Hence, we introduce the radial displacement of the cluster center $R_{off}$, while the off-centered density profile is the average of the centered profile over a circle drawn around the incorrect center \citep{doi:10.1111/j.1365-2966.2006.11091.x, 2007ApJ...656...27J}
\begin{equation}
\Sigma_{off}(R|R_{off}) = \frac{1}{2\pi} \int_0^{2\pi} \Sigma_{cen} \left( \sqrt{R^2 + R_{off}^2 + 2RR_{off} \cos{\theta}} \right) d\theta \ .
\label{eq:Sigma off single}
\end{equation}
This term holds for an isolated galaxy cluster. We extend the profile to a global population of galaxy clusters so that the off-centered contribution is given by
\begin{equation}
\overline{\Sigma}_{off}(R|\sigma_{off}) = \int_0^{\infty} P(R_{off}, \sigma_{off}) \Sigma_{off}(R|R_{off}) dR_{off} \ ,
\label{eq:Sigma off}
\end{equation}
where the displaced distances $R_{off}$ follows a Rayleigh distribution with parameter $\sigma_{off}^2$ \citep{2017MNRAS.466.3103S, 2017MNRAS.469.4899M}
\begin{equation}
P(R_{off}, \sigma_{off}) = \frac{R_{off}}{\sigma_{off}^2} \exp{ \left[ - \frac{1}{2} \left( \frac{R_{off}}{\sigma_{off}} \right)^2 \right]} \ .
\label{eq:Probability Roff}
\end{equation}

Considering $f_{off}$ as the fraction of the off-centered population, the total miscentered density profile can be modelled as
\begin{equation}
\Sigma_{mis}(R|\sigma_{off}, f_{off}) = (1-f_{off}) \Sigma_{cen}(R) + f_{off} \overline{\Sigma}_{off}(R|\sigma_{off}) \ .
\label{eq:Sigma mis}
\end{equation}
Since this mainly impacts the central region of the halo profile, we reduce the correction to the one-halo component of the model. The miscentering effect is illustrated in Figure~\ref{fig:halo model} with the two elements of the above sum. From the figure, we can also see that the miscentering parameters are degenerate with the halo concentration.

\subsection{Correlated matter component}
\label{sec:Correlated matter component}

On large scales, the lensing signal of the halo is dominated by correlated matter, e.g. neighbouring halos or filaments, rather than its own matter content. The two-halo term usually contributes to the whole profile at $R \gtrsim 10 \ Mpc/\textit{h}$. Following the standard approach, this signal is proportional to the matter-matter correlation function $\xi_m$ through the halo bias $b_h$
\begin{equation}
\xi_{2h} = b_h \xi_m \ .
\label{eq:2 halo}
\end{equation}
We derive the matter correlation function at radius $r$ from the Fourier transform of the dimensionless matter power spectrum $\Delta^2(k) \equiv P(k) k^3 / \left( 2\pi^2 \right)$, and the first-order spherical Bessel function $j_0(x) = \sin{x}/x$
\begin{equation}
\xi_m = \int_0^{\infty} \frac{\Delta^2(k)}{k} j_0(kr) dk \ .
\label{eq:correlation function}
\end{equation}
We illustrate the second term of the surface mass density profile in Figure~\ref{fig:halo model} assuming bias $b_h = 1$. We also display results given by the linear matter power spectrum \citep{1998ApJ...496..605E, 1999ApJ...511....5E} and by the non-linear matter power spectrum computed assuming the so-called halofit model \citep{2012ApJ...761..152T}. A halo mass of $M_{200c} = 10^{14} \ M_{\odot} / \textit{h}$ and concentration of $c_{200c} = 4$ contribute 15\% and 25\%, resptively, to the whole profile at the intermediate scale $R=3.16$ Mpc/\textit{h}, considering the BMO miscentered profile as the one-halo term. We focus on the linear version, since we provide a comparative analysis with theoretical mass-bias relations \citep[e.g.][]{2010ApJ...724..878T} derived from simulations, where results are given in terms of ``peak height" in the linear density field. However, it is important to keep in mind that the non-linear version of the power spectrum shows a non-negligible contribution of mass fluctuations at small and intermediate scales. 
The second term of the halo model is parameterized in terms of a degenerate value of the halo bias with $\sigma_8^2$. This parameter defines the rms fluctuations $\sigma (M)$ for a mass enclosed in a comoving sphere of radius $8 \ Mpc/\textit{h}$. This actually corresponds to the typical scale for the formation of galaxy clusters. The parameter $\sigma_8^2$ also derives from the matter power spectrum as a normalization factor and permits cosmological inference of the product $b_h\sigma_8^2$.

\subsection{Total halo model}
\label{sec:Total halo model}

The total surface mass density profile is modelled with the following terms and their associated marginalized parameters
\begin{equation}
\Sigma_{tot} = \Sigma_{\substack{1h \\ \BMO \\ mis}}(M_{200c}, c_{200c}, \sigma_{off}, f_{off}) + \Sigma_{\substack{2h \\ lin}}(b_h\sigma_8^2) \ .
\label{eq:Sigma tot}
\end{equation}
Mass and bias are the two most critical variables among the five free parameters since they both act on the amplitudes of the one-halo and two-halo terms, respectively. For example, 
Figure~\ref{fig:halo model} shows Equation~\eqref{eq:Sigma tot} in blue with $z_l = 0.2$, $z_s = 1$, $M_{200c} = 10^{14} \ M_{\odot} / \textit{h}$, $c_{200c} = 4$, $\sigma_{off} = 0.25 \ Mpc / \textit{h}$, $f_{off} = 0.25$ and $b_h\sigma_8^2 = 0.83^2$.

In Section~\ref{sec:MCMC method}, we describe the numerical method used to assess the posteriors and best estimates given by the data derived in Sections~\ref{sec:Data} and~\ref{sec:Method} with the model described in this Section. Bayesian inference allows us to correlate the different halo parameters together and completes the cosmological study.

\section{MCMC method}
\label{sec:MCMC method} 

In Bayesian statistics, the Monte Carlo Markov Chain (MCMC) method is commonly used to sample posterior distributions. The best parameters are found with the maximum likelihood distribution, giving the highest probability of the sample (also given by minimizing the $\chi^2$-distribution). 
In this specific study, the likelihood function is the joint probability of getting the measurement $\widetilde{\Delta\Sigma}$ with the parameters $\theta = [\log_{10} M_{200c}, c_{200c}, \sigma_{off}, f_{off}, b_h \sigma_8^2]$ given the model $\Delta \Sigma$. This probability distribution is assumed to be normal and multiplied over the radial bins $i,j$ of the profile to provide a global approximation of the variable
\begin{equation}
\mathcal{L} \left( \theta \right) \equiv p \left( \widetilde{\Delta\Sigma} \vert \theta \right) \propto \exp \left( - \frac{\chi^2}{2} \right) \ ,
\label{eq:likelihood}
\end{equation}
where
\begin{equation}
\chi^2 = \sum\limits_{i, j} \left( \widetilde{\Delta\Sigma}_i -  \Delta \Sigma_i \right) C_{ij}^{-1} \left( \widetilde{\Delta\Sigma}_j -  \Delta \Sigma_j \right) \ ,
\label{eq:chi square}
\end{equation}
and $C_{ij}$ is the covariance matrix described in Appendix~\ref{sec:Covariances}. 

The $\chi^2$ parameter is a good indicator of the goodness of fit of a statistical model. Its probability distribution depends on the degree of freedom which is the difference between the number of observations considered in the analysis and the number of variables in the halo model, here $df = 26 - 5 = 21$. In a goodness-of-fit test, the null hypothesis assumes that there is no significant difference between the observed and the expected values. Considering a significance level of $\alpha=0.01$ defining the critical $\chi^2$ values on the left and right tails of the distribution, the null hypothesis is verified if $8.9 < \chi^2 < 38.9$.

The likelihood is defined in the prior uniform distribution of the halo parameters having the following conservative bounds \citep[][]{2019MNRAS.484.1598B}:
\begin{description}
  \item[$\bullet$] $\log_{10} \left( M_{200c} /\left( M_{\odot} / \textit{h} \right) \right) \in [12.5, 15.5]$
  \item[$\bullet$] $c_{200c} \in [1, 20]$
  \item[$\bullet$] $\sigma_{off} \in [0, 0.5] \ Mpc / \textit{h}$
  \item[$\bullet$] $f_{off} \in [0, 0.5]$
  \item[$\bullet$] $b_h \sigma_8^2 \in [0, 20]$
\end{description}
We based the Bayesian inference on the \texttt{emcee}\footnote{\url{https://emcee.readthedocs.io/}} algorithm \citep[]{2013PASP..125..306F}, which uses an affine-invariant sampling method initially introduced in \cite{2010CAMCS...5...65G}. The cosmological parameters are defined for the fit as in Section~\ref{sec:Introduction}.

\begin{figure}
	\includegraphics[width=\columnwidth]{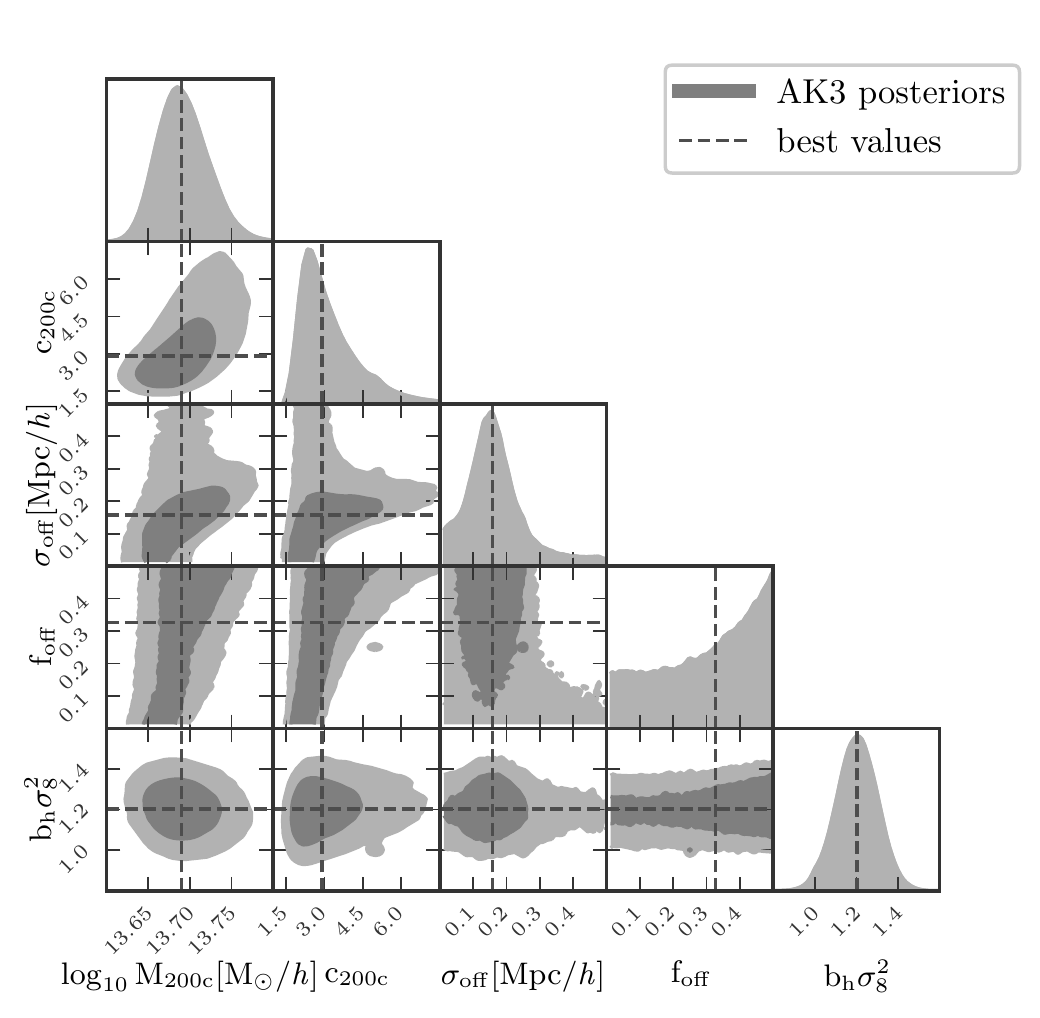}
    \caption{Posterior distributions arising from the halo model and the density profile derived in this study. The median of the marginalized distribution of the mass, concentration, off-centering parameters and bias are displayed as dashed lines. The 2D posterior distributions also show the 68\% and 95\% confidence regions in shaded grey regions.}
    \label{fig:MCMC}
\end{figure}

We adopted an ensemble sampler with 32 walkers over a chain of 10,000 steps, giving a total size of 320,000 walkers to sample the posterior distribution. This scheme was already adopted in \cite{2019MNRAS.482.1352M}. We define the burn-in phase as being twice the integrated autocorrelation time $\tau_f$ of our chain $f$. In addition, we tested the convergence of the MCMC by running the potential scale reduction factor $\hat{R}$ \citep[see][]{1992StaSc...7..457G}. 
Convergence is reached if the criterion $\hat{R} < 1.1$ is satisfied. 

In Figure~\ref{fig:MCMC}, we show the joint posterior distributions given by the sampler for the total profile shown in Figure~\ref{fig:full profile}. In the case of a normal PDF (as for the halo mass and bias), the 16\textit{th}-84\textit{th} and 2\textit{th}-98\textit{th} percentiles highlight $1\sigma$ and $2\sigma$ confidence regions forming ellipsoids in the 2D parameter space. 
In the opposite case, the percentiles show distorted ellipsoidal regions which define the errors on the parameter. For example the $f_{off}$ posterior distribution gives errors larger than the prior boundaries, while we expect the posterior of the parameter to follow a Gaussian-like distribution within the limits defined by the prior function. This effect suggests that the parameter is imprecisely constrained. Nevertheless, the sampler distributions of the parameters of interest (i.e. mass, concentration and bias) converge significantly, which makes it possible to consistently exploit their relation. For the following, we define the error on the parameters as the $1 \sigma$ confidence interval, specifically approximated here with the region where 68\% of walkers lie around the mean.

\section{Results}
\label{sec:Results}

\begin{figure*}
    \centering
    \includegraphics[width=\textwidth]{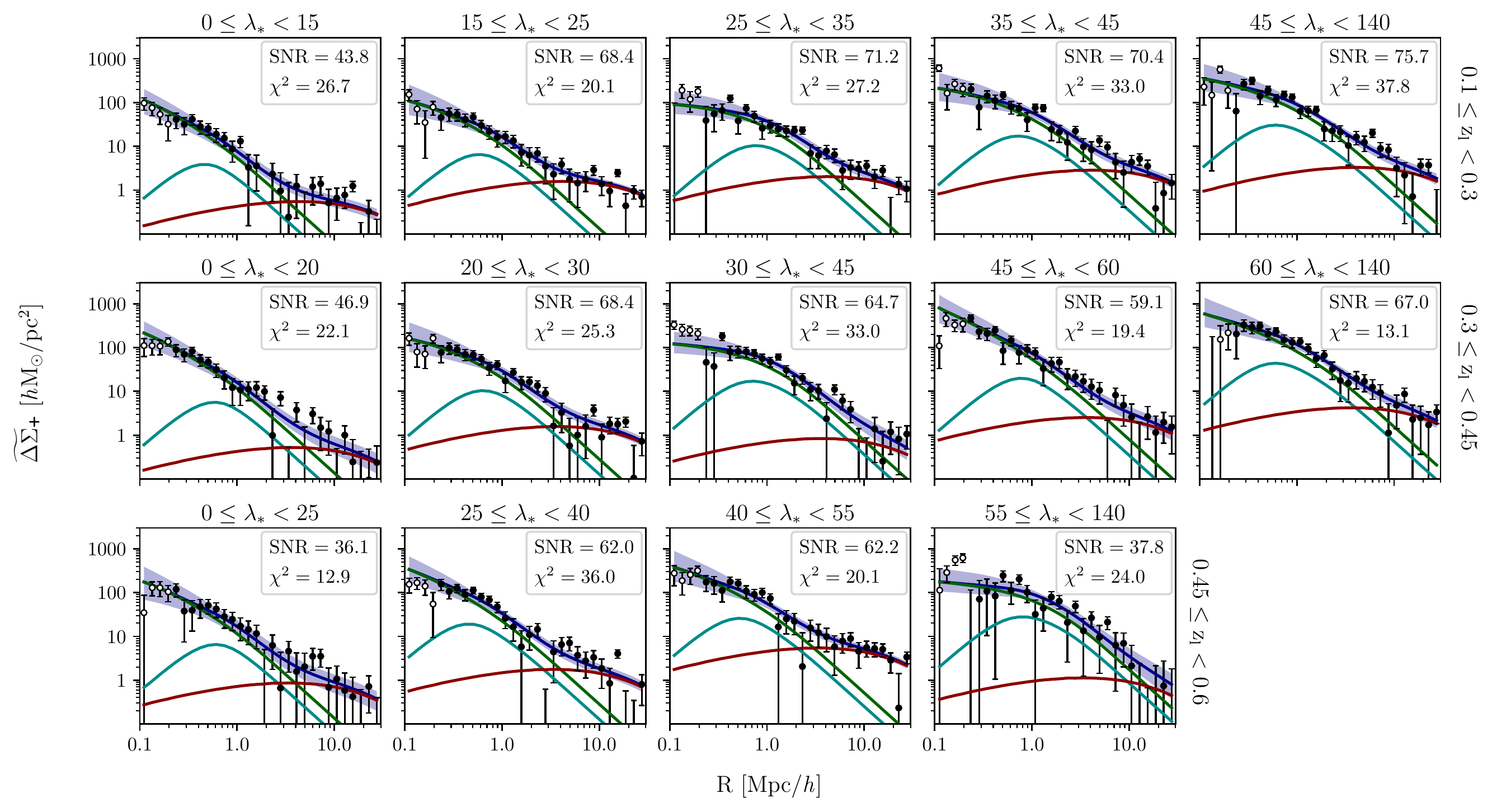}
    \caption{The stacked shear profiles and the halo model (blue) corresponding to the fitted parameters, with the $1 \sigma$ confidence interval (blue region). Each row corresponds to a redshift bin, while each panel corresponds to an associated richness bin. The top right legends show the SNR, computed from each radial bin and summed over the $[0.2, 30] \ Mpc/\textit{h}$ radial range, and the $\chi^2$ computed as in Equation~\eqref{eq:chi square} given by the 50\textit{th} percentile parameters. The model components: the main halo term (green), the off-centered contribution (cyan), and the correlated matter term (red). Empty points show the first four radial bins not considered in the fit.}
    \label{fig:fitted shears}
\end{figure*}

\begin{table*}
	\centering
	\caption{Mass, concentration and bias resulting from the fit with their errors given in separate rows as different redshift and richness bins. These values correspond to the 16\textit{th}, 50\textit{th} and 84\textit{th} percentiles of the posterior distributions. We also show the mass measurement in the radial range $[0.2, 3.16] \ Mpc/\textit{h}$ in brackets. Mean richness ($\bar{\lambda}_{\ast}$), lens redshift ($\bar{z}_l$) and source redshift ($\bar{z}_s$) are computed from Equations~\eqref{eq:cluster bin stacking} and~\eqref{eq:effective redshift} and their errors are  assumed to be the rms weighted sample deviation. We report both the number of clusters $N_l$ and the fraction of clusters relative to the full selected cluster sample in each redshift-richness bin (column 6).
	}
	\begin{tabular}{ccccccccc}
		\hline \hline
		$z_l$ & $\lambda_{\ast}$ & $\bar{z}_l$ & $\bar{\lambda}_{\ast}$ & $\bar{z}_s$ & $N_l$ & $\log_{10} \left( M_{200c} /\left( M_{\odot} / \textit{h} \right) \right)$ & $c_{200c}$ & $b_h \sigma_8^2$ \\
		\hline
		$\left[ 0.1, 0.6 \right[$ & $\left[ 0, 140 \right[$ & $0.372 \pm 0.005$ & $19.92 \pm 0.50$ & $0.763 \pm 0.004$ & $6961 \left( 100.0 \% \right)$ & $13.69\tiny{\substack{+0.03 \\ -0.03}} \ \left( 13.68\tiny{\substack{+0.03 \\ -0.03}} \right)$ & $2.90\tiny{\substack{+1.43 \\ -0.70}}$ & $1.20\tiny{\substack{+0.10 \\ -0.10}}$ \\
		\hline
		$\left[ 0.1, 0.3 \right[$ & $\left[ 0, 15 \right[$ & $0.192 \pm 0.004$ & $10.25 \pm 0.21$ & $0.700 \pm 0.004$ & $1246 \left( 17.9 \% \right)$ & $13.24\tiny{\substack{+0.08 \\ -0.08}} \ \left( 13.23\tiny{\substack{+0.08 \\ -0.08}} \right)$ & $9.27\tiny{\substack{+6.85 \\ -5.05}}$ & $0.60\tiny{\substack{+0.18 \\ -0.18}}$ \\
		$\left[ 0.1, 0.3 \right[$ & $\left[ 15, 25 \right[$ & $0.216 \pm 0.005$ & $18.94 \pm 0.28$ & $0.726 \pm 0.006$ & $683 \left( 9.8 \% \right)$ & $13.56\tiny{\substack{+0.08 \\ -0.08}} \ \left( 13.58\tiny{\substack{+0.08 \\ -0.07}} \right)$ & $4.25\tiny{\substack{+5.18 \\ -2.05}}$ & $1.71\tiny{\substack{+0.24 \\ -0.25}}$ \\
		$\left[ 0.1, 0.3 \right[$ & $\left[ 25, 35 \right[$ & $0.226 \pm 0.009$ & $29.09 \pm 0.51$ & $0.742 \pm 0.011$ & $209 \left( 3.0 \% \right)$ & $14.01\tiny{\substack{+0.07 \\ -0.07}} \ \left( 14.04\tiny{\substack{+0.07 \\ -0.07}} \right)$ & $1.64\tiny{\substack{+1.00 \\ -0.46}}$ & $2.19\tiny{\substack{+0.46 \\ -0.46}}$ \\
		$\left[ 0.1, 0.3 \right[$ & $\left[ 35, 45 \right[$ & $0.232 \pm 0.017$ & $39.61 \pm 0.83$ & $0.740 \pm 0.020$ & $83 \left( 1.2 \% \right)$ & $14.29\tiny{\substack{+0.06 \\ -0.07}} \ \left( 14.30\tiny{\substack{+0.06 \\ -0.07}} \right)$ & $3.17\tiny{\substack{+2.23 \\ -1.10}}$ & $3.07\tiny{\substack{+0.76 \\ -0.77}}$ \\
		$\left[ 0.1, 0.3 \right[$ & $\left[ 45, 140 \right[$ & $0.228 \pm 0.019$ & $56.05 \pm 5.86$ & $0.747 \pm 0.022$ & $44 \left( 0.6 \% \right)$ & $14.53\tiny{\substack{+0.05 \\ -0.06}} \ \left( 14.52\tiny{\substack{+0.06 \\ -0.06}} \right)$ & $3.95\tiny{\substack{+2.25 \\ -1.21}}$ & $3.56\tiny{\substack{+1.01 \\ -1.04}}$ \\
		\hline
		$\left[ 0.3, 0.45 \right[$ & $\left[ 0, 20 \right[$ & $0.374 \pm 0.005$ & $15.13 \pm 0.38$ & $0.860 \pm 0.002$ & $1110 \left( 15.9 \% \right)$ & $13.60\tiny{\substack{+0.08 \\ -0.08}} \ \left( 13.60\tiny{\substack{+0.08 \\ -0.08}} \right)$ & $9.31\tiny{\substack{+6.57 \\ -4.58}}$ & $0.52\tiny{\substack{+0.28 \\ -0.26}}$ \\
		$\left[ 0.3, 0.45 \right[$ & $\left[ 20, 30 \right[$ & $0.388 \pm 0.005$ & $24.16 \pm 0.39$ & $0.863 \pm 0.003$ & $769 \left( 11.0 \% \right)$ & $13.87\tiny{\substack{+0.07 \\ -0.07}} \ \left( 13.93\tiny{\substack{+0.07 \\ -0.07}} \right)$ & $3.65\tiny{\substack{+3.71 \\ -1.54}}$ & $1.57\tiny{\substack{+0.36 \\ -0.35}}$ \\
		$\left[ 0.3, 0.45 \right[$ & $\left[ 30, 45 \right[$ & $0.390 \pm 0.008$ & $35.94 \pm 0.94$ & $0.863 \pm 0.004$ & $320 \left( 4.6 \% \right)$ & $14.20\tiny{\substack{+0.06 \\ -0.06}} \ \left( 14.19\tiny{\substack{+0.06 \\ -0.06}} \right)$ & $1.63\tiny{\substack{+0.82 \\ -0.43}}$ & $0.83\tiny{\substack{+0.52 \\ -0.47}}$ \\
		$\left[ 0.3, 0.45 \right[$ & $\left[ 45, 60 \right[$ & $0.393 \pm 0.015$ & $50.94 \pm 1.86$ & $0.866 \pm 0.008$ & $87 \left( 1.2 \% \right)$ & $14.40\tiny{\substack{+0.08 \\ -0.08}} \ \left( 14.39\tiny{\substack{+0.07 \\ -0.08}} \right)$ & $10.65\tiny{\substack{+5.73 \\ -4.52}}$ & $2.51\tiny{\substack{+1.02 \\ -1.02}}$ \\
		$\left[ 0.3, 0.45 \right[$ & $\left[ 60, 140 \right[$ & $0.381 \pm 0.022$ & $75.81 \pm 9.29$ & $0.860 \pm 0.012$ & $45 \left( 0.6 \% \right)$ & $14.64\tiny{\substack{+0.06 \\ -0.06}} \ \left( 14.66\tiny{\substack{+0.06 \\ -0.06}} \right)$ & $5.11\tiny{\substack{+3.15 \\ -1.62}}$ & $4.20\tiny{\substack{+1.42 \\ -1.43}}$ \\
		\hline
		$\left[ 0.45, 0.6 \right[$ & $\left[ 0, 25 \right[$ & $0.498 \pm 0.006$ & $19.76 \pm 0.53$ & $0.887 \pm 0.003$ & $1107 \left( 15.9 \% \right)$ & $13.60\tiny{\substack{+0.10 \\ -0.11}} \ \left( 13.58\tiny{\substack{+0.10 \\ -0.11}} \right)$ & $6.53\tiny{\substack{+7.74 \\ -3.97}}$ & $0.82\tiny{\substack{+0.40 \\ -0.39}}$ \\
		$\left[ 0.45, 0.6 \right[$ & $\left[ 25, 40 \right[$ & $0.518 \pm 0.008$ & $30.75 \pm 0.74$ & $0.888 \pm 0.003$ & $952 \left( 13.7 \% \right)$ & $13.94\tiny{\substack{+0.06 \\ -0.06}} \ \left( 13.93\tiny{\substack{+0.06 \\ -0.06}} \right)$ & $8.43\tiny{\substack{+6.54 \\ -3.76}}$ & $1.68\tiny{\substack{+0.47 \\ -0.46}}$ \\
		$\left[ 0.45, 0.6 \right[$ & $\left[ 40, 55 \right[$ & $0.513 \pm 0.018$ & $46.14 \pm 1.54$ & $0.888 \pm 0.006$ & $232 \left( 3.3 \% \right)$ & $14.19\tiny{\substack{+0.07 \\ -0.08}} \ \left( 14.23\tiny{\substack{+0.07 \\ -0.08}} \right)$ & $6.18\tiny{\substack{+5.77 \\ -2.65}}$ & $5.16\tiny{\substack{+0.89 \\ -0.91}}$ \\
		$\left[ 0.45, 0.6 \right[$ & $\left[ 55, 140 \right[$ & $0.516 \pm 0.028$ & $66.69 \pm 8.22$ & $0.888 \pm 0.012$ & $74 \left( 1.1 \% \right)$ & $14.56\tiny{\substack{+0.08 \\ -0.10}} \ \left( 14.54\tiny{\substack{+0.10 \\ -0.11}} \right)$ & $1.50\tiny{\substack{+0.77 \\ -0.36}}$ & $1.07\tiny{\substack{+1.21 \\ -0.75}}$ \\
		\hline
	\end{tabular}
	\label{tab:halo parameters}
\end{table*}

We obtain the stacked radial shear profiles for the AMICO KiDS-DR3 galaxy clusters split into 14 redshift-richness bins, from 0.2 to 30 Mpc/\textit{h}. We use the MCMC method presented in Section~\ref{sec:MCMC method} to fit the profiles with the halo model discussed in Section~\ref{sec:Halo model}. Data and fitted models are shown in Figure~\ref{fig:fitted shears}. The SNR is computed as $\widetilde{\Delta\Sigma}_j / \sigma_{\widetilde{\Delta\Sigma}_j}$ from Equations~\eqref{eq:weighted radius delta sigma} and~\eqref{eq:error} and summed over the radial bins $j$.

Table~\ref{tab:halo parameters} shows the best fit values for the halo mass, the concentration and the halo bias in each cluster bin with the 68\% confidence bounds. The parameters computed over the stacked profile of the full catalog are also displayed in the first row, and correspond to the dashed values shown in Figure~\ref{fig:MCMC} with $\chi^2 = 29.8$, which suggests that the goodness-of-fit test has been passed, as for the other bins. The mean redshift and the mean richness of the lenses are computed as in Equation~\eqref{eq:cluster bin stacking}, while the mean redshift of the sources is the effective redshift $z_{back}$ in Equation~\eqref{eq:effective redshift}. We additionally measure the mass from a fitting in the radial range $[0.2, 3.16] \ Mpc/\textit{h}$ assuming the same priors for the full profile, unlike the bias derived from \cite{2010ApJ...724..878T}. These measurements are in good agreements with \cite{2019MNRAS.484.1598B} and show for the two lower redshift bins a relative percentage difference within $\sim 5\%$ (see Figure~\ref{fig:Mvslambda}). This variation could be explained by the different choice for the radial bins within 3.16 \ Mpc/\textit{h}: 14 logarithmically equispaced annuli were used in the previous study, while in this work we selected the radial bins within 3.16 \ Mpc/\textit{h} over the full radial range of the shear profile. These two definitions make the profiles and the derived measurements of the mass slightly different.

In the following, we investigate the correlations of the mass with the cluster richness (see Section~\ref{sec:Halo mass-richness relation}), with the concentration (see Section~\ref{sec:Halo mass-concentration relation}) and with the bias (see Section~\ref{sec:Halo mass-bias relation}).

\subsection{Halo mass-richness relation}
\label{sec:Halo mass-richness relation}

\begin{figure}
	\includegraphics[width=\columnwidth]{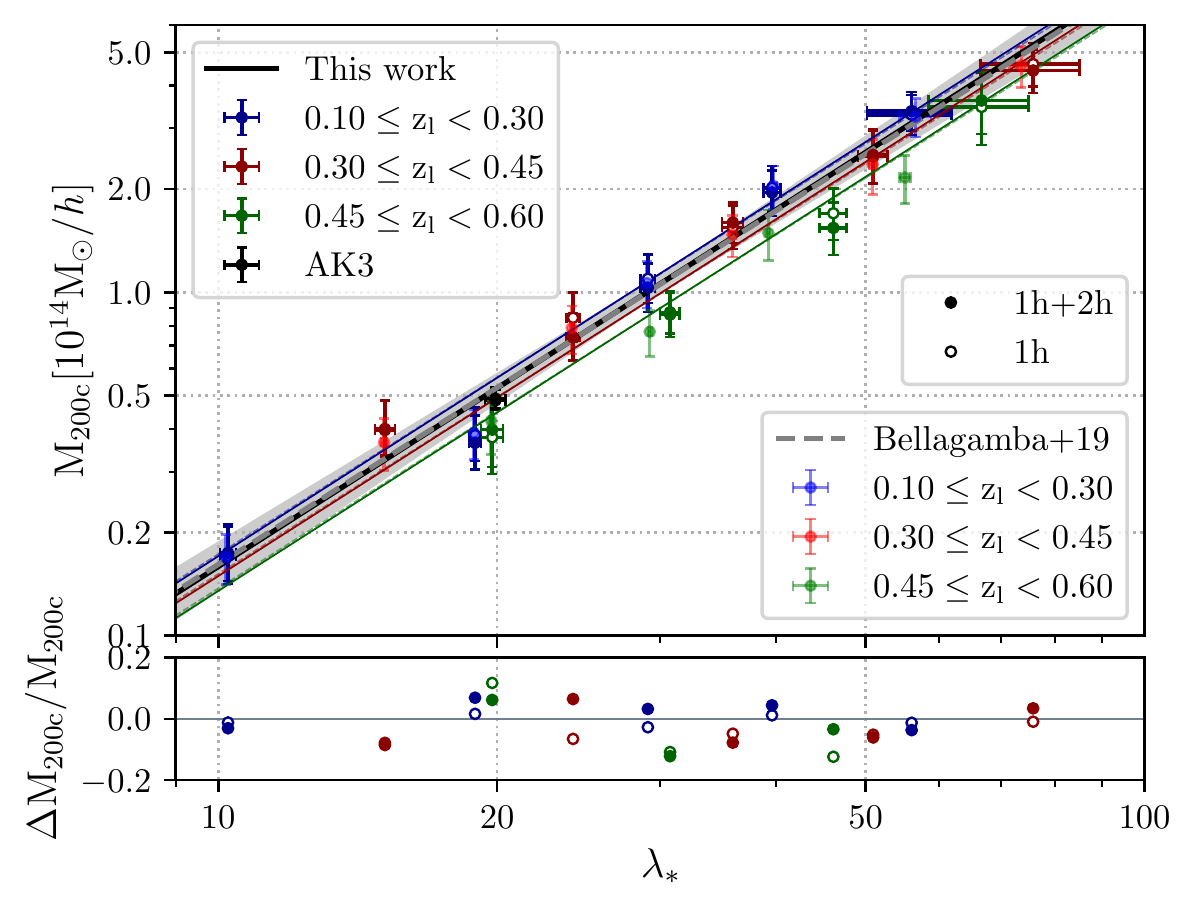}
    \caption{Mass-richness scaling relation for the full catalog (black) and for the low (blue), intermediate (red) and high (green) redshift bins. The thick line corresponds to the model formulated in Equation~\eqref{eq:Mvslambda}. Full and empty data points represent the measurements over the whole radial profile and over the central region of the halo, respectively. We compared our results with those presented in \protect\cite{2019MNRAS.484.1598B}. The fainter colored points represent the data and the dashed lines represent the model. The relative change with respect to the results of this work is displayed in the bottom panel.}
    \label{fig:Mvslambda}
\end{figure}

The average redshift and richness of the lenses in each redshift bin are shown in Figure~\ref{fig:distribution}, and follow the global trend given by the removal of low mass clusters at high redshift for AK3 clusters with $SNR < 3.5$.
Figure~\ref{fig:fitted shears} shows that the differential density at a given radius increases with  richness, suggesting a clear correlation between cluster mass and richness. Figure~\ref{fig:Mvslambda} shows the relation between the mass and the effective richness of the cluster bins. We fit this relation assuming the following power law in logarithmic scale
\begin{equation}
\log_{10} \frac{M_{200c}}{M_{piv}} = \alpha + \beta \log_{10} \frac{\lambda_{\ast}}{\lambda_{piv}} + \gamma \log_{10} \frac{E(z)}{E(z_{piv})} \ ,
\label{eq:Mvslambda}
\end{equation}
where $E(z) \equiv H(z)/H_0$ and $M_{piv} = 10^{14} M_{\odot} / \textit{h}$, $\lambda_{piv} = 30$, and $z_{piv} = 0.35$ corresponding to the median values for AK3 \citep[][]{2019MNRAS.484.1598B}. We estimate the parameters of this multi-linear function applying an orthogonal distance regression method (\texttt{ODR}\footnote{\url{https://docs.scipy.org/doc/scipy/reference/odr.html}}), involving mass, richness and redshift uncertainties. The fit gives
\begin{description}
  \item[$\bullet$] $\alpha = 0.007 \pm 0.019$
  \item[$\bullet$] $\beta =  1.72 \pm 0.09$
  \item[$\bullet$] $\gamma = -1.35 \pm 0.70$.
\end{description}
As Figure~\ref{fig:Mvslambda} shows, these results are in remarkable agreement with \cite{2019MNRAS.484.1598B} despite the different definition of richness bins at high redshifts and the different fitting method. In addition, they are also perfectly consistent with \cite{2020arXiv201212273L} and \cite{2020MNRAS.497..894S}, regardless of the different approaches employed to fit the scaling relation.

The positive correlation between shear signal and richness is shown in Figure~\ref{fig:fitted shears} at large radii and implies a strong correlation between the bias and the mass. The SNR of individual radial bins at large scales is relatively low due to the poor quality of the shear produced by low mass clusters, and increases with the richness. 
The highest redshift-richness bin shows a particularly low SNR with a low amplitude for the shear profile, where usually we expect the signal amplitudes at small and large scales to be high in large richness bins. The poor quality of the lensing signal in this specific bin also impacts the halo mass and bias with a downward trend.

\subsection{Halo mass-concentration relation}
\label{sec:Halo mass-concentration relation}

\begin{figure}
	\includegraphics[width=\columnwidth]{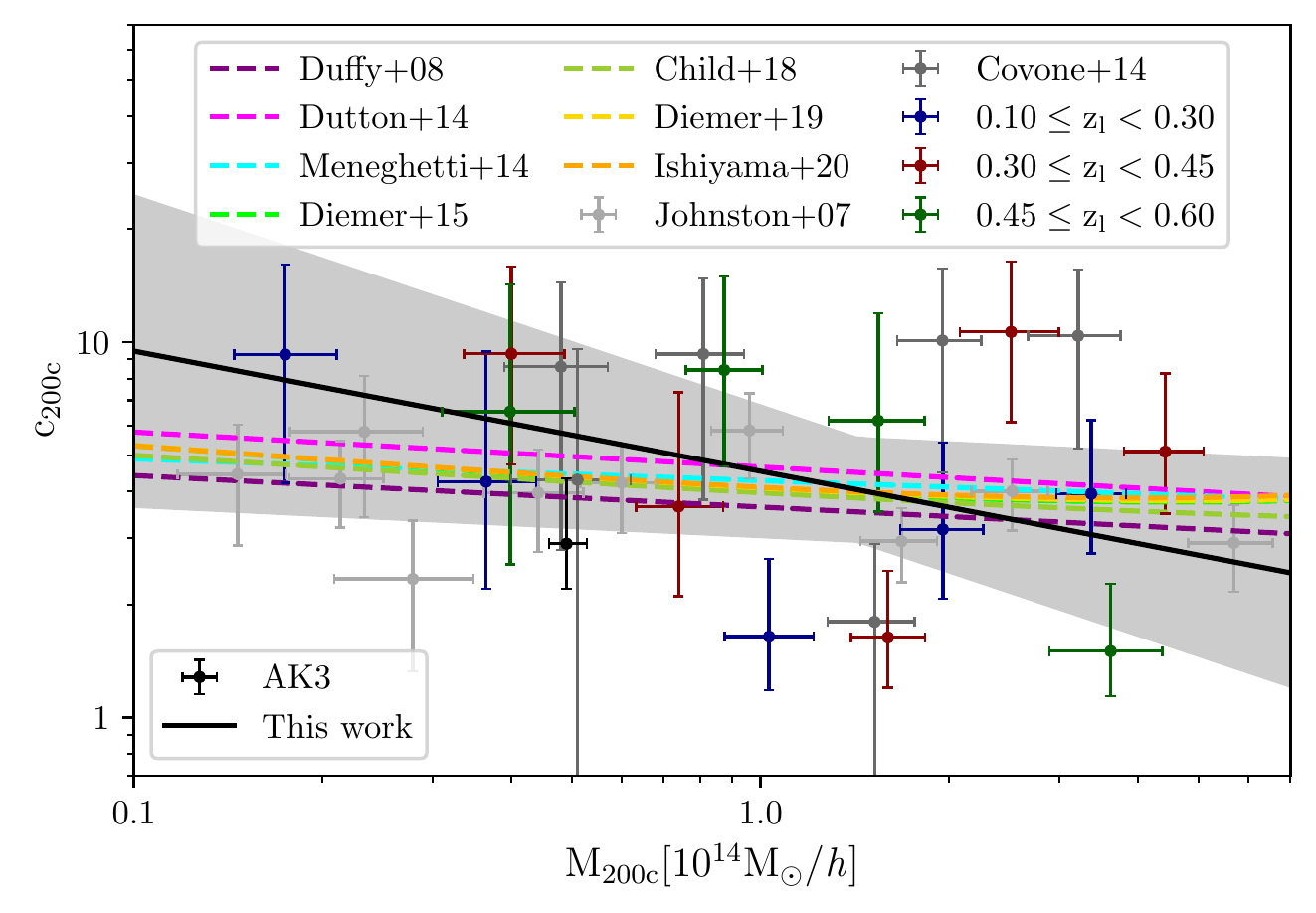}
    \caption{The relation between the mass and the halo concentration for the full catalog (black) and for the low (blue), intermediate (red) and high (green) redshift bins. The results on the concentration are compared with calibrated data from a stacked WL analysis on SDSS and CFHTLenS galaxy clusters \citep[][]{2007arXiv0709.1159J, 2014ApJ...784L..25C}. The thick black line reports the best estimate of the linear regression for Equation~\eqref{eq:cvsM} with its $1\sigma$ confidence region. The relation is contrasted with results given by different theoretical analyses \citep[][]{2008MNRAS.390L..64D, 2014MNRAS.441.3359D, 2014ApJ...797...34M, 2015ApJ...799..108D, 2018ApJ...859...55C, 2019ApJ...871..168D, 2020arXiv200714720I}.}
    \label{fig:CvsM}
\end{figure}

Halo concentration is determined by the mean density of the Universe at the epoch of halo formation \citep{neto07,giocoli12b}. Thus, clusters that assemble later are expected to have a lower concentration than older clusters, formed when the mean density was higher. This determines a clear correlation with the halo mass in such a way that the halo concentration is expected to be a decreasing function of the halo mass. This is supported by our results shown in Figure~\ref{fig:CvsM}. We compare the results with the concentration and mass measured with stacked WL data from 130,000 SDSS galaxy groups and clusters \citep{2007arXiv0709.1159J} and 1176 CFHTLenS galaxy clusters \citep{, 2014ApJ...784L..25C}. These analyses are consistent within $1\sigma$. 
The large and asymmetric error bars for the concentration reflect the high sensitivity of this parameter to the inner region, which is poorly covered by our WL analysis. 
\cite{2013MNRAS.434..878S}, \cite{2014ApJ...795..163U} and \cite{2015MNRAS.449.2024S} discussed the effects stemming from the different choices and forms of the priors, and found a log-uniform prior might underestimate the concentration. 
As done for the redshift-mass-richness relation, we fitted the redshift-concentration-mass relation with a power-law function \citep[][]{2008MNRAS.390L..64D}, given as
\begin{equation}
\log_{10} c_{200c} = \alpha + \beta \log_{10} \frac{M_{200c}}{M_{piv}} + \gamma \log_{10} \frac{1+z}{1+z_{piv}} \ .
\label{eq:cvsM}
\end{equation}
We assume the pivot mass and redshift have the same values as in Equation~\eqref{eq:Mvslambda}, while the multi-linear regression is processed with the \texttt{ODR} routine over the full sample. We find
\begin{description}
  \item[$\bullet$] $\alpha = 0.62 \pm 0.10$
  \item[$\bullet$] $\beta = -0.32 \pm 0.24$
  \item[$\bullet$] $\gamma = 0.71 \pm 2.51$.
\end{description}
The large error on $\gamma$ suggests a weak constraint of the redshift evolution due to the sparse number of data points
\citep[][]{2017MNRAS.472.1946S}. 
The black line in Figure~\ref{fig:CvsM} shows the fitted power law with the $1\sigma$ uncertainty interval, assumed as the range defined by the standard deviations of the estimated parameters and derived from the diagonal terms of the asymptotic form of the covariance matrix \citep[see][]{1987mem..book.....F}. 
Because of the small set of data points, the fit in each redshift bin does not provide consistent results for the coefficients. 
In Figure~\ref{fig:CvsM}, we also show the theoretical relations between mass and concentration given by six different analyses of numerical simulations \citep[][]{2008MNRAS.390L..64D, 2014MNRAS.441.3359D, 2014ApJ...797...34M, 2015ApJ...799..108D, 2018ApJ...859...55C, 2019ApJ...871..168D, 2020arXiv200714720I}. 
In the corresponding mass range, our results are in good agreement with the theoretical predictions, but have a steeper and lower relation with respect to the results obtained by 
\cite{2017MNRAS.472.1946S} on the PSZ2LenS sample. The average concentration for the full AK3 catalog seems to show a lower value than Equation~\eqref{eq:cvsM} and the theoretical expectations, but still remains in the $1\sigma$ confidence interval.

\subsection{Halo mass-bias relation}
\label{sec:Halo mass-bias relation}

\begin{figure}
	\includegraphics[width=\columnwidth]{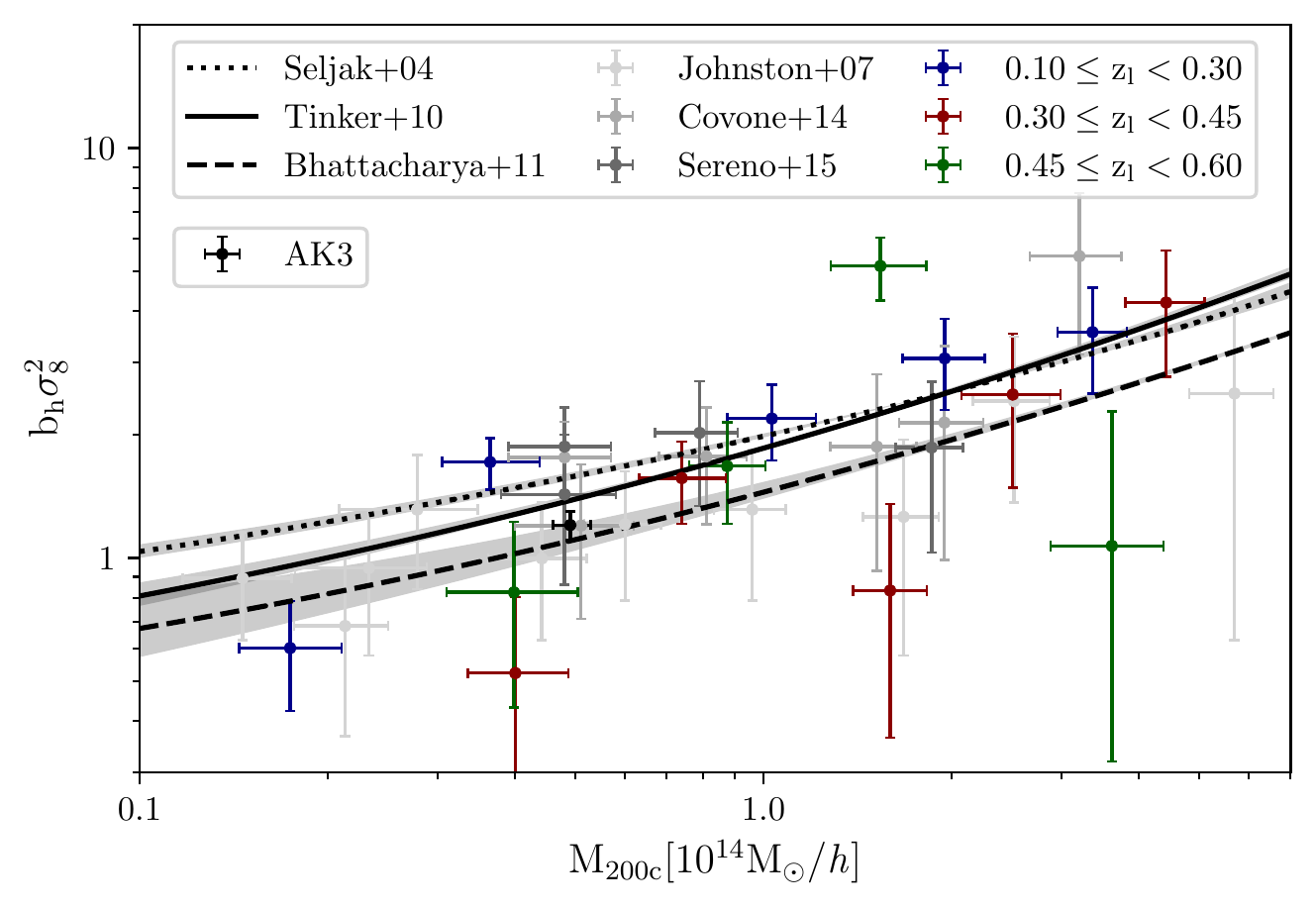}
    \caption{Halo bias-mass relation for the full catalog (black) and for the low (blue), intermediate (red) and high (green) redshift bins. The results on the halo bias are compared with calibrated data from a stacked WL analysis on SDSS and CFHTLenS galaxy clusters \citep[][]{2007arXiv0709.1159J, 2014ApJ...784L..25C, 2015MNRAS.449.4147S}. Theoretical relations are derived from \protect\cite{2004MNRAS.355..129S, 2010ApJ...724..878T, 2011ApJ...732..122B} and respectively displayed as dotted, thick and dashed lines. These functions are computed within their confidence interval using the values of $\sigma_8$ reported in Table~\ref{tab:sigma8}.}
    \label{fig:bvsM}
\end{figure}

In Figure~\ref{fig:bvsM} we show the correlation between the cluster mass and the halo bias for the different redshift bins. The corresponding values are also reported in Table~\ref{tab:halo parameters}. 
These results are also in good agreement with previous results based on stacked WL studies on SDSS \citep[][]{2007arXiv0709.1159J} and CFHTLens \citep[][]{2014ApJ...784L..25C, 2015MNRAS.449.4147S} galaxy clusters. 
As expected with the fourth richness bin at the highest redshift, the Bayesian inference of the halo bias shows a low SNR consistent with the poor quality of the lensing signal at large scales. 

\cite{2010ApJ...724..878T} calibrated the dependence of the large-scale bias on the mass by analysing the clustering of dark matter halos based on dark-matter only cosmological simulations, and obtained a 6\% scatter from simulation to simulation. 
Alternatively, \cite{2004MNRAS.355..129S} and \cite{2011ApJ...732..122B} also derived the average halo bias relation as a function of the cluster mass from \textit{N}-body simulations. 
These bias-mass theoretical relations are reported in Figure~\ref{fig:bvsM} using the corresponding values of $\sigma_8$ in Table~\ref{tab:sigma8}. 
Due to the limited number of points, the data in each redshift bin do not exhibit a strong correlation with the theoretical bias given at the effective redshift of the bin. 
The black lines present an agreement within $2\sigma$ with all our measurements except the third richness point for the high redshift bin, which agrees within $3\sigma$ due to its high amplitude. 
We attribute this statistical fluctuation to the low number of clusters in this region of richness-redshift space, since the few and uneven number of objects results in a poorer statistical measurement of the stacked lensing signal.

\subsection{Constraint on \texorpdfstring{$\sigma_8$}{sigma8}}
\label{sec:Constraint on sigma8}

Since the halo bias is degenerate with $\sigma_8^2$, it is important to obtain independent constraints on this cosmological parameter within a $\Lambda CDM$ framework. 
Here we let $\sigma_8$ be a free parameter in the theoretical mass-bias relation and fit the $b_h \sigma_8^2$ results with the method described in Section~\ref{sec:MCMC method}, assuming a uniform prior $\sigma_8 \in [0.2, 2.0]$. We use a diagonal covariance matrix, where the variance terms are the square of the errors on the bias defined by the 68\% confidence regions. We do not account for the errors on the mass, hence accurate mass measurements are essential to constrain $\sigma_8$.

\begin{table}
	\centering
	\caption{Median, 16\textit{th} and 84\textit{th} percentiles of the posterior distribution for $\sigma_8$. We also show the difference, $\Delta\sigma_8$, between $\sigma_8$ measured on the median mass values, and $\sigma_8$ measured on the mass 16\textit{th} and 84\textit{th} percentile values.} The cosmological parameter is given for three relations derived from numerical simulations.
	\begin{tabular}{lcc} 
		\hline \hline
		simulation & $\sigma_8$ & $\Delta\sigma_8$ \\
		\hline
		\cite{2004MNRAS.355..129S} & $1.01 \tiny{\substack{+0.05 \\ -0.05}}$ & $0.02$ \\
		\cite{2010ApJ...724..878T} & $0.63 \tiny{\substack{+0.11 \\ -0.10}}$ & $0.01$ \\
		\cite{2011ApJ...732..122B} & $0.66 \tiny{\substack{+0.19 \\ -0.27}}$ & $0.12$ \\
		\hline
	\end{tabular}
	\label{tab:sigma8}
\end{table}

The resulting best fit values 
are shown in Table~\ref{tab:sigma8}. \cite{2011ApJ...732..122B} used the ``peak-background split" approach of \cite{1999MNRAS.308..119S} to fit the parameters of the mass function. The authors note that the bias function does not match the numerical results as well as direct calibrations, which could explain the discrepancy with respect to the results obtained with the two other relations. In order to estimate the effect of the mass uncertainty on cosmological inference, we measured $\sigma_8$ at masses corresponding to the 16\textit{th} and 84\textit{th} percentiles and noticed a difference with the median masses smaller than the statistical uncertainty of the parameter (see Table~\ref{tab:sigma8}).

\begin{figure}
	\includegraphics[width=\columnwidth]{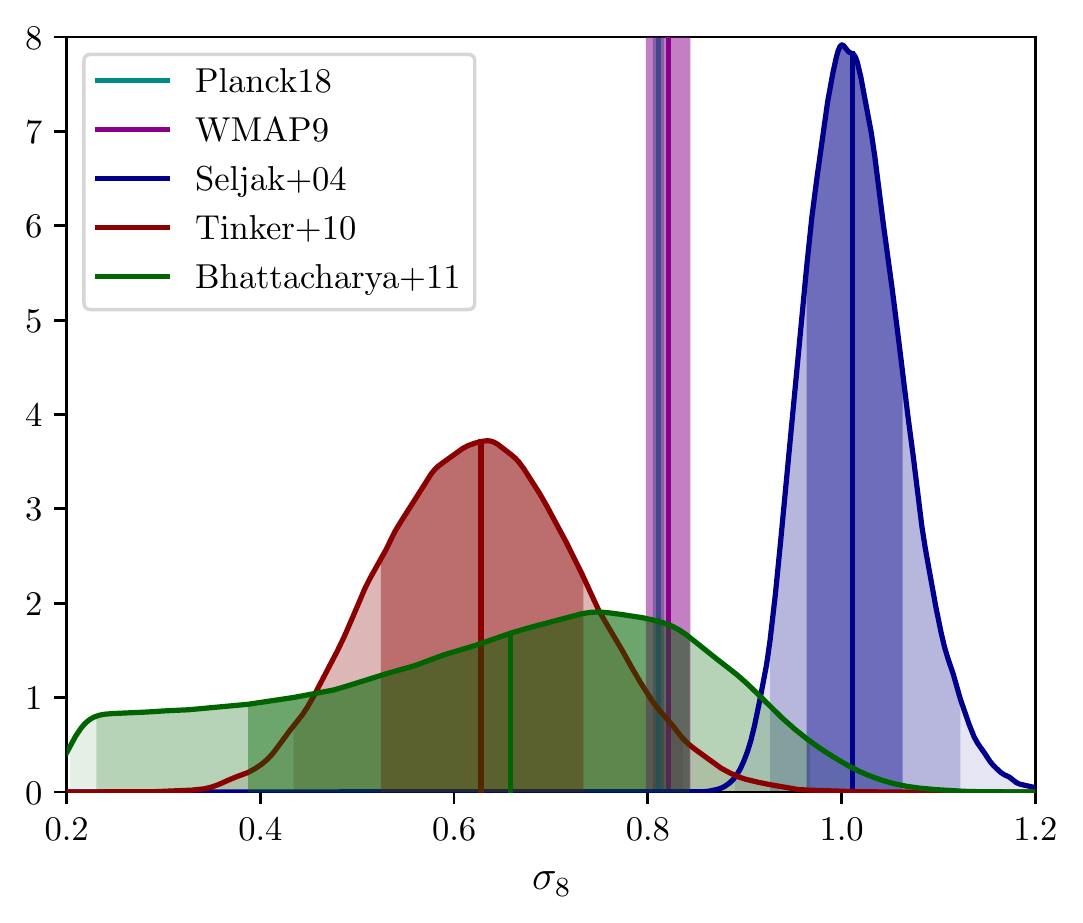}
    \caption{Posterior distribution for $\sigma_8$. The probability function is shown for three halo bias-mass relations, i.e. \protect\cite{2004MNRAS.355..129S}, \protect\cite{2010ApJ...724..878T} and \protect\cite{2011ApJ...732..122B}, shown in blue, red and green, respectively. The dark to light shaded regions correspond to the $1-2-3\sigma$ intervals. We compare these distributions with the median values of Planck \citep[cyan,][Table 2, TT,TE,EE+lowE+lensing]{2020A&A...641A...6P} and WMAP \citep[magenta,][Table 3, WMAP-only Nine-year]{2013ApJS..208...19H}.}
    \label{fig:posterior_sigma8}
    \end{figure}
    
Figure~\ref{fig:posterior_sigma8} shows the three posterior distributions for $\sigma_8$ obtained in this work compared with the results from the cosmic microwave backround measurements by Planck \citep[][Table 2, TT,TE,EE+lowE+lensing]{2020A&A...641A...6P} and WMAP \citep[][Table 3, WMAP-only Nine-year]{2013ApJS..208...19H}. 
Our constraint on $\sigma_8$ with the \cite{2004MNRAS.355..129S} model, which has a sharp posterior that peaks around $\sigma_8 \sim 1$, highlights a discrepancy larger than $3\sigma$ with CMB values. The posteriors given by the \cite{2010ApJ...724..878T} and \cite{2011ApJ...732..122B} models overlap within $2\sigma$ and $1\sigma$ with the CMB data, respectively, but the \cite{2011ApJ...732..122B} posterior is clearly different from a normal distribution. Because of the small size of the sample and the poor quality of the bias-mass measurements in some bins, our results yield quite broad posteriors that are necessarily in agreement with WMAP and Planck median values.

\begin{figure}
	\includegraphics[width=\columnwidth]{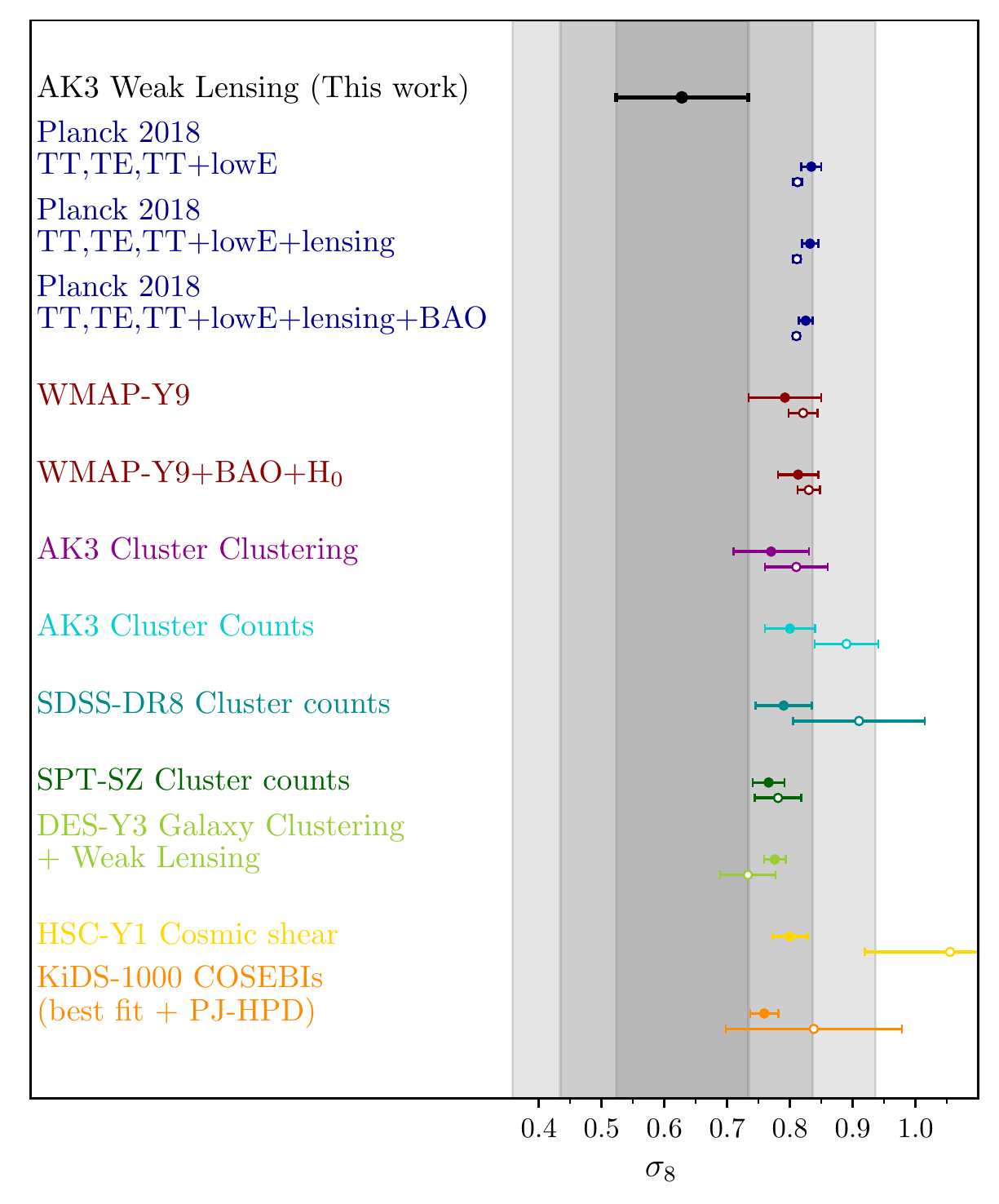}
    \caption{Comparison with literature results. Our reference $\sigma_8$ value is obtained assuming the \protect\cite{2010ApJ...724..878T} model. We show the median, 16\textit{th} and 84\textit{th} percentiles. We present from top to bottom results obtained in this work (black), \protect\cite{2020A&A...641A...6P} (blue), \protect\cite{2013ApJS..208...19H} (red), \protect\cite{nanni} (magenta), \protect\cite{2020arXiv201212273L} (cyan), \protect\cite{2019MNRAS.488.4779C} (turquoise), \protect\cite{2019ApJ...878...55B} (green), \protect\cite{2021arXiv210513549D} (light green), \protect\cite{2019PASJ...71...43H} (yellow) and \protect\cite{2021A&A...645A.104A} (orange). We show the relative constraints on $\sigma_8$ in a free cosmology (empty dots) and assuming $\Omega_m = 0.3$ (filled dots). The shaded regions correspond to the 99.7\%, 95\% and 68\% confidence intervals.}
    \label{fig:sigma8_relative}
\end{figure}

Finally in Figure~\ref{fig:sigma8_relative} we present our reference result from \cite{2010ApJ...724..878T} in the broader context of recent measurements of $\sigma_8$. This model was calibrated for a range of overdensities with respect to the mean density of the universe and can easily be converted to overdensities with respect to the critical density, which makes the bias more reliable for the mass definition $M_{200c}$. In addition, our $b_h \sigma_8^2$ results given by the \cite{2010ApJ...724..878T} relation are more reliable in comparative terms, since studies referenced in this paper base their analyses on this relation.
In particular, we display the results from clustering and cluster counts studies based on the AK3 galaxy clusters sample \citep[][]{nanni, 2020arXiv201212273L}, from cluster counts analyses done on SDSS-DR8 and $2500 \ deg^2$ SPT-SZ Survey data \citep[][]{2019MNRAS.488.4779C, 2019ApJ...878...55B}, from galaxy clustering and weak lensing in DES-Y3 \citep[][]{2021arXiv210513549D}, and from cosmic shear analysis based on the  HSC-Y1 and KiDS-DR4 catalogs \citep[][respectively]{2019PASJ...71...43H, 2021A&A...645A.104A}. 
We also show the results from Planck \citep[][Table 2]{2020A&A...641A...6P} and WMAP \citep[][Table 3]{2013ApJS..208...19H} measurements. 

Since the amplitude of the matter power spectrum correlates with the mean matter density, all these studies derived the combined parameter $S_8 \equiv \sigma_8 \sqrt{\Omega_m / 0.3}$. In this work we computed a direct measurement of $\sigma_8$, dependent on the specific cosmological model assumed in our analysis. 
In the figure, we indicate with different symbols the measurements of $\sigma_8$ obtained without assuming specific values of the cosmological parameters (empty dots) and those assuming $\Omega_m = 0.3$ (filled dots). 
Our results are closer to those obtained fixing $\Omega_m=0.3$, as a low inference of $\Omega_m$ induces a higher estimate of $\sigma_8$, and vice versa. 
For example, \cite{2020A&A...641A...6P} results show a posterior mean slightly higher than $\Omega_m=0.3$, while for cosmic shear studies it is slightly lower, hence when fixing $\Omega_m$ to $0.3$ there is a shift in $\sigma_8$ to larger values for \cite{2020A&A...641A...6P} and lower values for cosmic shear surveys.
However, the $2-3\sigma$ regions for the posteriors of the three theoretical relations agree with the results of these external references, regardless of the cosmological dependencies considered, but still have to be taken carefully into consideration because of the poor constraint.
The gap of $\sigma_8$ results from \cite{2004MNRAS.355..129S} to \cite{2010ApJ...724..878T} or \cite{2011ApJ...732..122B} also stresses the importance of the theoretical model when constraining cosmological parameters in a stacked WL analysis.

\section{Summary and discussion}
\label{sec:Summary and discussion} 

We investigated the halo bias from a revised stacked WL analysis presented in \cite{2019MNRAS.484.1598B} on 6961 AMICO galaxy clusters identified in the recent KiDS-DR3 field. We divided the catalog into 14 bins in redshift and richness and for each of them we derived the excess surface mass density profiles. 
We selected sources from their photometric redshifts or \textit{gri}-colors.
We compared the two color-color selections presented in \cite{2010MNRAS.405..257M} and \cite{2012MNRAS.420.3213O} with COSMOS accurate photometric redshifts in order to carry the most effective cut out for KiDS sources. 
The final WL profiles are obtained by subtracting the signals given by a large number of random lenses. We computed the covariances by applying the bootstrap technique to the cluster and random shears, and added together the matrices to assess the uncertainties of the final profiles. We performed the Bayesian inference of the halo parameters with a MCMC method run over a radial range from 0.2 to 30 Mpc/\textit{h}.

We modelled the  WL signal from galaxy clusters by including the contribution of a truncated version of the NFW profile, which includes a correction for the off-centered galaxy clusters and a correlated matter term originating from the linear matter power spectrum. 

Our measurements of the halo mass within 3.16 Mpc/\textit{h} agree with the results obtained by \cite{2019MNRAS.484.1598B} with a relative difference estimated on the order of 5\%. From the full radial range, we obtained halo masses and derived the mass-richness relation given by Equation~\eqref{eq:Mvslambda}
with $\alpha = 0.007 \pm 0.019$, $\beta = 1.72 \pm 0.09$ and $\gamma = -1.35 \pm 0.70$, in remarkable agreement with \cite{2019MNRAS.484.1598B}. We also studied the halo mass-concentration relation modelled as in Equation~\eqref{eq:cvsM}.
We obtained $\alpha = 0.62 \pm 0.10$, $\beta = -0.32 \pm 0.24$ and $\gamma = 0.71 \pm 2.51$. The constraints show a steeper but consistent relation with respect to theoretical results derived from the analysis of numerical simulations. 

Our results on the halo bias are consistent with previous measurements and with simulations in a $\Lambda$CDM framework. Some data points are affected by a relatively low SNR, as the number of galaxy clusters in the given redshift-richness bins is limited. These effects and the small number of richness bins prohibited the detection of any trend for the halo bias with the effective redshift of the clusters in each redshift bin. The measurements over the stacked profile of the full AK3 catalog give $b_h \sigma_8^2 = 1.2 \pm 0.1$ located at $M_{200c} = 4.9 \pm 0.3 \times 10^{13} M_{\odot}/\textit{h}$, in good agreement with $\Lambda$CDM predictions.

In the fitting procedure, the halo bias parameter is degenerate with the amplitude of the power spectrum $\sigma_8$. This last cosmological parameter is fitted with the theoretical mass-bias relations given in \cite{2004MNRAS.355..129S}, \cite{2010ApJ...724..878T} and \cite{2011ApJ...732..122B}. 
Assuming a flat $\Lambda$CDM cosmological model with $\Omega_m=1-\Omega_{\Lambda}=0.3$, we found $\sigma_8 = 1.01 \tiny{\substack{+0.05 \\ -0.05}}; \ 0.63 \tiny{\substack{+0.11 \\ -0.10}}; \ 0.66 \tiny{\substack{+0.19 \\ -0.27}}$ for the three above mentioned relations. These results present slight deviations with respect to the latest WMAP or Planck $\sigma_8$ estimates, but agree within $2\sigma$, with the exception of the results based on the \cite{2004MNRAS.355..129S} posterior, which shows a sharper distribution centered on a larger value of $\sigma_8$. Other works, based on cluster clustering, cluster counts and cosmic shear analyses, report values of $\sigma_8$ in agreement with our estimates within $2\sigma$, either assuming $\Omega_m$ fixed or free. The importance of the choice of the theoretical model for the halo bias also highlights the difficulty in constraining this cosmological parameter in a WL analysis.

For future work, we are interested in combining the inference on $\sigma_8$ with $\Omega_m$ to constrain the parameter $S_8 \equiv \sigma_8 \sqrt{\Omega_m / 0.3}$, which would compliment the study on $\sigma_8$ in this paper and $\Omega_m$ in \cite{2021arXiv210305653G}. 
Specifically, \cite{2021arXiv210305653G} provided a similar analysis on the AK3 galaxy clusters with a stacked shear profile up to 35 Mpc/\textit{h} and recovered consistent mass measurements with respect to \cite{2019MNRAS.484.1598B} and this paper. 
The binning scheme differs from this work since the cluster amplitude as a binning property was favored, while we opted for richness. This mainly affects the scaling relation between the mass and the cluster richness or amplitude. The impact of the truncation radius has been deeply investigated in \cite{2021arXiv210305653G}, here we performed a robust analysis of the covariances and cross-covariances and studied the effects of the lensing signal systematics in each patch of the field through the random signal. Both studies were carried out with independent numerical pipelines and followed a process of cross-validation among the KiDS collaboration. 

The methodology used in this work will constitute a baseline for future KiDS Data Releases \citep[][]{2019A&A...625A...2K} and similar but larger data sets that combine cluster and shear catalogs. Upcoming surveys, such as \textit{Euclid} \citep[][]{2019A&A...627A..23E} and LSST \citep[][]{2012arXiv1211.0310L}, will provide promising data sets allowing for further statistical analyses in deeper and wider fields. These data sets will be fundamental for the study of the halo properties such as mass and bias with stacked WL analyses, and will allow robust estimates of the main cosmological parameters.

\section*{Acknowledgements}
The authors acknowledge Shahab Joudaki for the thorough review of the manuscript. 
LI is grateful to Nicolas Martinet for a relevant discussion on the multiplicative bias and BPZ PDFs, and Peter Melchior for his fruitful comments on the cross-covariance.

Based on data products from observations made with ESO Telescopes at
the La Silla Paranal Observatory under programme IDs 177.A-3016,
177.A-3017 and 177.A-3018, and on data products produced by
Target/OmegaCEN, INAF-OACN, INAF-OAPD and the KiDS production team, on
behalf of the KiDS consortium. OmegaCEN and the KiDS production team
acknowledge support by NOVA and NWO-M grants. Members of INAF-OAPD and INAF-OACN also acknowledge the support from the Department of Physics \& Astronomy of the University of Padova, and of the Department of Physics of Univ. Federico II (Naples).  We acknowledge the KiDS collaboration for the public data realises and the various scientists working within it for the fruitful and helpful discussions.  
MS acknowledges financial contribution from contract ASI-INAF n.2017-14-H.0 and contract INAF mainstream project 1.05.01.86.10.
CG, FM and LM acknowledge the grants ASI n.I/023/12/0, ASI-INAF n.  2018-23-HH.0 and PRIN MIUR 2015 Cosmology and Fundamental Physics: illuminating the
Dark Universe with Euclid".  CG and LM are also supported by PRIN-MIUR 2017 WSCC32 ``Zooming into dark matter and proto-galaxies with massive lensing clusters''.  CG acknowledges support from the Italian Ministry of Foreign Affairs and International Cooperation, Directorate General for Country Promotion.

This paper makes use of the astronomical data analysis software \texttt{TOPCAT} \citep[][]{2005ASPC..347...29T}, and packages available in the Python's open scientific ecosystem, including \texttt{numpy} \citep[][]{2020NumPy-Array}, \texttt{scipy} \citep[][]{2020SciPy-NMeth}, \texttt{matplotlib} \citep[][]{Hunter:2007}, \texttt{astropy} \citep[][]{astropy:2018}, \texttt{emcee} \citep[][]{2013PASP..125..306F}, \texttt{ray} \citep[][]{DBLP:journals/corr/abs-1712-09381, DBLP:journals/corr/abs-1712-05889}, \texttt{colossus} \citep[][]{2018ApJS..239...35D} and \texttt{cluster toolkit}\footnote{\url{https://github.com/tmcclintock/cluster_toolkit/}}. 
Data analysis has been carried out with Fornax Physics Department computing cluster of the University Federico II of Naples.



\appendix

\section{Covariances}
\label{sec:Covariances}

Stacked WL signals are a comprehensive assessment of the profile given by a galaxy cluster population, but possible deviations arise due to statistical uncertainties and systematic biases. While the systematic noise can be efficiently corrected for using the random fields (see Appendix~\ref{sec:Random fields}), the statistical uncertainty of the stacked shear is essentially described by its covariance matrix. It can be decomposed into the contributions of large intrinsic variations of the shapes of galaxies \citep[shape noise, e.g.][]{2013MNRAS.432.1544M, 2015MNRAS.450.3675S, 2015MNRAS.452.3529V}, correlated and uncorrelated structures \citep[e.g.][]{2001A&A...370..743H, 2003MNRAS.339.1155H, 2011ApJ...726...48H, 2011ApJ...738...41U, 2015MNRAS.449.4264G}, and intrinsic scatter of the mass measurement \citep[e.g.][]{2001ApJ...547..560M, 2011MNRAS.416.1392G, 2011ApJ...740...25B, 2015MNRAS.449.4264G}. The statistical uncertainty is dominated by the shape noise of the sources \citep{2019MNRAS.482.1352M}, which has already been accounted for in Equation~\eqref{eq:error}. However, since galaxies contribute to the signal in different radial and redshift-richness bins, we may expect covariance terms to be significant between radii in identical and distinct stacked profiles. We therefore construct the covariance matrix from each pair of radial bins $ij$ over $N=1000$ bootstrap realizations of the source catalog,
\begin{equation}
C_{ij} = \frac{\sum_{n \in N} \left( \widetilde{\Delta\Sigma}_{i,n} - \overline{\widetilde{\Delta\Sigma}}_i \right) \left( \widetilde{\Delta\Sigma}_{j,n} - \overline{\widetilde{\Delta\Sigma}}_j \right)}{N-1} \ ,
\label{eq:covariance}
\end{equation}
with $\overline{\widetilde{\Delta\Sigma}} = \sum_{n \in N} \widetilde{\Delta\Sigma}_n / N$.

\begin{figure}
\includegraphics[width=\columnwidth]{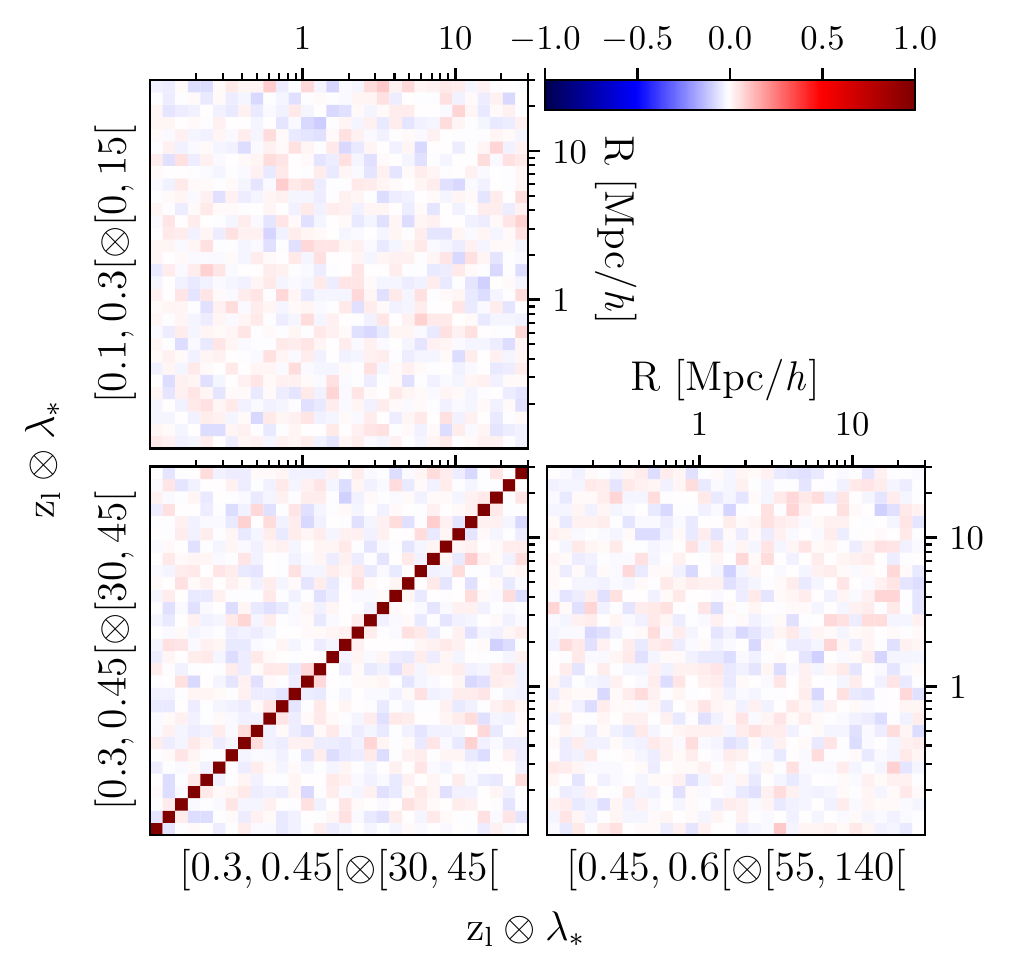}
\caption{Bootstrap correlation matrix of $\widetilde{\Delta\Sigma}$, computed from $z_l \otimes \lambda_{\ast}$ selected bins. Here, we investigate the bin $[0.3,0.45[ \otimes [30,45[$ correlated with itself (bottom left panel), with the bin $[0.1,0.3[ \otimes [0,15[$ (top panel) and with the bin $[0.45,0.6[ \otimes [55,140[$ (bottom right panel). The statistical uncertainty is mainly provided by the diagonal terms, while the off-diagonal terms are nearly consistent with zero, suggesting that radial and redshift-richness bins do not correlate.}
\label{fig:covariance}
\end{figure}

Figure~\ref{fig:covariance} displays the correlation matrices $R_{ij} = C_{ij}/\sqrt{C_{ii}C_{jj}}$ derived from the covariance profile for the cluster bin $z_l \otimes \lambda_{\ast} = [0.3,0.45[ \otimes [30,45[$ and the cross-covariances with the low and high redshift-richness bins $[0.1,0.3[ \otimes [0,15[$ and $[0.45,0.6[ \otimes [55,140[$. The correlation matrix does not show any strong contribution from off-diagonal terms, while the diagonal components encompass the majority of the statistical noise. We still consider the full covariance of each individual cluster bin to quantify the statistical uncertainty of the stacked WL signal, in order to account for the dependency between the radii of the bin when fitting the data. Furthermore, we combine uncertainties of the galaxy cluster signal and the random signal detailed in Appendix~\ref{sec:Random fields} by summing their covariances. These matrices are used when measuring the halo parameters in Section~\ref{sec:MCMC method}.

\section{Random fields}
\label{sec:Random fields} 

We performed stacked shear analysis around random lens points following the same process used in Section~\ref{sec:Method}. This spurious signal characterizes the residual systematic effects, usually coming either from the edges of the detector \citep[]{2015ApJ...806....1M}, the imperfect correction of optical distortion \citep[]{2005MNRAS.361.1287M} or the incorrect estimation of the redshift \citep[]{2019MNRAS.482.1352M}. If none of these effects impact the profile, the random stacked shear should vanish, while it deviates from zero as soon as the systematic bias is apparent \citep[]{2015ApJ...806....1M}. The random signal is finally subtracted from the shear profiles of the stacked bins to correct for these uncertainties.

We built a random catalog over the full [RA,Dec] sources range considering the K450 footprint of masked areas. Each equatorial random position is uniformly sampled over a \texttt{Nside=2048} pixel \textsc{healpix} map and associated to a redshift random position. We sample random redshift from an inverse transform method, assuming AK3 redshifts to follow a Weibull distribution \citep[e.g.][]{2003MNRAS.346..994P},
\begin{equation}
n \equiv \frac{\beta}{\Gamma \left(\frac{1+\alpha}{\beta} \right)} \frac{1}{z_0} \left( \frac{z_l}{z_0} \right)^{\alpha} \exp \left[ - \left( \frac{z_l}{z_0} \right)^{\beta} \right] \ .
\label{eq:Weibull nz}
\end{equation}
The parameters $\alpha$, $\beta$ and $z_0$ are marginalized and constrained to track the real distribution of AK3 redshifts. We find
\begin{description}
  \item[$\bullet$] $\alpha = 1.06$
  \item[$\bullet$] $\beta = 4.81$
  \item[$\bullet$] $z_0 = 0.59$
\end{description}

\begin{figure}
	\includegraphics[width=\columnwidth]{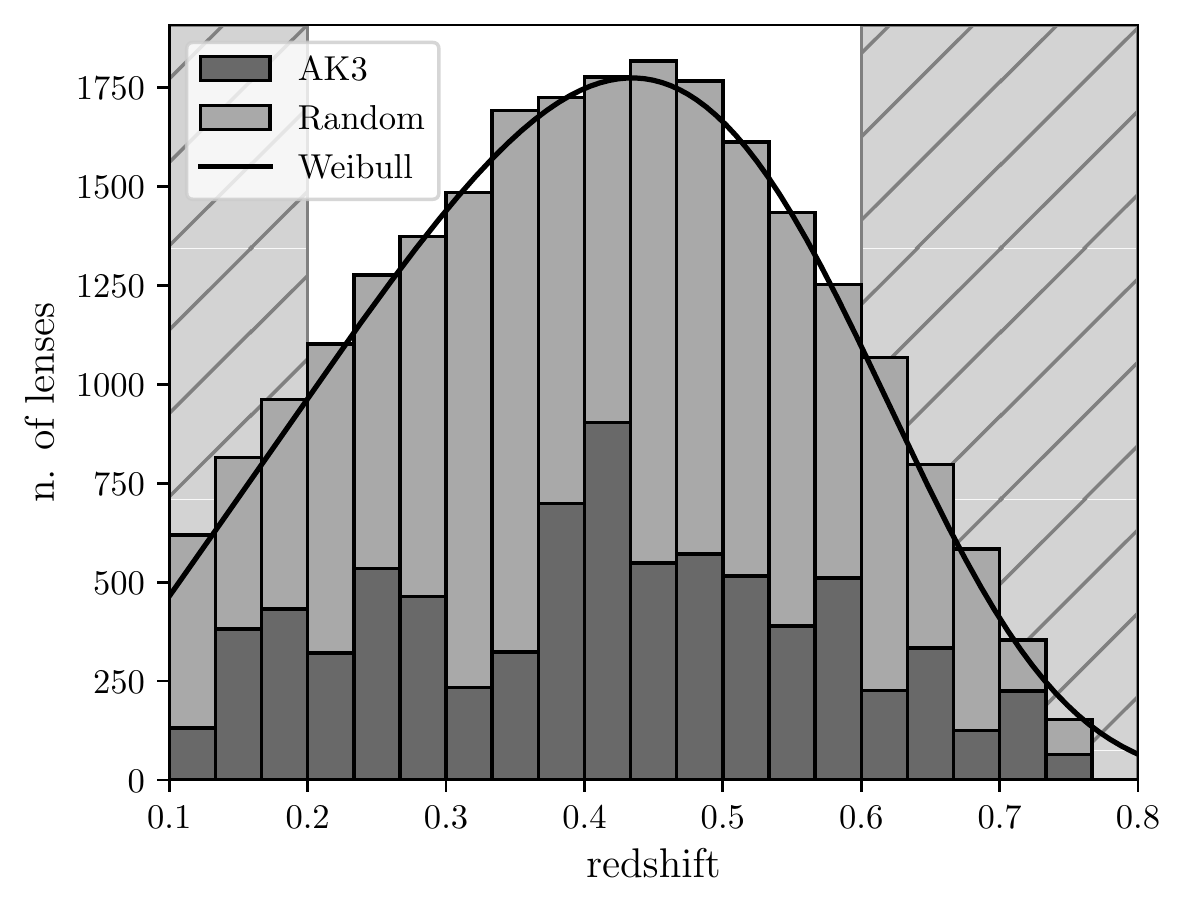}
    \caption{The redshift distribution of AK3 and random lenses. The random redshifts are sampled with an inverse transform method from the PDF described in Equation~\eqref{eq:Weibull nz}. The black curve describes this function and aims at simulating the distribution of the AK3 redshifts. The shaded regions delineate the redshift range of selected clusters discussed in Section~\ref{sec:Foreground clusters}.}
    \label{fig:randomz}
\end{figure}

Figure~\ref{fig:randomz} shows the distribution of random and AK3 lenses. Random redshifts follow the Weibull distribution and tend to recover the same distribution as clusters of galaxies for a more realistic representation of the random signal. 

\cite{2016ApJS..224....1R} suggest an efficient way to generate a random richness component from a depth map of the source catalog. However, their study is based on the redMaPPer algorithm for cluster detection which considerably differs from AMICO. 
Moreover, due to the absence of a depth map in K450  we cannot assign richness parameters to our random catalog. Still, the presence of random redshifts is a robust feature for the random catalog as we can associate the stacked random signal to each redshift bin. Finally, the number of random points exceed the number of real galaxy clusters by 15976 lenses in order to fully cover the 3D field of AK3 lenses.

\begin{figure*}
    \centering
    \includegraphics[width=\textwidth]{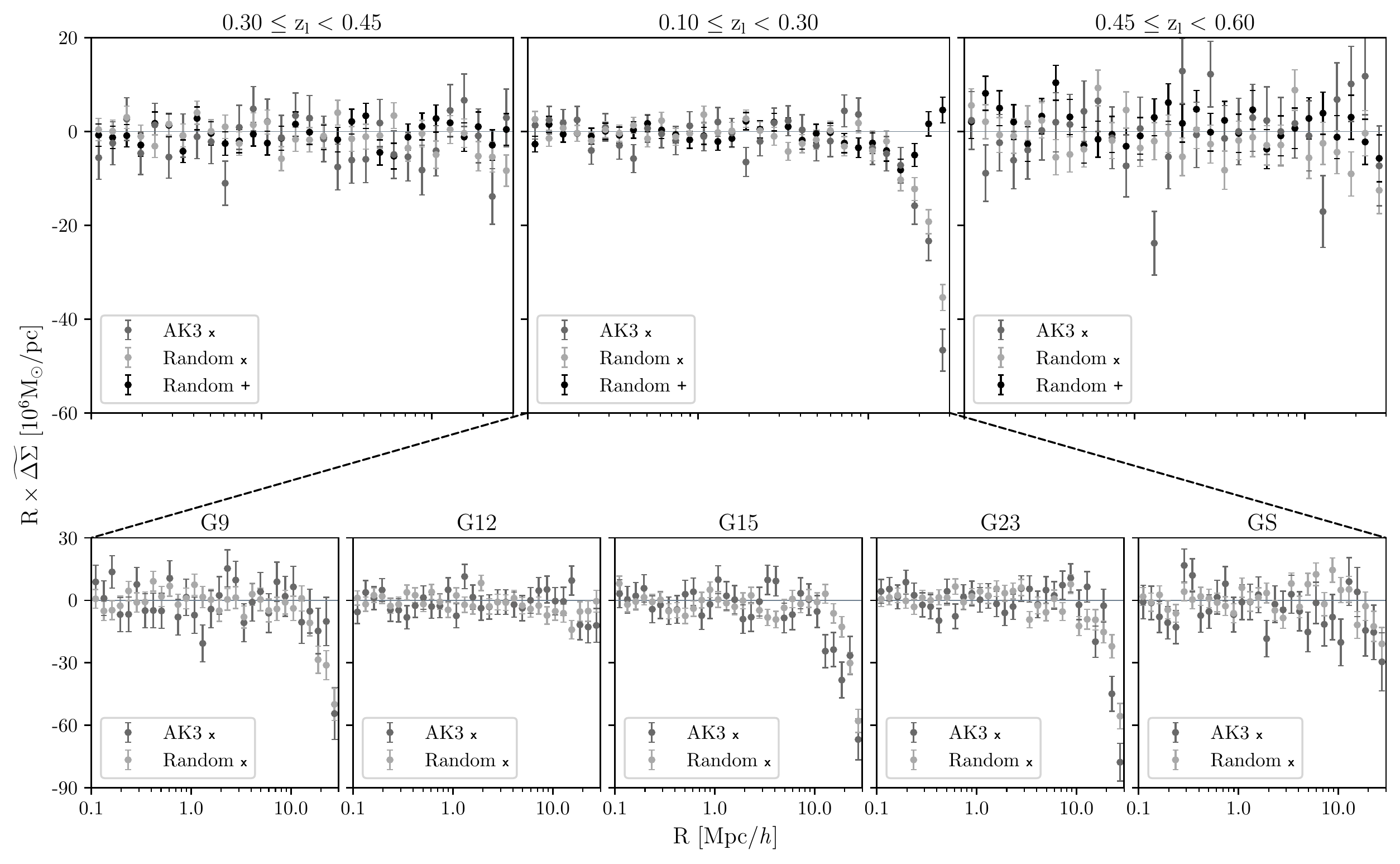}
    \caption{Differential density profiles of the cross component of AK3 lenses and the cross and tangential components of random lenses in the three redshift bins. The cross signals in the five KiDS DR3 patches of the lower redshift bin are also displayed. A significant deviation from the zero horizontal line indicates the presence of a systematic effect. It reveals an incomplete correction of the PSF of galaxies located close to the FoV edge.}
    \label{fig:random}
\end{figure*}

A simple test to check the correct processing of the subtraction of the systematics is to look at the tangential and cross stacked shear profile of the random lenses. The top panels of Figure~\ref{fig:random} presents three different profiles derived from AK3 cross, random cross and random tangential signals. While the tangential component of random points remain consistent with zero, the cross signal of the lower redshift bin reveals that systematics largely impact the shear in the last radial bins, and consequently might distort the estimation of the halo bias if no correction is applied. Looking deeply in the cross signal of the five KiDS DR3 patches, we observe that only three of them are significantly affected. 
We relate this systematic to the geometry of the field, which at some point is irregular in those specific patches. Indeed, since the lower redshift bin needs a larger field of view (FoV) to compute stacked shear over a fixed large radial profile, the resulting signal is much more sensitive to the discontinuities of the field (e.g. isolated tiles). \cite{2013PASJ...65..104H} suggest that the point spread function (PSF) in the shape measurement of galaxies located at the edge of the FoV is imperfectly corrected. This biased PSF anisotropy sensitively impacts the shear of galaxies, which consequently breaks the symmetry of the intrinsic ellipticity and leads to a non-zero cross component. However, since the subtraction of the two signals gives a signal consistent with zero, the correction suppresses this systematic effect and the final version of the data is ready for the analysis (see Section~\ref{sec:MCMC method}).

\section{Color-color selections}
\label{sec:Colour-colour selections}

In this section we compare the color-color selection discussed in Section~\ref{sec:Background galaxies} with an additional \textit{gri}-CC cut. More specifically, Figure~3 in \cite{2010MNRAS.405..257M} lays out in colored areas various populations of sources for three different Subaru clusters (e.g. cluster members in green). 
The paper shows in particular \textit{gri}-CC selected sources in the galaxy cluster A1703, for which \cite{2008ApJ...685L...9B} and \cite{2009ApJ...699.1038O} initially performed a WL analysis. Two singular areas in the color-color plane are clearly identified as background sources of A1703, efficiently selected at $z_s \gtrsim 0.6$ and displayed in blue/red in the figure of the study. They present the following segmentation
\begin{equation}
\begin{aligned}
\left[ \left( g-r < 2.17(r-i)-0.37 \right) \wedge \left( g-r < 1.85-0.6(r-i) \right) \right] \\
\vee \left( g-r < 0.47-0.4(r-i) \right) \vee \left( r-i < -0.06 \right) \ .
\end{aligned}
\label{eq:medezinski selection}
\end{equation}

\begin{figure}
    \includegraphics[width=\columnwidth]{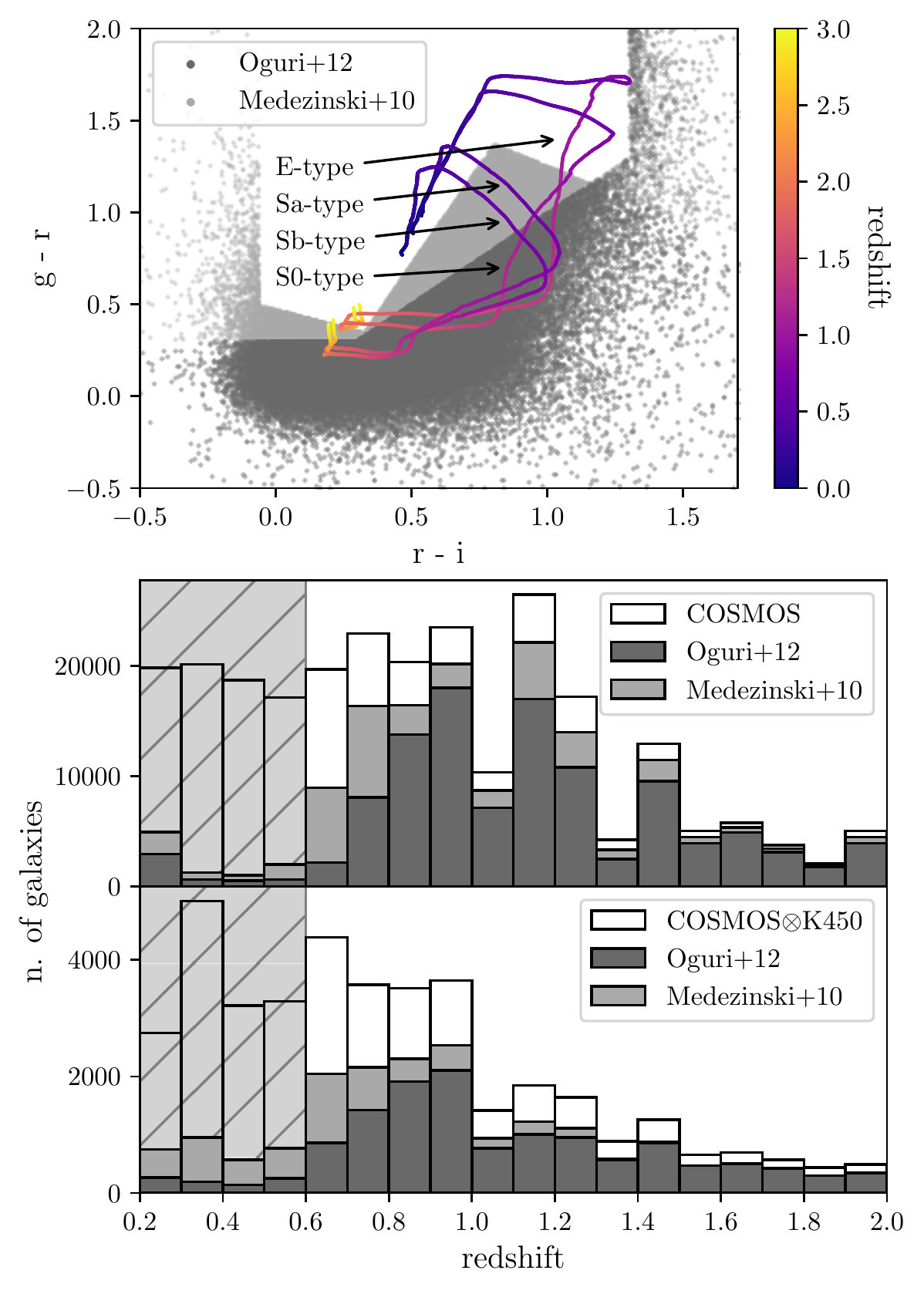}
    \caption{\textit{Top panel}: (\textit{r}-\textit{i}) vs (\textit{g}-\textit{r}) diagram. We show the selections discussed in this article, following previous complementary works \citep[Equations~\ref{eq:oguri selection} and~\ref{eq:medezinski selection},][respectively]{2012MNRAS.420.3213O, 2010MNRAS.405..257M}. We additionally show the evolving tracks of spiral, lenticular and elliptical galaxies in the \textit{gri}-CC plane obtained using the \textsc{galev} code \citep{2009MNRAS.396..462K}. \textit{Bottom panels}: COSMOS \citep{2009ApJ...690.1236I} and COSMOS$\otimes$K450 photometric redshift distributions for the full samples and for their dedicated \textit{gri}-CC selections. The shaded region highlights the contamination area, which corresponds to the cluster redshift range $\left[ 0.1, 0.6 \right[$ covered in Section~\ref{sec:Foreground clusters}.}
    \label{fig:cosmos}
\end{figure}

In order to evaluate the efficiency of the \textit{gri}-CC cuts explored in this study (Equations~\ref{eq:oguri selection} and~\ref{eq:medezinski selection}), we are interested in testing them over the COSMOS 30-Bands photometric catalog\footnote{\url{https://irsa.ipac.caltech.edu/data/COSMOS/}} \citep{2009ApJ...690.1236I}.
The full sample consists of 385,065 galaxies with very accurate photometric redshifts reliable up to magnitude $\textit{i} < 25$. In Figure~\ref{fig:cosmos}, we present the COSMOS sources selected with the two \textit{gri}-CC criteria. As a comparison, we generate evolving tracks using the \textsc{galev}\footnote{\url{http://www.galev.org/}} code \citep{2009MNRAS.396..462K} for the Hubble - de Vaucouleurs galaxy morphological types (Non-barred spiral Sa-type, Barred-spiral Sb-type, Lenticular S0-type and Elliptical E-type). We are interested in the contamination of objects belonging to the redshift range of $0.2 < z_s < 0.6$, in agreement with the selection of clusters done in Section~\ref{sec:Foreground clusters}. The cut shown in \cite{2012MNRAS.420.3213O} encompasses 125,754 galaxies with 96.2\% background sources for the corresponding redshift threshold $z_s \geq 0.6$. On the other hand, the selection done by \cite{2010MNRAS.405..257M} counts 170,429 (35.5\% more), and 94.6\% of them lie over $z_s \geq 0.6$. Both cuts efficiently remove contaminating members, we see a higher number of background COSMOS galaxies for Equation~\eqref{eq:medezinski selection}, while the contamination fraction given by Equation~\eqref{eq:oguri selection} is fully consistent with \cite{2017MNRAS.472.1946S}. 

Besides this observation, a more reliable analysis would be to consider the cross-matched catalog COSMOS$\otimes$K450 as the 47,619 COSMOS sources within K450. The lower panel of the figure provides this distribution and shows 14,857 of them to be filtered by \cite{2012MNRAS.420.3213O} and 20,540 by \cite{2010MNRAS.405..257M}. Respectively, 94.3\% and 85.2\% of selected sources appear to be uncontaminated. These statistics highlights a higher contamination from Equation~\eqref{eq:medezinski selection} and a more efficient removal of contaminated K450 sources for Equation~\eqref{eq:oguri selection}, but still has some drawbacks due to the limited number of objects. Another explanation for the main difference between the two cuts is the unequal reduction of galaxies from COSMOS to the cross match data set. K450 sources in COSMOS are few at $z_s > 1$, where \cite{2010MNRAS.405..257M} is consequently selecting more sources than \cite{2012MNRAS.420.3213O} in COSMOS only, while the proportion of galaxies at $z_s < 1$ remains high in both catalogs. In that sense, we prefer to retain Equation~\eqref{eq:oguri selection} as the principal \textit{gri}-CC selection for this work. 

\bibliographystyle{mnras}
\bibliography{biblio.bib} 

\begin{thebibliography}{}
\makeatletter
\relax
\def\mn@urlcharsother{\let\do\@makeother \do\$\do\&\do\#\do\^\do\_\do\%\do\~}
\def\mn@doi{\begingroup\mn@urlcharsother \@ifnextchar [ {\mn@doi@}
  {\mn@doi@[]}}
\def\mn@doi@[#1]#2{\def\@tempa{#1}\ifx\@tempa\@empty \href
  {http://dx.doi.org/#2} {doi:#2}\else \href {http://dx.doi.org/#2} {#1}\fi
  \endgroup}
\def\mn@eprint#1#2{\mn@eprint@#1:#2::\@nil}
\def\mn@eprint@arXiv#1{\href {http://arxiv.org/abs/#1} {{\tt arXiv:#1}}}
\def\mn@eprint@dblp#1{\href {http://dblp.uni-trier.de/rec/bibtex/#1.xml}
  {dblp:#1}}
\def\mn@eprint@#1:#2:#3:#4\@nil{\def\@tempa {#1}\def\@tempb {#2}\def\@tempc
  {#3}\ifx \@tempc \@empty \let \@tempc \@tempb \let \@tempb \@tempa \fi \ifx
  \@tempb \@empty \def\@tempb {arXiv}\fi \@ifundefined
  {mn@eprint@\@tempb}{\@tempb:\@tempc}{\expandafter \expandafter \csname
  mn@eprint@\@tempb\endcsname \expandafter{\@tempc}}}

\bibitem[\protect\citeauthoryear{{Angelinelli}, {Vazza}, {Giocoli}, {Ettori},
  {Jones}, {Brunetti}, {Br{\"u}ggen}  \& {Eckert}}{{Angelinelli}
  et~al.}{2020}]{angelinelli20}
{Angelinelli} M.,  {Vazza} F.,  {Giocoli} C.,  {Ettori} S.,  {Jones} T.~W.,
  {Brunetti} G.,  {Br{\"u}ggen} M.,   {Eckert} D.,  2020, \mn@doi [\mnras]
  {10.1093/mnras/staa975}, \href
  {https://ui.adsabs.harvard.edu/abs/2020MNRAS.495..864A} {495, 864}

\bibitem[\protect\citeauthoryear{{Asgari} et~al.,}{{Asgari}
  et~al.}{2021}]{2021A&A...645A.104A}
{Asgari} M.,  et~al., 2021, \mn@doi [\aap] {10.1051/0004-6361/202039070}, \href
  {https://ui.adsabs.harvard.edu/abs/2021A&A...645A.104A} {645, A104}

\bibitem[\protect\citeauthoryear{{Astropy Collaboration} et~al.,}{{Astropy
  Collaboration} et~al.}{2018}]{astropy:2018}
{Astropy Collaboration} et~al., 2018, \mn@doi [\aj] {10.3847/1538-3881/aabc4f},
  \href {https://ui.adsabs.harvard.edu/abs/2018AJ....156..123A} {156, 123}

\bibitem[\protect\citeauthoryear{{Baltz}, {Marshall}  \& {Oguri}}{{Baltz}
  et~al.}{2009}]{2009JCAP...01..015B}
{Baltz} E.~A.,  {Marshall} P.,   {Oguri} M.,  2009, \mn@doi [\jcap]
  {10.1088/1475-7516/2009/01/015}, \href
  {https://ui.adsabs.harvard.edu/abs/2009JCAP...01..015B} {2009, 015}

\bibitem[\protect\citeauthoryear{{Bartelmann} \& {Schneider}}{{Bartelmann} \&
  {Schneider}}{2001}]{2001PhR...340..291B}
{Bartelmann} M.,  {Schneider} P.,  2001, \mn@doi [physrep]
  {10.1016/S0370-1573(00)00082-X}, \href
  {http://adsabs.harvard.edu/abs/2001PhR...340..291B} {340, 291}

\bibitem[\protect\citeauthoryear{{Becker} \& {Kravtsov}}{{Becker} \&
  {Kravtsov}}{2011}]{2011ApJ...740...25B}
{Becker} M.~R.,  {Kravtsov} A.~V.,  2011, \mn@doi [\apj]
  {10.1088/0004-637X/740/1/25}, \href
  {https://ui.adsabs.harvard.edu/abs/2011ApJ...740...25B} {740, 25}

\bibitem[\protect\citeauthoryear{{Bellagamba}, {Maturi}, {Hamana},
  {Meneghetti}, {Miyazaki}  \& {Moscardini}}{{Bellagamba}
  et~al.}{2011}]{2011MNRAS.413.1145B}
{Bellagamba} F.,  {Maturi} M.,  {Hamana} T.,  {Meneghetti} M.,  {Miyazaki} S.,
   {Moscardini} L.,  2011, \mn@doi [mnras] {10.1111/j.1365-2966.2011.18202.x},
  \href {http://adsabs.harvard.edu/abs/2011MNRAS.413.1145B} {413, 1145}

\bibitem[\protect\citeauthoryear{{Bellagamba}, {Roncarelli}, {Maturi}  \&
  {Moscardini}}{{Bellagamba} et~al.}{2018}]{2018MNRAS.473.5221B}
{Bellagamba} F.,  {Roncarelli} M.,  {Maturi} M.,   {Moscardini} L.,  2018,
  \mn@doi [\mnras] {10.1093/mnras/stx2701}, \href
  {https://ui.adsabs.harvard.edu/abs/2018MNRAS.473.5221B} {473, 5221}

\bibitem[\protect\citeauthoryear{{Bellagamba} et~al.,}{{Bellagamba}
  et~al.}{2019}]{2019MNRAS.484.1598B}
{Bellagamba} F.,  et~al., 2019, \mn@doi [\mnras] {10.1093/mnras/stz090}, \href
  {https://ui.adsabs.harvard.edu/abs/2019MNRAS.484.1598B} {484, 1598}

\bibitem[\protect\citeauthoryear{{Ben\'{i}tez}}{{Ben\'{i}tez}}{2000}]{2000ApJ...536..571B}
{Ben\'{i}tez} N.,  2000, \mn@doi [\apj] {10.1086/308947}, \href
  {https://ui.adsabs.harvard.edu/abs/2000ApJ...536..571B} {536, 571}

\bibitem[\protect\citeauthoryear{{Bhattacharya}, {Heitmann}, {White},
  {Luki{\'c}}, {Wagner}  \& {Habib}}{{Bhattacharya}
  et~al.}{2011}]{2011ApJ...732..122B}
{Bhattacharya} S.,  {Heitmann} K.,  {White} M.,  {Luki{\'c}} Z.,  {Wagner} C.,
   {Habib} S.,  2011, \mn@doi [\apj] {10.1088/0004-637X/732/2/122}, \href
  {https://ui.adsabs.harvard.edu/abs/2011ApJ...732..122B} {732, 122}

\bibitem[\protect\citeauthoryear{{Bocquet} et~al.,}{{Bocquet}
  et~al.}{2019}]{2019ApJ...878...55B}
{Bocquet} S.,  et~al., 2019, \mn@doi [\apj] {10.3847/1538-4357/ab1f10}, \href
  {https://ui.adsabs.harvard.edu/abs/2019ApJ...878...55B} {878, 55}

\bibitem[\protect\citeauthoryear{{Borgani}}{{Borgani}}{2008}]{2008LNP...740..287B}
{Borgani} S.,  2008, {Cosmology with Clusters of Galaxies}.
Springer Netherlands, p.~24, \mn@doi{10.1007/978-1-4020-6941-3_9}

\bibitem[\protect\citeauthoryear{Broadhurst, Takada, Umetsu, Kong, Arimoto,
  Chiba  \& Futamase}{Broadhurst et~al.}{2005}]{1538-4357-619-2-L143}
Broadhurst T.,  Takada M.,  Umetsu K.,  Kong X.,  Arimoto N.,  Chiba M.,
  Futamase T.,  2005, The Astrophysical Journal Letters, 619, L143

\bibitem[\protect\citeauthoryear{{Broadhurst}, {Umetsu}, {Medezinski}, {Oguri}
  \& {Rephaeli}}{{Broadhurst} et~al.}{2008}]{2008ApJ...685L...9B}
{Broadhurst} T.,  {Umetsu} K.,  {Medezinski} E.,  {Oguri} M.,   {Rephaeli} Y.,
  2008, \mn@doi [\apjl] {10.1086/592400}, \href
  {https://ui.adsabs.harvard.edu/abs/2008ApJ...685L...9B} {685, L9}

\bibitem[\protect\citeauthoryear{{Bullock}, {Kolatt}, {Sigad}, {Somerville},
  {Kravtsov}, {Klypin}, {Primack}  \& {Dekel}}{{Bullock}
  et~al.}{2001}]{2001MNRAS.321..559B}
{Bullock} J.~S.,  {Kolatt} T.~S.,  {Sigad} Y.,  {Somerville} R.~S.,  {Kravtsov}
  A.~V.,  {Klypin} A.~A.,  {Primack} J.~R.,   {Dekel} A.,  2001, \mn@doi
  [\mnras] {10.1046/j.1365-8711.2001.04068.x}, \href
  {https://ui.adsabs.harvard.edu/abs/2001MNRAS.321..559B} {321, 559}

\bibitem[\protect\citeauthoryear{{Capaccioli} \& {Schipani}}{{Capaccioli} \&
  {Schipani}}{2011}]{2011Msngr.146....2C}
{Capaccioli} M.,  {Schipani} P.,  2011, The Messenger, \href
  {https://ui.adsabs.harvard.edu/abs/2011Msngr.146....2C} {146, 2}

\bibitem[\protect\citeauthoryear{{Child}, {Habib}, {Heitmann}, {Frontiere},
  {Finkel}, {Pope}  \& {Morozov}}{{Child} et~al.}{2018}]{2018ApJ...859...55C}
{Child} H.~L.,  {Habib} S.,  {Heitmann} K.,  {Frontiere} N.,  {Finkel} H.,
  {Pope} A.,   {Morozov} V.,  2018, \mn@doi [\apj] {10.3847/1538-4357/aabf95},
  \href {https://ui.adsabs.harvard.edu/abs/2018ApJ...859...55C} {859, 55}

\bibitem[\protect\citeauthoryear{{Coe} et~al.,}{{Coe} et~al.}{2012}]{coe12}
{Coe} D.,  et~al., 2012, \mn@doi [\apj] {10.1088/0004-637X/757/1/22}, \href
  {http://adsabs.harvard.edu/abs/2012ApJ...757...22C} {757, 22}

\bibitem[\protect\citeauthoryear{{Cole} \& {Kaiser}}{{Cole} \&
  {Kaiser}}{1989}]{1989MNRAS.237.1127C}
{Cole} S.,  {Kaiser} N.,  1989, \mn@doi [\mnras] {10.1093/mnras/237.4.1127},
  \href {https://ui.adsabs.harvard.edu/abs/1989MNRAS.237.1127C} {237, 1127}

\bibitem[\protect\citeauthoryear{{Costanzi} et~al.,}{{Costanzi}
  et~al.}{2019}]{2019MNRAS.488.4779C}
{Costanzi} M.,  et~al., 2019, \mn@doi [\mnras] {10.1093/mnras/stz1949}, \href
  {https://ui.adsabs.harvard.edu/abs/2019MNRAS.488.4779C} {488, 4779}

\bibitem[\protect\citeauthoryear{{Covone}, {Sereno}, {Kilbinger}  \&
  {Cardone}}{{Covone} et~al.}{2014}]{2014ApJ...784L..25C}
{Covone} G.,  {Sereno} M.,  {Kilbinger} M.,   {Cardone} V.~F.,  2014, \mn@doi
  [apjl] {10.1088/2041-8205/784/2/L25}, \href
  {http://adsabs.harvard.edu/abs/2014ApJ...784L..25C} {784, L25}

\bibitem[\protect\citeauthoryear{{DES Collaboration} et~al.,}{{DES
  Collaboration} et~al.}{2021}]{2021arXiv210513549D}
{DES Collaboration} et~al., 2021, arXiv e-prints, \href
  {https://ui.adsabs.harvard.edu/abs/2021arXiv210513549D} {p. arXiv:2105.13549}

\bibitem[\protect\citeauthoryear{{Despali}, {Giocoli}  \& {Tormen}}{{Despali}
  et~al.}{2014}]{despali14}
{Despali} G.,  {Giocoli} C.,   {Tormen} G.,  2014, \mn@doi [\mnras]
  {10.1093/mnras/stu1393}, \href
  {http://adsabs.harvard.edu/abs/2014MNRAS.443.3208D} {443, 3208}

\bibitem[\protect\citeauthoryear{{Despali}, {Giocoli}, {Angulo}, {Tormen},
  {Sheth}, {Baso}  \& {Moscardini}}{{Despali} et~al.}{2016}]{despali16}
{Despali} G.,  {Giocoli} C.,  {Angulo} R.~E.,  {Tormen} G.,  {Sheth} R.~K.,
  {Baso} G.,   {Moscardini} L.,  2016, \mn@doi [\mnras]
  {10.1093/mnras/stv2842}, \href
  {http://adsabs.harvard.edu/abs/2016MNRAS.456.2486D} {456, 2486}

\bibitem[\protect\citeauthoryear{{Diemer}}{{Diemer}}{2018}]{2018ApJS..239...35D}
{Diemer} B.,  2018, \mn@doi [\apjs] {10.3847/1538-4365/aaee8c}, \href
  {https://ui.adsabs.harvard.edu/abs/2018ApJS..239...35D} {239, 35}

\bibitem[\protect\citeauthoryear{{Diemer} \& {Joyce}}{{Diemer} \&
  {Joyce}}{2019}]{2019ApJ...871..168D}
{Diemer} B.,  {Joyce} M.,  2019, \mn@doi [\apj] {10.3847/1538-4357/aafad6},
  \href {https://ui.adsabs.harvard.edu/abs/2019ApJ...871..168D} {871, 168}

\bibitem[\protect\citeauthoryear{{Diemer} \& {Kravtsov}}{{Diemer} \&
  {Kravtsov}}{2015}]{2015ApJ...799..108D}
{Diemer} B.,  {Kravtsov} A.~V.,  2015, \mn@doi [\apj]
  {10.1088/0004-637X/799/1/108}, \href
  {https://ui.adsabs.harvard.edu/abs/2015ApJ...799..108D} {799, 108}

\bibitem[\protect\citeauthoryear{{Donahue} et~al.,}{{Donahue}
  et~al.}{2016}]{donahue16}
{Donahue} M.,  et~al., 2016, \mn@doi [\apj] {10.3847/0004-637X/819/1/36}, \href
  {http://adsabs.harvard.edu/abs/2016ApJ...819...36D} {819, 36}

\bibitem[\protect\citeauthoryear{{Driver} et~al.,}{{Driver}
  et~al.}{2011}]{2011MNRAS.413..971D}
{Driver} S.~P.,  et~al., 2011, \mn@doi [\mnras]
  {10.1111/j.1365-2966.2010.18188.x}, \href
  {https://ui.adsabs.harvard.edu/abs/2011MNRAS.413..971D} {413, 971}

\bibitem[\protect\citeauthoryear{{Duffy}, {Schaye}, {Kay}  \& {Dalla
  Vecchia}}{{Duffy} et~al.}{2008}]{2008MNRAS.390L..64D}
{Duffy} A.~R.,  {Schaye} J.,  {Kay} S.~T.,   {Dalla Vecchia} C.,  2008, \mn@doi
  [\mnras] {10.1111/j.1745-3933.2008.00537.x}, \href
  {https://ui.adsabs.harvard.edu/abs/2008MNRAS.390L..64D} {390, L64}

\bibitem[\protect\citeauthoryear{{Dutton} \& {Macci{\`o}}}{{Dutton} \&
  {Macci{\`o}}}{2014}]{2014MNRAS.441.3359D}
{Dutton} A.~A.,  {Macci{\`o}} A.~V.,  2014, \mn@doi [\mnras]
  {10.1093/mnras/stu742}, \href
  {https://ui.adsabs.harvard.edu/abs/2014MNRAS.441.3359D} {441, 3359}

\bibitem[\protect\citeauthoryear{{Eisenstein} \& {Hu}}{{Eisenstein} \&
  {Hu}}{1998}]{1998ApJ...496..605E}
{Eisenstein} D.~J.,  {Hu} W.,  1998, \mn@doi [\apj] {10.1086/305424}, \href
  {https://ui.adsabs.harvard.edu/abs/1998ApJ...496..605E} {496, 605}

\bibitem[\protect\citeauthoryear{{Eisenstein} \& {Hu}}{{Eisenstein} \&
  {Hu}}{1999}]{1999ApJ...511....5E}
{Eisenstein} D.~J.,  {Hu} W.,  1999, \mn@doi [\apj] {10.1086/306640}, \href
  {https://ui.adsabs.harvard.edu/abs/1999ApJ...511....5E} {511, 5}

\bibitem[\protect\citeauthoryear{{Euclid Collaboration} et~al.,}{{Euclid
  Collaboration} et~al.}{2019}]{2019A&A...627A..23E}
{Euclid Collaboration} et~al., 2019, \mn@doi [\aap]
  {10.1051/0004-6361/201935088}, \href
  {https://ui.adsabs.harvard.edu/abs/2019A&A...627A..23E} {627, A23}

\bibitem[\protect\citeauthoryear{{Fenech Conti}, {Herbonnet}, {Hoekstra},
  {Merten}, {Miller}  \& {Viola}}{{Fenech Conti}
  et~al.}{2017}]{2017MNRAS.467.1627F}
{Fenech Conti} I.,  {Herbonnet} R.,  {Hoekstra} H.,  {Merten} J.,  {Miller} L.,
    {Viola} M.,  2017, \mn@doi [mnras] {10.1093/mnras/stx200}, \href
  {http://adsabs.harvard.edu/abs/2017MNRAS.467.1627F} {467, 1627}

\bibitem[\protect\citeauthoryear{{Foreman-Mackey}, {Hogg}, {Lang}  \&
  {Goodman}}{{Foreman-Mackey} et~al.}{2013}]{2013PASP..125..306F}
{Foreman-Mackey} D.,  {Hogg} D.~W.,  {Lang} D.,   {Goodman} J.,  2013, \mn@doi
  [\pasp] {10.1086/670067}, \href
  {https://ui.adsabs.harvard.edu/abs/2013PASP..125..306F} {125, 306}

\bibitem[\protect\citeauthoryear{{Fuller}}{{Fuller}}{1987}]{1987mem..book.....F}
{Fuller} W.~A.,  1987, {Measurement error models}.
Wiley-Interscience

\bibitem[\protect\citeauthoryear{{Gelman} \& {Rubin}}{{Gelman} \&
  {Rubin}}{1992}]{1992StaSc...7..457G}
{Gelman} A.,  {Rubin} D.~B.,  1992, \mn@doi [Statistical Science]
  {10.1214/ss/1177011136}, \href
  {https://ui.adsabs.harvard.edu/abs/1992StaSc...7..457G} {7, 457}

\bibitem[\protect\citeauthoryear{{George} et~al.,}{{George}
  et~al.}{2012}]{2012ApJ...757....2G}
{George} M.~R.,  et~al., 2012, \mn@doi [apj] {10.1088/0004-637X/757/1/2}, \href
  {http://adsabs.harvard.edu/abs/2012ApJ...757....2G} {757, 2}

\bibitem[\protect\citeauthoryear{{Giocoli}, {Tormen}, {Sheth}  \& {van den
  Bosch}}{{Giocoli} et~al.}{2010a}]{giocoli10a}
{Giocoli} C.,  {Tormen} G.,  {Sheth} R.~K.,   {van den Bosch} F.~C.,  2010a,
  \mn@doi [\mnras] {10.1111/j.1365-2966.2010.16311.x}, \href
  {http://adsabs.harvard.edu/abs/2010MNRAS.404..502G} {404, 502}

\bibitem[\protect\citeauthoryear{{Giocoli}, {Bartelmann}, {Sheth}  \&
  {Cacciato}}{{Giocoli} et~al.}{2010b}]{giocoli10b}
{Giocoli} C.,  {Bartelmann} M.,  {Sheth} R.~K.,   {Cacciato} M.,  2010b,
  \mn@doi [\mnras] {10.1111/j.1365-2966.2010.17108.x}, \href
  {http://adsabs.harvard.edu/abs/2010MNRAS.408..300G} {408, 300}

\bibitem[\protect\citeauthoryear{{Giocoli}, {Tormen}  \& {Sheth}}{{Giocoli}
  et~al.}{2012}]{giocoli12b}
{Giocoli} C.,  {Tormen} G.,   {Sheth} R.~K.,  2012, \mn@doi [\mnras]
  {10.1111/j.1365-2966.2012.20594.x}, \href
  {http://adsabs.harvard.edu/abs/2012MNRAS.422..185G} {422, 185}

\bibitem[\protect\citeauthoryear{{Giocoli} et~al.,}{{Giocoli}
  et~al.}{2021}]{2021arXiv210305653G}
{Giocoli} C.,  et~al., 2021, arXiv e-prints, \href
  {https://ui.adsabs.harvard.edu/abs/2021arXiv210305653G} {p. arXiv:2103.05653}

\bibitem[\protect\citeauthoryear{{Golse} \& {Kneib}}{{Golse} \&
  {Kneib}}{2002}]{2002A&A...390..821G}
{Golse} G.,  {Kneib} J.~P.,  2002, \mn@doi [\aap] {10.1051/0004-6361:20020639},
  \href {https://ui.adsabs.harvard.edu/abs/2002A&A...390..821G} {390, 821}

\bibitem[\protect\citeauthoryear{{Goodman} \& {Weare}}{{Goodman} \&
  {Weare}}{2010}]{2010CAMCS...5...65G}
{Goodman} J.,  {Weare} J.,  2010, \mn@doi [Communications in Applied
  Mathematics and Computational Science] {10.2140/camcos.2010.5.65}, \href
  {https://ui.adsabs.harvard.edu/abs/2010CAMCS...5...65G} {5, 65}

\bibitem[\protect\citeauthoryear{{Gruen}, {Bernstein}, {Lam}  \&
  {Seitz}}{{Gruen} et~al.}{2011}]{2011MNRAS.416.1392G}
{Gruen} D.,  {Bernstein} G.~M.,  {Lam} T.~Y.,   {Seitz} S.,  2011, \mn@doi
  [\mnras] {10.1111/j.1365-2966.2011.19135.x}, \href
  {https://ui.adsabs.harvard.edu/abs/2011MNRAS.416.1392G} {416, 1392}

\bibitem[\protect\citeauthoryear{{Gruen}, {Seitz}, {Becker}, {Friedrich}  \&
  {Mana}}{{Gruen} et~al.}{2015}]{2015MNRAS.449.4264G}
{Gruen} D.,  {Seitz} S.,  {Becker} M.~R.,  {Friedrich} O.,   {Mana} A.,  2015,
  \mn@doi [\mnras] {10.1093/mnras/stv532}, \href
  {https://ui.adsabs.harvard.edu/abs/2015MNRAS.449.4264G} {449, 4264}

\bibitem[\protect\citeauthoryear{{Hamana}, {Miyazaki}, {Okura}, {Okamura}  \&
  {Futamase}}{{Hamana} et~al.}{2013}]{2013PASJ...65..104H}
{Hamana} T.,  {Miyazaki} S.,  {Okura} Y.,  {Okamura} T.,   {Futamase} T.,
  2013, \mn@doi [\pasj] {10.1093/pasj/65.5.104}, \href
  {https://ui.adsabs.harvard.edu/abs/2013PASJ...65..104H} {65, 104}

\bibitem[\protect\citeauthoryear{Harris et~al.,}{Harris
  et~al.}{2020}]{2020NumPy-Array}
Harris C.~R.,  et~al., 2020, \mn@doi [Nature] {10.1038/s41586-020-2649-2}, 585,
  357–362

\bibitem[\protect\citeauthoryear{{Heymans} et~al.,}{{Heymans}
  et~al.}{2012}]{2012MNRAS.427..146H}
{Heymans} C.,  et~al., 2012, \mn@doi [mnras]
  {10.1111/j.1365-2966.2012.21952.x}, \href
  {http://adsabs.harvard.edu/abs/2012MNRAS.427..146H} {427, 146}

\bibitem[\protect\citeauthoryear{{Hikage} et~al.,}{{Hikage}
  et~al.}{2019}]{2019PASJ...71...43H}
{Hikage} C.,  et~al., 2019, \mn@doi [\pasj] {10.1093/pasj/psz010}, \href
  {https://ui.adsabs.harvard.edu/abs/2019PASJ...71...43H} {71, 43}

\bibitem[\protect\citeauthoryear{{Hildebrandt} et~al.,}{{Hildebrandt}
  et~al.}{2012}]{2012MNRAS.421.2355H}
{Hildebrandt} H.,  et~al., 2012, \mn@doi [mnras]
  {10.1111/j.1365-2966.2012.20468.x}, \href
  {http://adsabs.harvard.edu/abs/2012MNRAS.421.2355H} {421, 2355}

\bibitem[\protect\citeauthoryear{{Hildebrandt} et~al.,}{{Hildebrandt}
  et~al.}{2017}]{2017MNRAS.465.1454H}
{Hildebrandt} H.,  et~al., 2017, \mn@doi [mnras] {10.1093/mnras/stw2805}, \href
  {http://adsabs.harvard.edu/abs/2017MNRAS.465.1454H} {465, 1454}

\bibitem[\protect\citeauthoryear{{Hinshaw} et~al.,}{{Hinshaw}
  et~al.}{2013}]{2013ApJS..208...19H}
{Hinshaw} G.,  et~al., 2013, \mn@doi [\apjs] {10.1088/0067-0049/208/2/19},
  \href {https://ui.adsabs.harvard.edu/abs/2013ApJS..208...19H} {208, 19}

\bibitem[\protect\citeauthoryear{{Hoekstra}}{{Hoekstra}}{2001}]{2001A&A...370..743H}
{Hoekstra} H.,  2001, \mn@doi [\aap] {10.1051/0004-6361:20010293}, \href
  {https://ui.adsabs.harvard.edu/abs/2001A&A...370..743H} {370, 743}

\bibitem[\protect\citeauthoryear{{Hoekstra}}{{Hoekstra}}{2003}]{2003MNRAS.339.1155H}
{Hoekstra} H.,  2003, \mn@doi [\mnras] {10.1046/j.1365-8711.2003.06264.x},
  \href {https://ui.adsabs.harvard.edu/abs/2003MNRAS.339.1155H} {339, 1155}

\bibitem[\protect\citeauthoryear{{Hoekstra}, {Donahue}, {Conselice}, {McNamara}
   \& {Voit}}{{Hoekstra} et~al.}{2011}]{2011ApJ...726...48H}
{Hoekstra} H.,  {Donahue} M.,  {Conselice} C.~J.,  {McNamara} B.~R.,   {Voit}
  G.~M.,  2011, \mn@doi [apj] {10.1088/0004-637X/726/1/48}, \href
  {http://adsabs.harvard.edu/abs/2011ApJ...726...48H} {726, 48}

\bibitem[\protect\citeauthoryear{Hunter}{Hunter}{2007}]{Hunter:2007}
Hunter J.~D.,  2007, \mn@doi [Computing in Science \& Engineering]
  {10.1109/MCSE.2007.55}, 9, 90

\bibitem[\protect\citeauthoryear{{Ilbert} et~al.,}{{Ilbert}
  et~al.}{2009}]{2009ApJ...690.1236I}
{Ilbert} O.,  et~al., 2009, \mn@doi [apj] {10.1088/0004-637X/690/2/1236}, \href
  {http://adsabs.harvard.edu/abs/2009ApJ...690.1236I} {690, 1236}

\bibitem[\protect\citeauthoryear{{Ishiyama} et~al.,}{{Ishiyama}
  et~al.}{2020}]{2020arXiv200714720I}
{Ishiyama} T.,  et~al., 2020, arXiv e-prints, \href
  {https://ui.adsabs.harvard.edu/abs/2020arXiv200714720I} {p. arXiv:2007.14720}

\bibitem[\protect\citeauthoryear{{Johnston} et~al.,}{{Johnston}
  et~al.}{2007a}]{2007arXiv0709.1159J}
{Johnston} D.~E.,  et~al., 2007a, arXiv e-prints, \href
  {https://ui.adsabs.harvard.edu/abs/2007arXiv0709.1159J} {p. arXiv:0709.1159}

\bibitem[\protect\citeauthoryear{{Johnston}, {Sheldon}, {Tasitsiomi},
  {Frieman}, {Wechsler}  \& {McKay}}{{Johnston}
  et~al.}{2007b}]{2007ApJ...656...27J}
{Johnston} D.~E.,  {Sheldon} E.~S.,  {Tasitsiomi} A.,  {Frieman} J.~A.,
  {Wechsler} R.~H.,   {McKay} T.~A.,  2007b, \mn@doi [apj] {10.1086/510060},
  \href {http://adsabs.harvard.edu/abs/2007ApJ...656...27J} {656, 27}

\bibitem[\protect\citeauthoryear{{Kaiser}}{{Kaiser}}{1984}]{1984ApJ...284L...9K}
{Kaiser} N.,  1984, \mn@doi [\apjl] {10.1086/184341}, \href
  {https://ui.adsabs.harvard.edu/abs/1984ApJ...284L...9K} {284, L9}

\bibitem[\protect\citeauthoryear{{Kilbinger}}{{Kilbinger}}{2015}]{2015RPPh...78h6901K}
{Kilbinger} M.,  2015, \mn@doi [Reports on Progress in Physics]
  {10.1088/0034-4885/78/8/086901}, \href
  {https://ui.adsabs.harvard.edu/abs/2015RPPh...78h6901K} {78, 086901}

\bibitem[\protect\citeauthoryear{{Kitching}, {Miller}, {Heymans}, {van
  Waerbeke}  \& {Heavens}}{{Kitching} et~al.}{2008}]{2008MNRAS.390..149K}
{Kitching} T.~D.,  {Miller} L.,  {Heymans} C.~E.,  {van Waerbeke} L.,
  {Heavens} A.~F.,  2008, \mn@doi [\mnras] {10.1111/j.1365-2966.2008.13628.x},
  \href {https://ui.adsabs.harvard.edu/abs/2008MNRAS.390..149K} {390, 149}

\bibitem[\protect\citeauthoryear{{Kotulla}, {Fritze}, {Weilbacher}  \&
  {Anders}}{{Kotulla} et~al.}{2009}]{2009MNRAS.396..462K}
{Kotulla} R.,  {Fritze} U.,  {Weilbacher} P.,   {Anders} P.,  2009, \mn@doi
  [\mnras] {10.1111/j.1365-2966.2009.14717.x}, \href
  {https://ui.adsabs.harvard.edu/abs/2009MNRAS.396..462K} {396, 462}

\bibitem[\protect\citeauthoryear{{Kuijken}}{{Kuijken}}{2011}]{2011Msngr.146....8K}
{Kuijken} K.,  2011, The Messenger, \href
  {http://adsabs.harvard.edu/abs/2011Msngr.146....8K} {146, 8}

\bibitem[\protect\citeauthoryear{{Kuijken} et~al.,}{{Kuijken}
  et~al.}{2015}]{2015MNRAS.454.3500K}
{Kuijken} K.,  et~al., 2015, \mn@doi [mnras] {10.1093/mnras/stv2140}, \href
  {http://adsabs.harvard.edu/abs/2015MNRAS.454.3500K} {454, 3500}

\bibitem[\protect\citeauthoryear{{Kuijken} et~al.,}{{Kuijken}
  et~al.}{2019}]{2019A&A...625A...2K}
{Kuijken} K.,  et~al., 2019, \mn@doi [\aap] {10.1051/0004-6361/201834918},
  \href {https://ui.adsabs.harvard.edu/abs/2019A&A...625A...2K} {625, A2}

\bibitem[\protect\citeauthoryear{{LSST Dark Energy Science
  Collaboration}}{{LSST Dark Energy Science
  Collaboration}}{2012}]{2012arXiv1211.0310L}
{LSST Dark Energy Science Collaboration} 2012, arXiv e-prints, \href
  {https://ui.adsabs.harvard.edu/abs/2012arXiv1211.0310L} {p. arXiv:1211.0310}

\bibitem[\protect\citeauthoryear{{Lesci} et~al.,}{{Lesci}
  et~al.}{2020}]{2020arXiv201212273L}
{Lesci} G.~F.,  et~al., 2020, arXiv e-prints, \href
  {https://ui.adsabs.harvard.edu/abs/2020arXiv201212273L} {p. arXiv:2012.12273}

\bibitem[\protect\citeauthoryear{Liang, Liaw, Nishihara, Moritz, Fox, Gonzalez,
  Goldberg  \& Stoica}{Liang et~al.}{2017}]{DBLP:journals/corr/abs-1712-09381}
Liang E.,  Liaw R.,  Nishihara R.,  Moritz P.,  Fox R.,  Gonzalez J.,  Goldberg
  K.,   Stoica I.,  2017, CoRR, abs/1712.09381

\bibitem[\protect\citeauthoryear{{Mandelbaum} et~al.,}{{Mandelbaum}
  et~al.}{2005}]{2005MNRAS.361.1287M}
{Mandelbaum} R.,  et~al., 2005, \mn@doi [\mnras]
  {10.1111/j.1365-2966.2005.09282.x}, \href
  {https://ui.adsabs.harvard.edu/abs/2005MNRAS.361.1287M} {361, 1287}

\bibitem[\protect\citeauthoryear{{Mandelbaum}, {Slosar}, {Baldauf}, {Seljak},
  {Hirata}, {Nakajima}, {Reyes}  \& {Smith}}{{Mandelbaum}
  et~al.}{2013}]{2013MNRAS.432.1544M}
{Mandelbaum} R.,  {Slosar} A.,  {Baldauf} T.,  {Seljak} U.,  {Hirata} C.~M.,
  {Nakajima} R.,  {Reyes} R.,   {Smith} R.~E.,  2013, \mn@doi [mnras]
  {10.1093/mnras/stt572}, \href
  {http://adsabs.harvard.edu/abs/2013MNRAS.432.1544M} {432, 1544}

\bibitem[\protect\citeauthoryear{{Maturi}, {Meneghetti}, {Bartelmann}, {Dolag}
  \& {Moscardini}}{{Maturi} et~al.}{2005}]{2005A&A...442..851M}
{Maturi} M.,  {Meneghetti} M.,  {Bartelmann} M.,  {Dolag} K.,   {Moscardini}
  L.,  2005, \mn@doi [aap] {10.1051/0004-6361:20042600}, \href
  {http://adsabs.harvard.edu/abs/2005A%26A...442..851M} {442, 851}

\bibitem[\protect\citeauthoryear{{Maturi}, {Bellagamba}, {Radovich},
  {Roncarelli}, {Sereno}, {Moscardini}, {Bardelli}  \& {Puddu}}{{Maturi}
  et~al.}{2019}]{2019MNRAS.485..498M}
{Maturi} M.,  {Bellagamba} F.,  {Radovich} M.,  {Roncarelli} M.,  {Sereno} M.,
  {Moscardini} L.,  {Bardelli} S.,   {Puddu} E.,  2019, \mn@doi [\mnras]
  {10.1093/mnras/stz294}, \href
  {https://ui.adsabs.harvard.edu/abs/2019MNRAS.485..498M} {485, 498}

\bibitem[\protect\citeauthoryear{{McClintock} et~al.,}{{McClintock}
  et~al.}{2019}]{2019MNRAS.482.1352M}
{McClintock} T.,  et~al., 2019, \mn@doi [\mnras] {10.1093/mnras/sty2711}, \href
  {https://ui.adsabs.harvard.edu/abs/2019MNRAS.482.1352M} {482, 1352}

\bibitem[\protect\citeauthoryear{{Medezinski} et~al.,}{{Medezinski}
  et~al.}{2007}]{2007ApJ...663..717M}
{Medezinski} E.,  et~al., 2007, \mn@doi [apj] {10.1086/518638}, \href
  {http://adsabs.harvard.edu/abs/2007ApJ...663..717M} {663, 717}

\bibitem[\protect\citeauthoryear{{Medezinski}, {Broadhurst}, {Umetsu}, {Oguri},
  {Rephaeli}  \& {Ben{'{i}}tez}}{{Medezinski}
  et~al.}{2010}]{2010MNRAS.405..257M}
{Medezinski} E.,  {Broadhurst} T.,  {Umetsu} K.,  {Oguri} M.,  {Rephaeli} Y.,
  {Ben{'{i}}tez} N.,  2010, \mn@doi [mnras] {10.1111/j.1365-2966.2010.16491.x},
  \href {http://adsabs.harvard.edu/abs/2010MNRAS.405..257M} {405, 257}

\bibitem[\protect\citeauthoryear{{Melchior} et~al.,}{{Melchior}
  et~al.}{2017}]{2017MNRAS.469.4899M}
{Melchior} P.,  et~al., 2017, \mn@doi [mnras] {10.1093/mnras/stx1053}, \href
  {https://ui.adsabs.harvard.edu/\#abs/2017MNRAS.469.4899M} {469, 4899}

\bibitem[\protect\citeauthoryear{{Meneghetti}, {Rasia}, {Merten}, {Bellagamba},
  {Ettori}, {Mazzotta}, {Dolag}  \& {Marri}}{{Meneghetti}
  et~al.}{2010}]{2010A&A...514A..93M}
{Meneghetti} M.,  {Rasia} E.,  {Merten} J.,  {Bellagamba} F.,  {Ettori} S.,
  {Mazzotta} P.,  {Dolag} K.,   {Marri} S.,  2010, \mn@doi [\aap]
  {10.1051/0004-6361/200913222}, \href
  {https://ui.adsabs.harvard.edu/abs/2010A&A...514A..93M} {514, A93}

\bibitem[\protect\citeauthoryear{{Meneghetti} et~al.,}{{Meneghetti}
  et~al.}{2014}]{2014ApJ...797...34M}
{Meneghetti} M.,  et~al., 2014, \mn@doi [\apj] {10.1088/0004-637X/797/1/34},
  \href {https://ui.adsabs.harvard.edu/abs/2014ApJ...797...34M} {797, 34}

\bibitem[\protect\citeauthoryear{{Merten} et~al.,}{{Merten}
  et~al.}{2015a}]{merten15}
{Merten} J.,  et~al., 2015a, \mn@doi [\apj] {10.1088/0004-637X/806/1/4}, \href
  {http://adsabs.harvard.edu/abs/2015ApJ...806....4M} {806, 4}

\bibitem[\protect\citeauthoryear{{Merten} et~al.,}{{Merten}
  et~al.}{2015b}]{2015ApJ...806....4M}
{Merten} J.,  et~al., 2015b, \mn@doi [\apj] {10.1088/0004-637X/806/1/4}, \href
  {https://ui.adsabs.harvard.edu/abs/2015ApJ...806....4M} {806, 4}

\bibitem[\protect\citeauthoryear{{Metzler}, {White}  \& {Loken}}{{Metzler}
  et~al.}{2001}]{2001ApJ...547..560M}
{Metzler} C.~A.,  {White} M.,   {Loken} C.,  2001, \mn@doi [\apj]
  {10.1086/318406}, \href
  {https://ui.adsabs.harvard.edu/abs/2001ApJ...547..560M} {547, 560}

\bibitem[\protect\citeauthoryear{{Miller}, {Kitching}, {Heymans}, {Heavens}  \&
  {van Waerbeke}}{{Miller} et~al.}{2007}]{2007MNRAS.382..315M}
{Miller} L.,  {Kitching} T.~D.,  {Heymans} C.,  {Heavens} A.~F.,   {van
  Waerbeke} L.,  2007, \mn@doi [\mnras] {10.1111/j.1365-2966.2007.12363.x},
  \href {https://ui.adsabs.harvard.edu/\#abs/2007MNRAS.382..315M} {382, 315}

\bibitem[\protect\citeauthoryear{{Miller} et~al.,}{{Miller}
  et~al.}{2013}]{2013MNRAS.429.2858M}
{Miller} L.,  et~al., 2013, \mn@doi [\mnras] {10.1093/mnras/sts454}, \href
  {https://ui.adsabs.harvard.edu/\#abs/2013MNRAS.429.2858M} {429, 2858}

\bibitem[\protect\citeauthoryear{{Miyatake} et~al.,}{{Miyatake}
  et~al.}{2015}]{2015ApJ...806....1M}
{Miyatake} H.,  et~al., 2015, \mn@doi [\apj] {10.1088/0004-637X/806/1/1}, \href
  {https://ui.adsabs.harvard.edu/abs/2015ApJ...806....1M} {806, 1}

\bibitem[\protect\citeauthoryear{{Mo} \& {White}}{{Mo} \&
  {White}}{1996}]{1996MNRAS.282..347M}
{Mo} H.~J.,  {White} S.~D.~M.,  1996, \mn@doi [\mnras]
  {10.1093/mnras/282.2.347}, \href
  {https://ui.adsabs.harvard.edu/abs/1996MNRAS.282..347M} {282, 347}

\bibitem[\protect\citeauthoryear{{Mo}, {Jing}  \& {White}}{{Mo}
  et~al.}{1996}]{1996MNRAS.282.1096M}
{Mo} H.~J.,  {Jing} Y.~P.,   {White} S.~D.~M.,  1996, \mn@doi [\mnras]
  {10.1093/mnras/282.3.1096}, \href
  {https://ui.adsabs.harvard.edu/abs/1996MNRAS.282.1096M} {282, 1096}

\bibitem[\protect\citeauthoryear{Moritz et~al.,}{Moritz
  et~al.}{2017}]{DBLP:journals/corr/abs-1712-05889}
Moritz P.,  et~al., 2017, CoRR, abs/1712.05889

\bibitem[\protect\citeauthoryear{{Nanni}, {Marulli}  \& {Veropalumbo}}{{Nanni}
  et~al.}{prep}]{nanni}
{Nanni} F.,  {Marulli} L.~{Moscardini} F.,   {Veropalumbo} e.,  in prep., AMICO
  galaxy clusters in KiDS-DR3: constraints on cosmological parameters and on
  the mass-richness relation from the clustering of galaxy clusters, in
  preparation

\bibitem[\protect\citeauthoryear{{Navarro}, {Frenk}  \& {White}}{{Navarro}
  et~al.}{1996}]{1996ApJ...462..563N}
{Navarro} J.~F.,  {Frenk} C.~S.,   {White} S. D.~M.,  1996, \mn@doi [\apj]
  {10.1086/177173}, \href
  {https://ui.adsabs.harvard.edu/abs/1996ApJ...462..563N} {462, 563}

\bibitem[\protect\citeauthoryear{{Navarro}, {Frenk}  \& {White}}{{Navarro}
  et~al.}{1997}]{1997ApJ...490..493N}
{Navarro} J.~F.,  {Frenk} C.~S.,   {White} S. D.~M.,  1997, \mn@doi [\apj]
  {10.1086/304888}, \href
  {https://ui.adsabs.harvard.edu/abs/1997ApJ...490..493N} {490, 493}

\bibitem[\protect\citeauthoryear{{Neto} et~al.,}{{Neto} et~al.}{2007}]{neto07}
{Neto} A.~F.,  et~al., 2007, \mn@doi [\mnras]
  {10.1111/j.1365-2966.2007.12381.x}, \href
  {http://adsabs.harvard.edu/abs/2007MNRAS.381.1450N} {381, 1450}

\bibitem[\protect\citeauthoryear{{Oguri} \& {Hamana}}{{Oguri} \&
  {Hamana}}{2011}]{2011MNRAS.414.1851O}
{Oguri} M.,  {Hamana} T.,  2011, \mn@doi [\mnras]
  {10.1111/j.1365-2966.2011.18481.x}, \href
  {https://ui.adsabs.harvard.edu/abs/2011MNRAS.414.1851O} {414, 1851}

\bibitem[\protect\citeauthoryear{Oguri, Takada, Umetsu  \& Broadhurst}{Oguri
  et~al.}{2005}]{Oguri_2005}
Oguri M.,  Takada M.,  Umetsu K.,   Broadhurst T.,  2005, \mn@doi [The
  Astrophysical Journal] {10.1086/452629}, 632, 841

\bibitem[\protect\citeauthoryear{{Oguri} et~al.,}{{Oguri}
  et~al.}{2009}]{2009ApJ...699.1038O}
{Oguri} M.,  et~al., 2009, \mn@doi [\apj] {10.1088/0004-637X/699/2/1038}, \href
  {https://ui.adsabs.harvard.edu/abs/2009ApJ...699.1038O} {699, 1038}

\bibitem[\protect\citeauthoryear{{Oguri}, {Bayliss}, {Dahle}, {Sharon},
  {Gladders}, {Natarajan}, {Hennawi}  \& {Koester}}{{Oguri}
  et~al.}{2012}]{2012MNRAS.420.3213O}
{Oguri} M.,  {Bayliss} M.~B.,  {Dahle} H.,  {Sharon} K.,  {Gladders} M.~D.,
  {Natarajan} P.,  {Hennawi} J.~F.,   {Koester} B.~P.,  2012, \mn@doi [\mnras]
  {10.1111/j.1365-2966.2011.20248.x}, \href
  {https://ui.adsabs.harvard.edu/abs/2012MNRAS.420.3213O} {420, 3213}

\bibitem[\protect\citeauthoryear{{Peebles}}{{Peebles}}{1980}]{1980lssu.book.....P}
{Peebles} P.~J.~E.,  1980, {The large-scale structure of the universe}.
Princeton University Press

\bibitem[\protect\citeauthoryear{{Peebles}}{{Peebles}}{1993}]{1993ppc..book.....P}
{Peebles} P.~J.~E.,  1993, {Principles of Physical Cosmology}.
Princeton University Press

\bibitem[\protect\citeauthoryear{{Pen}, {Lu}, {van Waerbeke}  \&
  {Mellier}}{{Pen} et~al.}{2003}]{2003MNRAS.346..994P}
{Pen} U.-L.,  {Lu} T.,  {van Waerbeke} L.,   {Mellier} Y.,  2003, \mn@doi
  [\mnras] {10.1111/j.1365-2966.2003.07152.x}, \href
  {https://ui.adsabs.harvard.edu/abs/2003MNRAS.346..994P} {346, 994}

\bibitem[\protect\citeauthoryear{{Planck Collaboration} et~al.,}{{Planck
  Collaboration} et~al.}{2020}]{2020A&A...641A...6P}
{Planck Collaboration} et~al., 2020, \mn@doi [\aap]
  {10.1051/0004-6361/201833910}, \href
  {https://ui.adsabs.harvard.edu/abs/2020A&A...641A...6P} {641, A6}

\bibitem[\protect\citeauthoryear{{Postman} et~al.,}{{Postman}
  et~al.}{2012}]{2012ApJS..199...25P}
{Postman} M.,  et~al., 2012, \mn@doi [\apjs] {10.1088/0067-0049/199/2/25},
  \href {https://ui.adsabs.harvard.edu/abs/2012ApJS..199...25P} {199, 25}

\bibitem[\protect\citeauthoryear{{Press} \& {Schechter}}{{Press} \&
  {Schechter}}{1974}]{PS...1974ApJ...187..425P}
{Press} W.~H.,  {Schechter} P.,  1974, \mn@doi [\apj] {10.1086/152650}, \href
  {https://ui.adsabs.harvard.edu/abs/1974ApJ...187..425P} {187, 425}

\bibitem[\protect\citeauthoryear{{Radovich} et~al.,}{{Radovich}
  et~al.}{2017}]{2017A&A...598A.107R}
{Radovich} M.,  et~al., 2017, \mn@doi [aap] {10.1051/0004-6361/201629353},
  \href {https://ui.adsabs.harvard.edu/#abs/2017A&A...598A.107R} {598, A107}

\bibitem[\protect\citeauthoryear{{Rasia}, {Tormen}  \& {Moscardini}}{{Rasia}
  et~al.}{2004}]{rasia04}
{Rasia} E.,  {Tormen} G.,   {Moscardini} L.,  2004, \mn@doi [\mnras]
  {10.1111/j.1365-2966.2004.07775.x}, \href
  {http://adsabs.harvard.edu/abs/2004MNRAS.351..237R} {351, 237}

\bibitem[\protect\citeauthoryear{{Rasia} et~al.,}{{Rasia}
  et~al.}{2006}]{rasia06}
{Rasia} E.,  et~al., 2006, \mn@doi [\mnras] {10.1111/j.1365-2966.2006.10466.x},
  \href {http://adsabs.harvard.edu/abs/2006MNRAS.369.2013R} {369, 2013}

\bibitem[\protect\citeauthoryear{{Rykoff} et~al.,}{{Rykoff}
  et~al.}{2016}]{2016ApJS..224....1R}
{Rykoff} E.~S.,  et~al., 2016, \mn@doi [\apjs] {10.3847/0067-0049/224/1/1},
  \href {https://ui.adsabs.harvard.edu/abs/2016ApJS..224....1R} {224, 1}

\bibitem[\protect\citeauthoryear{Schneider}{Schneider}{2006}]{Schneider:2005ka}
Schneider P.,  2006, in {Proceedings, 33rd Advanced Saas Fee Course on
  Gravitational Lensing: Strong, Weak, and Micro: Les Diablerets, Switzerland,
  April 7-12, 2003}. pp 269--451 (\mn@eprint {arXiv} {astro-ph/0509252}),
  \mn@doi{10.1007/978-3-540-30310-7_3}

\bibitem[\protect\citeauthoryear{{Seitz} \& {Schneider}}{{Seitz} \&
  {Schneider}}{1997}]{1997A&A...318..687S}
{Seitz} C.,  {Schneider} P.,  1997, \aap, \href
  {https://ui.adsabs.harvard.edu/abs/1997A%26A...318..687S} {318, 687}

\bibitem[\protect\citeauthoryear{{Seljak} \& {Warren}}{{Seljak} \&
  {Warren}}{2004}]{2004MNRAS.355..129S}
{Seljak} U.,  {Warren} M.~S.,  2004, \mn@doi [\mnras]
  {10.1111/j.1365-2966.2004.08297.x}, \href
  {https://ui.adsabs.harvard.edu/abs/2004MNRAS.355..129S} {355, 129}

\bibitem[\protect\citeauthoryear{{Seljak} et~al.,}{{Seljak}
  et~al.}{2005}]{2005PhRvD..71d3511S}
{Seljak} U.,  et~al., 2005, \mn@doi [\prd] {10.1103/PhysRevD.71.043511}, \href
  {https://ui.adsabs.harvard.edu/abs/2005PhRvD..71d3511S} {71, 043511}

\bibitem[\protect\citeauthoryear{{Sereno} \& {Covone}}{{Sereno} \&
  {Covone}}{2013}]{2013MNRAS.434..878S}
{Sereno} M.,  {Covone} G.,  2013, \mn@doi [\mnras] {10.1093/mnras/stt1086},
  \href {https://ui.adsabs.harvard.edu/abs/2013MNRAS.434..878S} {434, 878}

\bibitem[\protect\citeauthoryear{{Sereno} \& {Ettori}}{{Sereno} \&
  {Ettori}}{2015}]{2015MNRAS.450.3675S}
{Sereno} M.,  {Ettori} S.,  2015, \mn@doi [mnras] {10.1093/mnras/stv814}, \href
  {http://adsabs.harvard.edu/abs/2015MNRAS.450.3675S} {450, 3675}

\bibitem[\protect\citeauthoryear{{Sereno} \& {Umetsu}}{{Sereno} \&
  {Umetsu}}{2011}]{2011MNRAS.416.3187S}
{Sereno} M.,  {Umetsu} K.,  2011, \mn@doi [\mnras]
  {10.1111/j.1365-2966.2011.19274.x}, \href
  {https://ui.adsabs.harvard.edu/abs/2011MNRAS.416.3187S} {416, 3187}

\bibitem[\protect\citeauthoryear{{Sereno}, {Giocoli}, {Ettori}  \&
  {Moscardini}}{{Sereno} et~al.}{2015a}]{2015MNRAS.449.2024S}
{Sereno} M.,  {Giocoli} C.,  {Ettori} S.,   {Moscardini} L.,  2015a, \mn@doi
  [\mnras] {10.1093/mnras/stv416}, \href
  {https://ui.adsabs.harvard.edu/abs/2015MNRAS.449.2024S} {449, 2024}

\bibitem[\protect\citeauthoryear{{Sereno}, {Veropalumbo}, {Marulli}, {Covone},
  {Moscardini}  \& {Cimatti}}{{Sereno} et~al.}{2015b}]{2015MNRAS.449.4147S}
{Sereno} M.,  {Veropalumbo} A.,  {Marulli} F.,  {Covone} G.,  {Moscardini} L.,
   {Cimatti} A.,  2015b, \mn@doi [\mnras] {10.1093/mnras/stv280}, \href
  {https://ui.adsabs.harvard.edu/abs/2015MNRAS.449.4147S} {449, 4147}

\bibitem[\protect\citeauthoryear{{Sereno}, {Covone}, {Izzo}, {Ettori}, {Coupon}
   \& {Lieu}}{{Sereno} et~al.}{2017}]{2017MNRAS.472.1946S}
{Sereno} M.,  {Covone} G.,  {Izzo} L.,  {Ettori} S.,  {Coupon} J.,   {Lieu} M.,
   2017, \mn@doi [\mnras] {10.1093/mnras/stx2085}, \href
  {https://ui.adsabs.harvard.edu/abs/2017MNRAS.472.1946S} {472, 1946}

\bibitem[\protect\citeauthoryear{{Sereno} et~al.,}{{Sereno}
  et~al.}{2018}]{2018NatAs...2..744S}
{Sereno} M.,  et~al., 2018, \mn@doi [Nature Astronomy]
  {10.1038/s41550-018-0508-y}, \href
  {https://ui.adsabs.harvard.edu/abs/2018NatAs...2..744S} {2, 744}

\bibitem[\protect\citeauthoryear{{Sereno} et~al.,}{{Sereno}
  et~al.}{2020}]{2020MNRAS.497..894S}
{Sereno} M.,  et~al., 2020, \mn@doi [\mnras] {10.1093/mnras/staa1902}, \href
  {https://ui.adsabs.harvard.edu/abs/2020MNRAS.497..894S} {497, 894}

\bibitem[\protect\citeauthoryear{{Sheldon} et~al.,}{{Sheldon}
  et~al.}{2004}]{2004AJ....127.2544S}
{Sheldon} E.~S.,  et~al., 2004, \mn@doi [aj] {10.1086/383293}, \href
  {http://adsabs.harvard.edu/abs/2004AJ....127.2544S} {127, 2544}

\bibitem[\protect\citeauthoryear{{Sheth} \& {Tormen}}{{Sheth} \&
  {Tormen}}{1999}]{1999MNRAS.308..119S}
{Sheth} R.~K.,  {Tormen} G.,  1999, \mn@doi [\mnras]
  {10.1046/j.1365-8711.1999.02692.x}, \href
  {https://ui.adsabs.harvard.edu/abs/1999MNRAS.308..119S} {308, 119}

\bibitem[\protect\citeauthoryear{{Sheth}, {Mo}  \& {Tormen}}{{Sheth}
  et~al.}{2001}]{2001MNRAS.323....1S}
{Sheth} R.~K.,  {Mo} H.~J.,   {Tormen} G.,  2001, \mn@doi [\mnras]
  {10.1046/j.1365-8711.2001.04006.x}, \href
  {https://ui.adsabs.harvard.edu/abs/2001MNRAS.323....1S} {323, 1}

\bibitem[\protect\citeauthoryear{{Simet}, {McClintock}, {Mandelbaum}, {Rozo},
  {Rykoff}, {Sheldon}  \& {Wechsler}}{{Simet}
  et~al.}{2017}]{2017MNRAS.466.3103S}
{Simet} M.,  {McClintock} T.,  {Mandelbaum} R.,  {Rozo} E.,  {Rykoff} E.,
  {Sheldon} E.,   {Wechsler} R.~H.,  2017, \mn@doi [mnras]
  {10.1093/mnras/stw3250}, \href
  {http://adsabs.harvard.edu/abs/2017MNRAS.466.3103S} {466, 3103}

\bibitem[\protect\citeauthoryear{{Singh}, {Mandelbaum}, {Seljak}, {Slosar}  \&
  {Vazquez Gonzalez}}{{Singh} et~al.}{2017}]{2017MNRAS.471.3827S}
{Singh} S.,  {Mandelbaum} R.,  {Seljak} U.,  {Slosar} A.,   {Vazquez Gonzalez}
  J.,  2017, \mn@doi [\mnras] {10.1093/mnras/stx1828}, \href
  {https://ui.adsabs.harvard.edu/abs/2017MNRAS.471.3827S} {471, 3827}

\bibitem[\protect\citeauthoryear{{Skibba} \& {Macci{\`o}}}{{Skibba} \&
  {Macci{\`o}}}{2011}]{2011MNRAS.416.2388S}
{Skibba} R.~A.,  {Macci{\`o}} A.~V.,  2011, \mn@doi [\mnras]
  {10.1111/j.1365-2966.2011.19218.x}, \href
  {https://ui.adsabs.harvard.edu/abs/2011MNRAS.416.2388S} {416, 2388}

\bibitem[\protect\citeauthoryear{{Springel}, {White}, {Tormen}  \&
  {Kauffmann}}{{Springel} et~al.}{2001}]{springel01b}
{Springel} V.,  {White} S.~D.~M.,  {Tormen} G.,   {Kauffmann} G.,  2001,
  \mn@doi [\mnras] {10.1046/j.1365-8711.2001.04912.x}, \href
  {http://adsabs.harvard.edu/abs/2001MNRAS.328..726S} {328, 726}

\bibitem[\protect\citeauthoryear{{Takada} \& {Jain}}{{Takada} \&
  {Jain}}{2003}]{2003MNRAS.340..580T}
{Takada} M.,  {Jain} B.,  2003, \mn@doi [\mnras]
  {10.1046/j.1365-8711.2003.06321.x}, \href
  {https://ui.adsabs.harvard.edu/abs/2003MNRAS.340..580T} {340, 580}

\bibitem[\protect\citeauthoryear{{Takahashi}, {Sato}, {Nishimichi}, {Taruya}
  \& {Oguri}}{{Takahashi} et~al.}{2012}]{2012ApJ...761..152T}
{Takahashi} R.,  {Sato} M.,  {Nishimichi} T.,  {Taruya} A.,   {Oguri} M.,
  2012, \mn@doi [\apj] {10.1088/0004-637X/761/2/152}, \href
  {https://ui.adsabs.harvard.edu/abs/2012ApJ...761..152T} {761, 152}

\bibitem[\protect\citeauthoryear{{Taylor}}{{Taylor}}{2005}]{2005ASPC..347...29T}
{Taylor} M.~B.,  2005, in {Shopbell} P.,  {Britton} M.,   {Ebert} R.,  eds,
  Astronomical Society of the Pacific Conference Series Vol. 347, Astronomical
  Data Analysis Software and Systems XIV. p.~29

\bibitem[\protect\citeauthoryear{{Tinker}, {Weinberg}, {Zheng}  \&
  {Zehavi}}{{Tinker} et~al.}{2005}]{2005ApJ...631...41T}
{Tinker} J.~L.,  {Weinberg} D.~H.,  {Zheng} Z.,   {Zehavi} I.,  2005, \mn@doi
  [\apj] {10.1086/432084}, \href
  {https://ui.adsabs.harvard.edu/abs/2005ApJ...631...41T} {631, 41}

\bibitem[\protect\citeauthoryear{{Tinker}, {Robertson}, {Kravtsov}, {Klypin},
  {Warren}, {Yepes}  \& {Gottl{\"o}ber}}{{Tinker}
  et~al.}{2010}]{2010ApJ...724..878T}
{Tinker} J.~L.,  {Robertson} B.~E.,  {Kravtsov} A.~V.,  {Klypin} A.,  {Warren}
  M.~S.,  {Yepes} G.,   {Gottl{\"o}ber} S.,  2010, \mn@doi [\apj]
  {10.1088/0004-637X/724/2/878}, \href
  {https://ui.adsabs.harvard.edu/abs/2010ApJ...724..878T} {724, 878}

\bibitem[\protect\citeauthoryear{{Tormen}}{{Tormen}}{1998}]{tormen98a}
{Tormen} G.,  1998, \mnras, \href
  {http://adsabs.harvard.edu/abs/1998MNRAS.297..648T} {297, 648}

\bibitem[\protect\citeauthoryear{{Umetsu}, {Broadhurst}, {Zitrin},
  {Medezinski}, {Coe}  \& {Postman}}{{Umetsu}
  et~al.}{2011}]{2011ApJ...738...41U}
{Umetsu} K.,  {Broadhurst} T.,  {Zitrin} A.,  {Medezinski} E.,  {Coe} D.,
  {Postman} M.,  2011, \mn@doi [\apj] {10.1088/0004-637X/738/1/41}, \href
  {https://ui.adsabs.harvard.edu/abs/2011ApJ...738...41U} {738, 41}

\bibitem[\protect\citeauthoryear{{Umetsu} et~al.,}{{Umetsu}
  et~al.}{2014}]{2014ApJ...795..163U}
{Umetsu} K.,  et~al., 2014, \mn@doi [\apj] {10.1088/0004-637X/795/2/163}, \href
  {https://ui.adsabs.harvard.edu/abs/2014ApJ...795..163U} {795, 163}

\bibitem[\protect\citeauthoryear{{Viola} et~al.,}{{Viola}
  et~al.}{2015}]{2015MNRAS.452.3529V}
{Viola} M.,  et~al., 2015, \mn@doi [\mnras] {10.1093/mnras/stv1447}, \href
  {https://ui.adsabs.harvard.edu/abs/2015MNRAS.452.3529V} {452, 3529}

\bibitem[\protect\citeauthoryear{Virtanen et~al.,}{Virtanen
  et~al.}{2020}]{2020SciPy-NMeth}
Virtanen P.,  et~al., 2020, \mn@doi [Nature Methods]
  {10.1038/s41592-019-0686-2}, \href {https://rdcu.be/b08Wh} {17, 261}

\bibitem[\protect\citeauthoryear{{Voit}}{{Voit}}{2005}]{2005RvMP...77..207V}
{Voit} G.~M.,  2005, \mn@doi [Reviews of Modern Physics]
  {10.1103/RevModPhys.77.207}, \href
  {https://ui.adsabs.harvard.edu/abs/2005RvMP...77..207V} {77, 207}

\bibitem[\protect\citeauthoryear{{White} \& {Rees}}{{White} \&
  {Rees}}{1978}]{1978MNRAS.183..341W}
{White} S.~D.~M.,  {Rees} M.~J.,  1978, \mn@doi [\mnras]
  {10.1093/mnras/183.3.341}, \href
  {https://ui.adsabs.harvard.edu/abs/1978MNRAS.183..341W} {183, 341}

\bibitem[\protect\citeauthoryear{Yang, Mo, Van Den~Bosch, Jing, Weinmann  \&
  Meneghetti}{Yang et~al.}{2006}]{doi:10.1111/j.1365-2966.2006.11091.x}
Yang X.,  Mo H.~J.,  Van Den~Bosch F.~C.,  Jing Y.~P.,  Weinmann S.~M.,
  Meneghetti M.,  2006, \mn@doi [Monthly Notices of the Royal Astronomical
  Society] {10.1111/j.1365-2966.2006.11091.x}, 373, 1159

\bibitem[\protect\citeauthoryear{{Zitrin}, {Broadhurst}, {Barkana}, {Rephaeli}
  \& {Ben{\'{\i}}tez}}{{Zitrin} et~al.}{2011a}]{zitrin11a}
{Zitrin} A.,  {Broadhurst} T.,  {Barkana} R.,  {Rephaeli} Y.,
  {Ben{\'{\i}}tez} N.,  2011a, \mn@doi [\mnras]
  {10.1111/j.1365-2966.2010.17574.x}, \href
  {http://adsabs.harvard.edu/abs/2011MNRAS.410.1939Z} {410, 1939}

\bibitem[\protect\citeauthoryear{{Zitrin}, {Broadhurst}, {Coe}, {Umetsu},
  {Postman}, {Ben{\'{\i}}tez}, {Meneghetti}  \& et al.}{{Zitrin}
  et~al.}{2011b}]{zitrin11b}
{Zitrin} A.,  {Broadhurst} T.,  {Coe} D.,  {Umetsu} K.,  {Postman} M.,
  {Ben{\'{\i}}tez} N.,  {Meneghetti} M.,   et al. M.,  2011b, \mn@doi [\apj]
  {10.1088/0004-637X/742/2/117}, \href
  {http://adsabs.harvard.edu/abs/2011ApJ...742..117Z} {742, 117}

\bibitem[\protect\citeauthoryear{{de Jong}, {Verdoes Kleijn}, {Kuijken}  \&
  {Valentijn}}{{de Jong} et~al.}{2013}]{2013ExA....35...25D}
{de Jong} J.~T.~A.,  {Verdoes Kleijn} G.~A.,  {Kuijken} K.~H.,   {Valentijn}
  E.~A.,  2013, \mn@doi [Experimental Astronomy] {10.1007/s10686-012-9306-1},
  \href {http://adsabs.harvard.edu/abs/2013ExA....35...25D} {35, 25}

\bibitem[\protect\citeauthoryear{{de Jong} et~al.,}{{de Jong}
  et~al.}{2015}]{2015A&A...582A..62D}
{de Jong} J. T.~A.,  et~al., 2015, \mn@doi [aap] {10.1051/0004-6361/201526601},
  \href {https://ui.adsabs.harvard.edu/\#abs/2015A&A...582A..62D} {582, A62}

\bibitem[\protect\citeauthoryear{{de Jong} et~al.,}{{de Jong}
  et~al.}{2017}]{2017A&A...604A.134D}
{de Jong} J. T.~A.,  et~al., 2017, \mn@doi [aap] {10.1051/0004-6361/201730747},
  \href {https://ui.adsabs.harvard.edu/\#abs/2017A&A...604A.134D} {604, A134}

\bibitem[\protect\citeauthoryear{{van Uitert}, {Gilbank}, {Hoekstra},
  {Semboloni}, {Gladders}  \& {Yee}}{{van Uitert}
  et~al.}{2016}]{2016A&A...586A..43V}
{van Uitert} E.,  {Gilbank} D.~G.,  {Hoekstra} H.,  {Semboloni} E.,  {Gladders}
  M.~D.,   {Yee} H. K.~C.,  2016, \mn@doi [\aap] {10.1051/0004-6361/201526719},
  \href {https://ui.adsabs.harvard.edu/abs/2016A&A...586A..43V} {586, A43}

\makeatother
\end{thebibliography}


\bsp	
\label{lastpage}
\end{document}